\documentclass[aps,preprintnumbers,amsmath,amssymb,citeautoscript,superscriptaddress,a4]{article}
\usepackage[dvips]{graphics}
\usepackage{graphicx}

\begin{document}

\title{Order in 2D nodal superconductors}

\author{Maria Hermanns}
\date{March 21, 2006}






\maketitle
\vfill

\begin{figure}[h]
\begin{center}
\includegraphics[angle=0,width=0.20\textwidth]{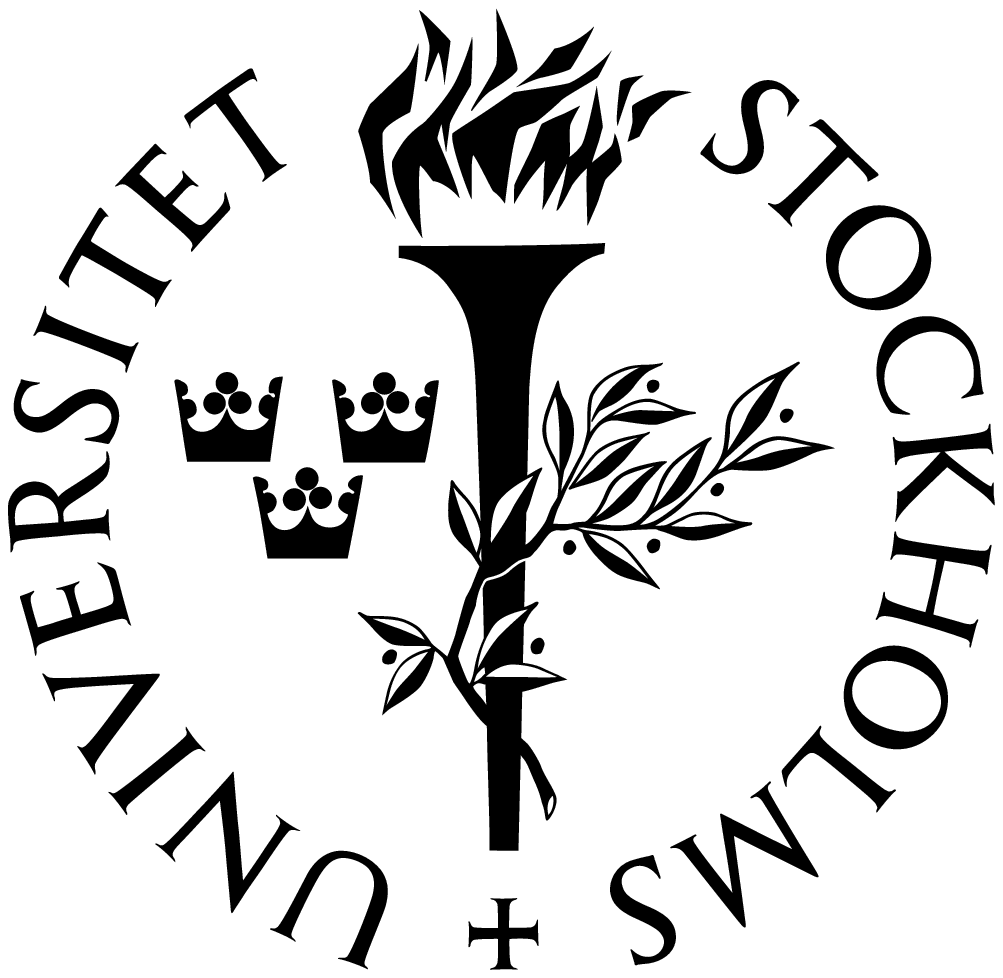}
\end{center}
\end{figure}

\vspace{0.2cm}
\normalsize
\noindent
\begin{center}
Stockholm University\\
Department of Physics\\
2006
\end{center}

\newpage

\begin{abstract}
A topological theory of d-wave superconductors is derived in this thesis. Ginzburg-Landau theory describes superconductivity by defining a complex order parameter and applying Landau's theory for phase transitions. However, there is no local, gauge invariant order parameter for a superconductor and classical order is no appropriate description. Topological order has proven to be a powerful tool for describing the Quantum Hall effect and it gives also an appropriate description of s-wave superconductors. For d-wave superconductors there are gapless excitations at four points on the Fermi surface. The topological theory for superconductors exhibits a similar structure for the ground state as found in the s-wave case. However, the gapless excitations destroy the topological degeneracy of the ground state and introduce an additional degeneracy in one of the flux sectors. In search for regularities which reflect the topological order, I include a  magnetic point impurity and focus on the effects it has on the Casimir energy. Indeed, it is shown that the energy shift reflects the topology to some extend. However, this is work in progress and further computations will be done. 
\end{abstract}

\newpage

\section*{Acknowledgments}

I would like to thank my supervisor Hans Hansson. He always found time for interesting discussions and has helped and guided me a lot during this project. During the last year I got introduced to an interesting field of physics that I didn't know much about. 
In addition, I would like to thank  Prof. Dr. Matthias Vojta for being my co-supervisor and, thus, making it possible for me to prolong my stay at Stockholm University.  
I also thank everyone in the Quantum and Field Theory group and in the Cosmology, Particle Astrophysics and String theory group for a lot of interesting discussions (not just) about physics and for creating such a nice working climate. 

\pagebreak


\newcommand{\be}[1]{\begin{eqnarray} \mbox{$\label{#1}$}   }



\newcommand{\ee}{\end{eqnarray}}
\newcommand{\pref}[1]{(\ref{#1})}

\newcommand\ie {{\it i.e. }}
\newcommand\eg {{\it e.g. }}
\newcommand\etc{{\it etc. }}
\newcommand\etcp{{\it etc.. }}
\newcommand\cf {{\it cf.  }}
\newcommand\grad{\vec\nabla}
\newcommand\half{\frac 1 2 }
\newcommand\halfe{\frac{1}{2e} }
\newcommand\noi{\noindent}

\newcommand\ver {\vec r}
\newcommand\vri [1] {\vec r_#1}
\newcommand\vra [1] {\vec r_#1}
\newcommand \deltar {\Delta_R}

\newcommand\psii [2] {\psi_{#1 , #2} }
\newcommand\psiidag [2] {\psi_{#1,#2}^\dagger}
\newcommand\psiu[1] {\psi_{#1,\uparrow}}
\newcommand\psid [1] {\psi_{ #1,\downarrow}}
\newcommand\psidag [2] {\psi^\dagger_{#1 , #2} }
\newcommand\psiudag [1] {\psi^\dagger_{#1, \uparrow }}
\newcommand\psiddag [1] {\psi^\dagger_{#1. \downarrow }}
\newcommand \ofr {(\vec r)} 
\newcommand \ofri [1] {(\vec r_#1)}
\newcommand \ofrii [2] {(\vec r_#1, \vec r_#2)}
\newcommand \ai [1] {\frac{\bar a_#1}{L_#1}}
\newcommand \bi [1] {\frac{\bar b_#1}{L_#1}}
\newcommand \veq {\vec q}
\newcommand \ofq {(\vec q)}
\newcommand \zero {|0\rangle}
\newcommand \zerod {\langle0|}
\newcommand \Tr {\mbox{Tr}}
\newcommand \ad [1] {a_{#1}^\dagger(\vec k)}
\newcommand \bd [1] {b_{#1}^\dagger(\vec k)}
\newcommand \appsection[1] {\renewcommand
\thesection{Appendix \Alph{section}}}
\tableofcontents

\newpage

\section{Introduction}

\subsection{Classical and Topological Order}
In classical physics we can describe an N-body system by a probability distribution $P(r_1...r_N)$. If we change the external conditions a little, the distribution will typically change continuously. However, if we change too much it might cause a discontinuous change in the properties of the system, a phase transition. Ginzburg and Landau developed a very successful theory that describes phase transitions in terms of a (local) order parameter $\phi$, which is zero in one phase but finite in the other. Order is a property of a (broken) symmetry of the Hamiltonian. The ground state breaks one or more symmetries of the Hamiltonian. Each broken symmetry generates a gapless particle, which  is called Goldstone boson. There is, however, a loophole for this mechanism. In presence of a gauge field, the Goldstone boson will combine with the gauge particle and form a massive, spin-1 particle. This is known as the Higgs Mechanism. 

With help of the order parameter $\phi$, one can write an expansion of the 
Landau free energy in terms of $\phi$. This expansion will look the same (except for the coefficients) for systems with similar symmetries. Therefore, systems with similar symmetry properties have the same critical behavior and are said to be in the same universality class.   

Although this theory is very successful for classical phase transitions\footnote{However, even at the classical level symmetry breaking cannot describe all kinds of phase transitions, see for example the Kosterlitz-Thouless phase transition} it fails to describe the Quantum Hall Effect or charged superconductors\footnote{Although Ginzburg-Landau theory, in principle, does not qualify for superconductors, it  gives nevertheless an accurate description of for example the H-T phase diagram}. In both cases no local, gauge invariant order parameter can be found \cite{Hans}. Therefore, it is more suitable to describe these with topological orders to explain their characteristics.
Topological order is a property of the ground state wave function $\psi(\vra 1 ...\vra N)$. Classical order is a function of the probability distribution 
$P(\vra 1 \ldots \vra N) = |\psi(\vra 1 \ldots \vra N)|^2$ (at finite temperature). It misses the phase of the wave function, which is a quantum effect. Therefore, it can only partially describe the internal structure of the ground state. 

Topological order is robust against local perturbations as for example impurities. The ground state degeneracy for example is not a consequence
of a symmetry of the Hamiltonian. Therefore, impurities, which in general destroy the symmetry, do not have an effect on topological orders. 
However, it is sensitive to the topology of the underlying manifold. For trivial topology the ground state is unique. The simplest example for a non trivial topology is described by a hole in the manifold. A closed path around this hole cannot be deformed continuously into a point, it is said to be  non-contractible. The paths can be labeled by how many times they wind around this hole. This will lead to degenerate ground states. Thus, topological orders can be characterized by analyzing the ground state degeneracy \cite{Nayak}

Another probe of topological order is to study its defects: the low energy 
excitations, so called quasiparticles. These can carry fractional charge and have fractional statistics. In contrast to the Goldstone Bosons of the classical 
order these particles are normally massive. Fractional statistics of point 
particles is generally a feature of two dimensions.Consider two identical particles. Taking one particle around the other, the wave function will acquire a phase. In three dimensions this path can be continuously deformed into a point. Thus the wave function cannot be changed by that path and the phase is a multiple of $2\pi$. Therefore one can only obtain two different values of the phase for exchanging the particles. For bosons this phase is zero corresponding to integer spin and Bose-Einstein statistics. For Fermions this phase is $\pi$ with half-integer spin and Fermi-statistics. In two dimensions this argument is not valid anymore. The encircled particle prohibits the deformation of the path to a point. Thus the phase can in general be arbitrary. The particles are therefore called anyons. They interpolate continuously between fermions and bosons.

As the wave functions are not single valued in two dimensions, one can
count how many times one particle circles around another. If the 
underlying manifold is not simply connected, there are also closed path
on this manifold which cannot be continuously deformed into a point and 
will put a phase to the wave function. This forms the so
called the braid group. For two particles on a simply connected 
manifold it is just the group of integers. If both particles are identical
exchanges are also allowed. This is described by including half-integers.
For N particles on a non-simply connected manifold the braid
group is generally more complicated and does not have to be abelian
\cite{Nayak}.

If one has a finite system with a boundary, there will also be 
edge excitations. Their structures are extremely rich and, in general, they
provide a more complete and a more practical measurement of topological
order than the bulk excitations. To get a better idea of these excitations
consider a finite droplet of a FQH liquid. Although it cannot be compressed 
changing its shape will not cost much energy. Thus one can visualize edge 
states as the surface waves on these droplets. In this case the edge states
will be gapless, \cite{Nayak}, in general they can also be gapped.   

The concept of topological order was first introduced in a study of spin 
liquids. The first experimentally observed state with non trivial 
topological orders was the superconducting state. In standard literature
superconductors are often described by breaking of a gauge symmetry. The gauge 
photons will acquire a mass. Thus, the magnetic field will be excluded from
the superconducting material. The order parameter is taken to be 
proportional to the pair wave function 
$\langle c_\sigma(\vec k)c_{-\sigma}(-\vec k)\rangle$ and is in general a 
complex-valued function.  However, it is not gauge invariant and, 
therefore, not a suitable candidate for the order parameter. 
On the other hand, gauge invariant objects which could serve as an order 
paramter are not local. So there is no local order parameter in the 
superconducting state \cite{Hans}.
The main success of topological orders came with the FQH liquids. Although
not all of the FQH states can be explained by it, it provides a good 
description for simple so called filling fractions $\nu=\frac{1}{m}$.

\subsection{Outline}

This diploma work analyzes whether a high-temperature superconductor can be described by a topological theory. 
It has already been shown that this is an appropriate description for conventional superconductors \cite{Hans}. The superconducting state exhibits various
features of topological order, i.e. its excitations are gapped and it is 
sensitive to the global topology of the underlying manifold. In high-temperature 
superconductors, the gap function is not of s-wave but of d-wave symmetry: 
$\Delta_k\sim k_x^2 +k_y^2$. In contrast to the s-wave symmetry of conventional superconductors, it vanishes at four lines in momentum space. At the points where these lines intersect with the Fermi surface, the energy depends linearly on the momentum and the excitations are gapless. It is an open question, how these gapless modes affect topological properties. 

In the first part of this thesis, I derive the low-energy theory,  starting  
from the BCS Hamiltonian. It is shown that the appropriate description is a 
Dirac Lagrangian for free, massless fermions, coupled to a topological gauge 
potential. This model is then probed by taking a torus to be the underlying 
manifold and the results are then compared to the ones obtained in the s-wave
case. 
Although the quasiparticle fields couple to a topological gauge potential, the model itself yields results, which differ in some important ways from conventional topological models.

Some aspects of topological order are destroyed by the presence of gapless modes. It can be shown, that the energy spectrum becomes sensitive to local perturbations. The question arises, whether these modes destroy all topological features or whether one can formulate some weaker statements, which also hold in the presence of gapless quasiparticles. More specifically, in the second part I look for other features which depend solely on the topology.  It is argued that the energy shift under local perturbations may be such a one. Although the ground state energies are not  constant under these perturbations, there is evidence that they at least behave in the same way. It can be shown that the energy shift vanishes for all ground states in first order perturbation theory. The expressions for the second order contribution are derived and regularized. For symmetry reasons, it can be argued that some of them show similar behaviour even in second order. 

In the introductory section, I will give some necessary background material. I begin with a short review on topological field theory. In particular, I give a short introduction in BF-theory and state two applications, the quantum Hall effect and superconductors. I proceed with a section on the Casimir energy and give some comments on regularization in quantum field theory (QFT).

\subsection{Topological Field Theory}

A field theory is called topological if neither the action nor the observables 
depend on the metric of the underlying manifold. The easiest way to achieve 
this is to require both quantities to be explicitly metric independent.
Examples are 
BF-Theory and Chern-Simons theory, both of which can be solved exactly for simple
cases. The 
observables for Chern-Simons gauge fields A are Wilson loops $e^{i\oint A}$.  
In this section, I start with describing BF-Theory and its application, first 
in the Quantum Hall Effect and then in the superconducting case. There is also 
a discussion, how the latter behaves on a torus.

\subsubsection{General Chern Simons Theory}

In contrast to BF-theory which can be formulated in any number of space time 
dimensions, Chern-Simons theory (CS) only works in an odd number.  Mostly
one is interested in $ d=2+1 $, where the CS Lagrangian is quadratic in
the fields. 
\be {general chern simons lagrangian}
\mathcal{L} & = & \frac \kappa 2 \varepsilon^{\mu\nu\lambda}
a_\mu\partial_\nu a_\lambda -a_\mu j^\mu
\ee
where $\kappa$ is a positive integer.

This Lagrangian is gauge invariant, if one can neglect boundary terms. 
This is for example possible on a compact manifold such as a torus.  
Varying (\ref{general chern simons lagrangian}) with respect to $a_\mu$, one 
obtains the Euler Lagrange equation. 
\be {eulerlag}
\kappa \epsilon^{\mu\nu\lambda}F_{\nu\lambda}&=& j^\mu
\ee
In contrast to Maxwell's equations there are no derivatives involved. The 
current conservation is ensured by the $\epsilon^{\mu\nu\lambda}$ tensor. In 
particular the lack of time derivatives implies that the fields are 
non-dynamical \cite{Ezawa}.
Without a current, the only degrees of freedom are topological ones.  
Also the different components of $a_\mu$ do not commute with each other
but form a canonically conjugate pair.

To see this, first note that $a_0$ can be regarded as a Lagrange multiplier
as there are no derivatives acting on it.  Thus it constrains $a_i$ to 
fulfill:
\be{gCS2}
\varepsilon^{ij}\partial_ia_j &=&j^0
\ee
Explicit solution will depend on both, the topology of the manifold and the 
density $j^0$. 
Calculating the Euler-Lagrange equations for the remaining terms
one finds that $a_2$ is the conjugate momentum to $a_1$, leading to 
the following commutation relations: 
\be {canconjpair}
[a_1(\vec x),a_2(\vec y)]&=&\frac{i}{\kappa}\delta(\vec x-\vec y)
\ee

As $a_j$ is a gauge field and thus not gauge invariant, it cannot 
serve as an observable. These are given by the Wilson loops around 
non-contractible paths $\gamma$. These are gauge invariant, global 
operators. 
$$ A_j = e^{i\oint_\gamma a_j}$$

\subsubsection{Quantum Hall Effect}
The best known example of a topological ordered physical system is the Quantum
Hall effect (QHE). Both the Integer Quantum Hall Effect (IQHE), discovered 1980, 
and the Fractional Quantum Hall Effect (FQHE), discovered two
years later, had a great impact on condensed matter physics. 

Classically the Hall resistivity should be a linear function of the electron 
density. In the quantum regime, i.e. at low temperatures and high magnetic
fields, the resistivity is not linear any more but develops plateaus at certain 
"magical" values of the filling fraction $\nu$, which is the corresponding 
quantity to the classical density:
$$ \nu=\frac{\mbox{Number of electrons}}{\mbox{Number of states N}} =
\frac{\rho\Phi}{B}$$
The IQHE describes the plateaus for $\nu$ being an integer number, fractional 
fillig factors can be described by the FQHE. Moreover, states with simple 
rational filling factors, i.e. $\nu=\frac 1 3$,  are more stable and easier to 
observe then states with more complex ones, i.e. $\nu=\frac 4 9$. These require
lower temperatures and cleaner samples. The IQHE on the other side is visible
even for rather dirty samples. 
The observables are also insensitive to the microscopic details  of the sample.
By measuring the Hall resistivity one obtains a very precise value for
the fine structure constant $\alpha$. If the resistivity depended on e.g.
the physical dimensions of the sample, a measurement of that accuracy 
would be impossible.  
But the quantization is not the only interesting property of this new
state. Because of the light electron mass, the 2-dim electron gas
cannot condense to a crystal but will instead form a quantum liquid. However,
it is much more rigid than a solid as it can not be compressed. Furthermore,
though it does not develop a spatial order, the movements of the electrons
are highly organized.

I want to consider a two dimensional electron gas with an additional 
magnetic field perpendicular to the plane. Classically one would say that 
the electrons  move in circles. 
In quantum mechanics the energies of the electrons will be quantized in 
terms of the cyclotron frequency $\omega_c$, so that each electrons can
have an energy of $E=(2n+1)\hbar\omega_c$. 
These levels are called Landau Levels (LL). The Lowest Landau Level (LLL)
contains all the states with $n=0$.  
Taking the spin into account,
it is filled at $\nu =2$. With increasing magnetic field both the 
number of states and the energy difference of the Landau Levels will 
increase. So for high enough magnetic fields and low enough temperatures
one can always restrict oneself to the LLL and ignore the higher energy 
levels. 

As stated earlier the QHE is described as a quantum liquid.  In the IQHE
all the states of the LL are filled. If one tries to compress the liquid one
would have to excite electrons to a higher energy level. However this 
requires a finite energy due to the energy gap between the LL. This
will prevent the compression and thus lead to the very low
compressibility. 

For the FQHE at filling fraction $\nu = \frac{1}{m}=\frac{1}{2p+1}$ the 
explanation is a little bit more subtle as, in that case, there are vacant 
states in the LLL. The energy gap is in this case provided by the large Coulomb
repulsion of the electrons. The system can be described as neutral plasma
which will try to spread out the electrons as homogeneously as possible. 
By compressing the plasma one has to overcome this large Coulomb repulsion. 

However the system can also be described with help of composite particles,
i.e. particles to which flux quanta are bound.  
This can be achieved by a Chern-Simons field. 
If it binds 2p flux quanta to each fermion, it will become a composite
fermion, moving in a reduced effective magnetic field:
$$ B = B-2p\Phi\rho$$
Thereby it will fill the LLL and the effect can be described as the IQHE of
these composite fermions. 
If it binds $2p+1$ flux quanta to the electron then a composite boson will
be formed. These bosons will experience
no effective magnetic field at all and therefore will condense into a 
Bose condensate. 
This condensate is then incompressible by a many-body Coulomb effect. 
\cite{Ezawa}

In addition it's interesting to see what happens when we insert or take away 
an electron: the system at filling $\nu = \frac{1}{m} $ will react as if m 
quasielectrons resp. quasiholes have been inserted, each having a fractional 
charge of
$e^*=\pm\frac{e}{m}$ and also fractional spin.  
This also emphasizes again the fact that the appropriate
description for the QHE is by a Chern-Simons theory.

\subsubsection{BF-theory for superconductors}

In my discussion I will use a Chern-Simons like theory, usually referred to as a
BF theory.  As I also consider vortices, two Chern-Simons fields are 
needed. The name is simply reflecting that the second gauge field 
normally is denoted by b. The two fields will couple to two different
currents. In my case $a^\mu$ will couple to the spin current $j^\mu$, 
while $ b^\mu$ will couple to the vortex current $\tilde\jmath^\mu$:

\be {scCSL} 
\mathcal{L}&=&\frac{1}{\pi} \varepsilon^{\mu\nu\lambda}
b_\mu\partial_\nu a_\lambda 
- a_\mu j^\mu - b_\mu \tilde\jmath^\mu
\ee 
The coefficients in the Lagrangian are chosen to give the correct braiding 
statistics. When a quasiparticle goes around a vertex, the wave function 
should pick up a minus sign. 
In this case the different components of $a_\mu$ commute with each 
other, the same for $b_\mu$. But one obtains nontrivial canonical 
commutation relations  between $a_i$ and $b_j$:

$$[ a_1(\vec x),b_2(\vec y) ] = i\pi\delta(\vec x -\vec y) $$
$$[ a_2(\vec y),b_1(\vec y) ] = -i\pi\delta(\vec x -\vec y) $$

Calculating the Euler-Lagrange equations for the Lagrangian
gives the current conservation for both currents:
\be{currentconservation}
j^\mu &=& \frac{1}{\pi} \varepsilon^{\mu\nu\lambda}\partial_\nu b_\lambda
\nonumber\\
\tilde\jmath^\mu &=& \frac{1}{\pi} \varepsilon^{\mu\nu\lambda}
\partial_\nu a_\lambda
\ee

After separating the imaginary time and space components and doing 
some partial integrations, the BF action can be written in the following 
form, 
\be{BFaction}
\mathcal{S}&=& \frac 1 \pi \int d^3x \,\epsilon^{ij}\dot{\bar{a_i}}b_j
+a_0\epsilon^{ij}\partial_ib_j + b_0\epsilon_{ij}\partial_ia_j
\ee

Analogously to the discussion of the general Chern-Simons theory, 
$a_0$ and $b_0$ can be regarded as 
Lagrange multipliers and the solutions for $a_i$ and $b_i$ have to fulfill:
\be{CSconstraints}
j^0 &=& \frac{1}{\pi} \varepsilon^{ij}\partial_ib_j\nonumber\\
\tilde \jmath^0 &=& \frac{1}{\pi} \varepsilon^{ij}\partial_ia_j 
\ee

\subsubsection{BF theory on the torus}
By imposing periodic boundary conditions, one turns the Euclidean plane
into a compact manifold, a torus. It can visualized by taking
a parallelogram and identifying opposite sides.  
It's straightforward to solve (\ref{CSconstraints}) on a torus:
\be{solveCSL}
a_i&=&\partial_i\Lambda_a +\ai{i} \nonumber\\
b_i &=& \partial_i\Lambda_b + \bi{i}
\ee
where $\Lambda_q$ and $\Lambda_b$ are periodic functions on the torus and 
$\bar{a}_i$ and $\bar{b}_i$ are spatially constant.  After reinserting 
(\ref{solveCSL}) into (\ref{BFaction}), the action is given by:
\be{solvingtorus}
\mathcal{S}&=& \frac 1 \pi \int d^3x\,
\epsilon^{ij}(\partial_t(\partial_i\Lambda_a)\partial_j\Lambda_b
+\frac{\dot{\bar{a}}_i}{L_i}\partial_j\Lambda_b +\partial_i\dot{\Lambda}_a\bi{j}
+\frac{\dot{\bar{a}}_i\bar{b}_j}{L_iL_j})
\ee
The second and the third term vanish, as they are total space derivatives
of periodic functions. The first term can be integrated by parts and written
as $\epsilon^{ij}(\partial_t\Lambda_a)(\partial_i\partial_j\Lambda_b)$. This 
term vanishes also, as the derivatives commute with each other.
The final Lagrangian is given in terms of the spatially constant functions 
$\bar a_i$ and $\bar b_i$:
\be{finalCSLagr}
\mathcal{L}&=& \epsilon^{ij}\frac{1}{\pi} 
\frac{\dot{\bar{a}}_i\bar{b}_i}{L_iL_j}
\ee

The quasiparticle and vortex number are quantized. This requires the 
corresponding gauge fields to be compact: $\bar{a_i}$ and $\bar{b_i}$ to be 
angular variables defined only modulo $2\pi$. In a 
superconductor, the flux is quantized in multiples of $\pi$. Therefore, both
gauge field can have only two different values, 0 (for no flux quanta present) 
and $\pi$ (for one flux quanta present).

One can also approach this problem with algebraic methods. For this purpose
 one defines 
operators which generate non-contractible loops. With only one type of 
quasiparticle present, there are three such operators, two which take the
quasiparticle j around the two cycles of the torus, $\tau_j$ and $\rho_j$, and 
one taking quasiparticle j around another one, $\sigma_j$. Any process can be 
written as a product of these basic ones. $\sigma_j$ commutes
with the other operators, but $\tau_j$ and $\rho_j$ don't. Assume you want to 
measure the flux through both cycles. 
By taking the quasiparticle around one cycle, you 
measure the enclosed flux. However, at the same time you create
a flux through the other one. Therefore, the two measuring processes don't 
commute with each other: a result of the non-zero commutation relations 
between the different spatial components of the gauge field. 

With two types of particles the braid group becomes more complex. In particular, 
one has much  more operators to deal with. Therefore, I restrict the discussion 
to just one quasiparticle on the torus. Then the only non-contractible paths left
are the ones around the cycles of the torus.

\begin{figure}[ht]
\begin{minipage}[b]{5cm}
\rotatebox{270}{\resizebox{!}{6cm}{\includegraphics{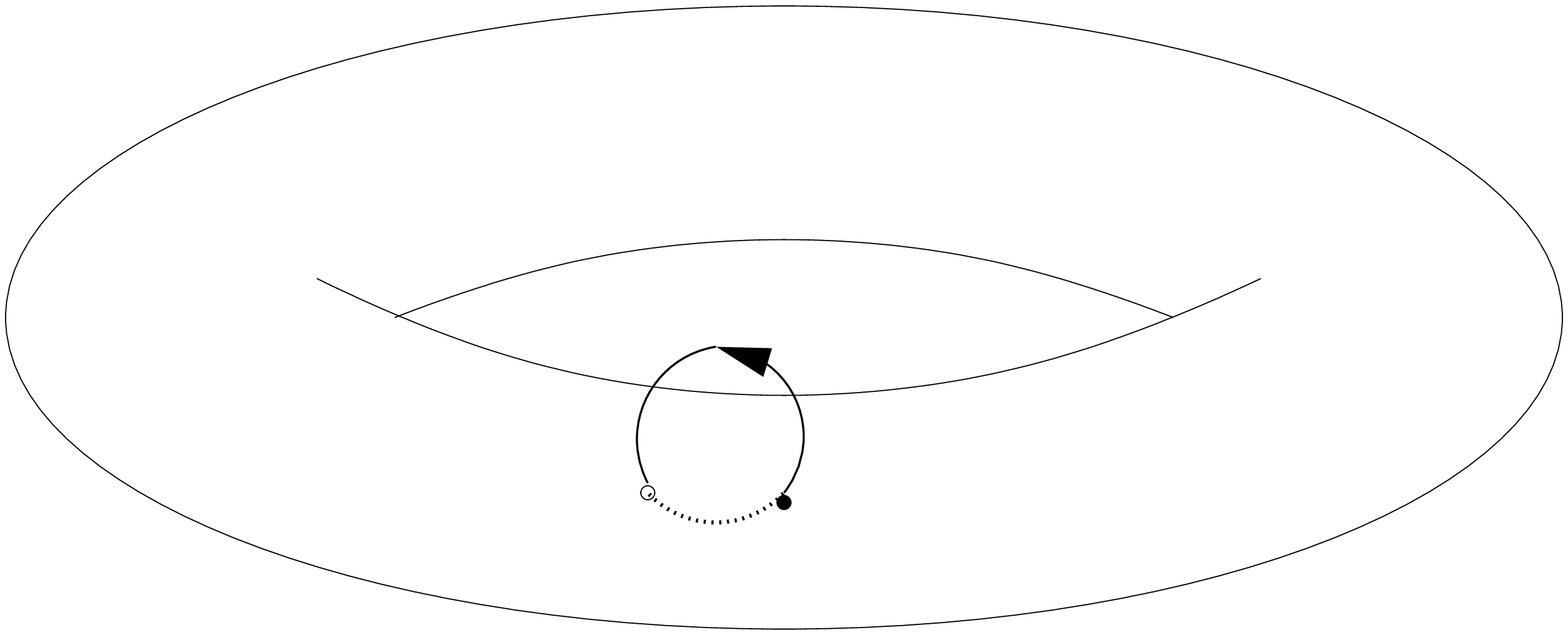}}}
\end{minipage}\hspace{1cm}
\begin{minipage}[b]{5cm}
\rotatebox{270}{\resizebox{!}{6cm}{\includegraphics{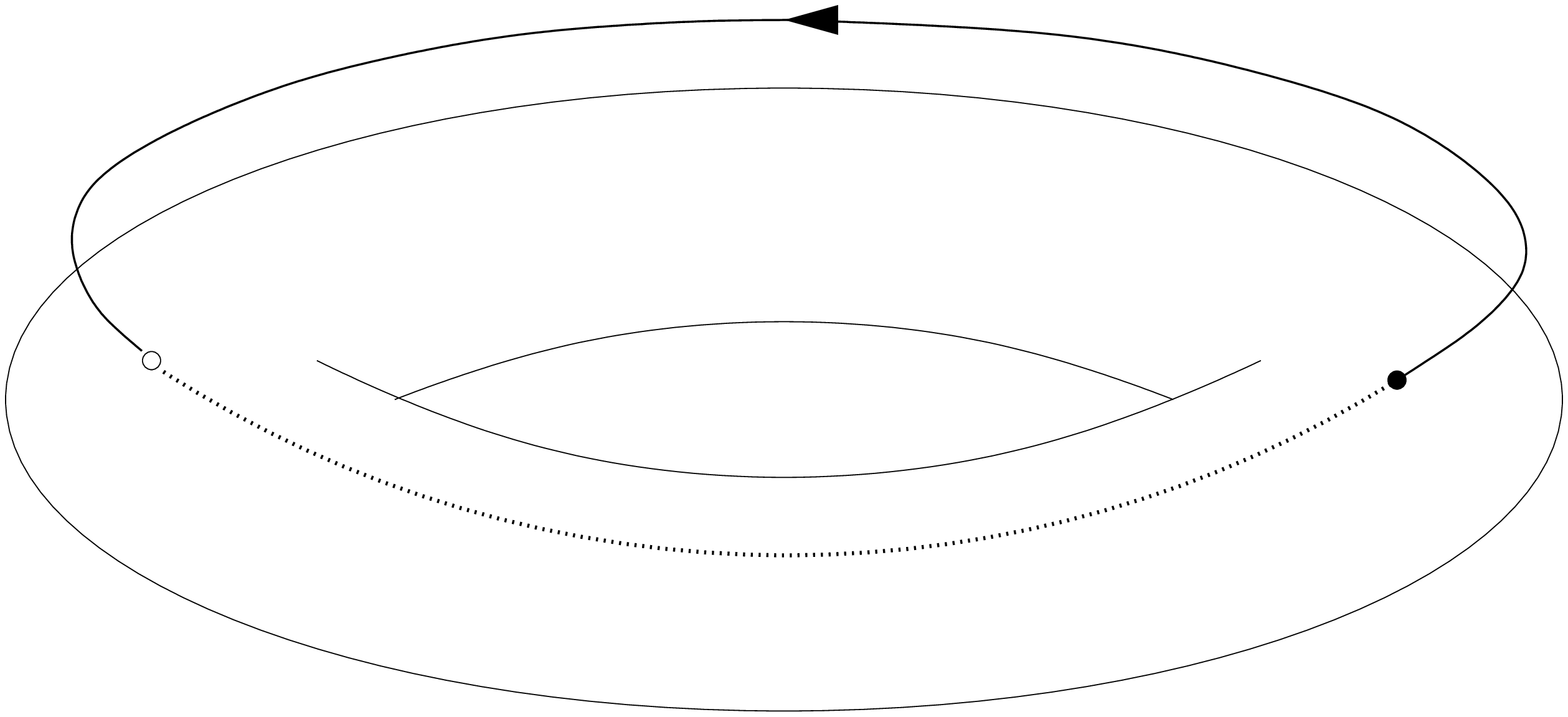}}}
\end{minipage}\\
\begin{minipage}[b]{5cm}
\rotatebox{270}{\resizebox{!}{6cm}{\includegraphics{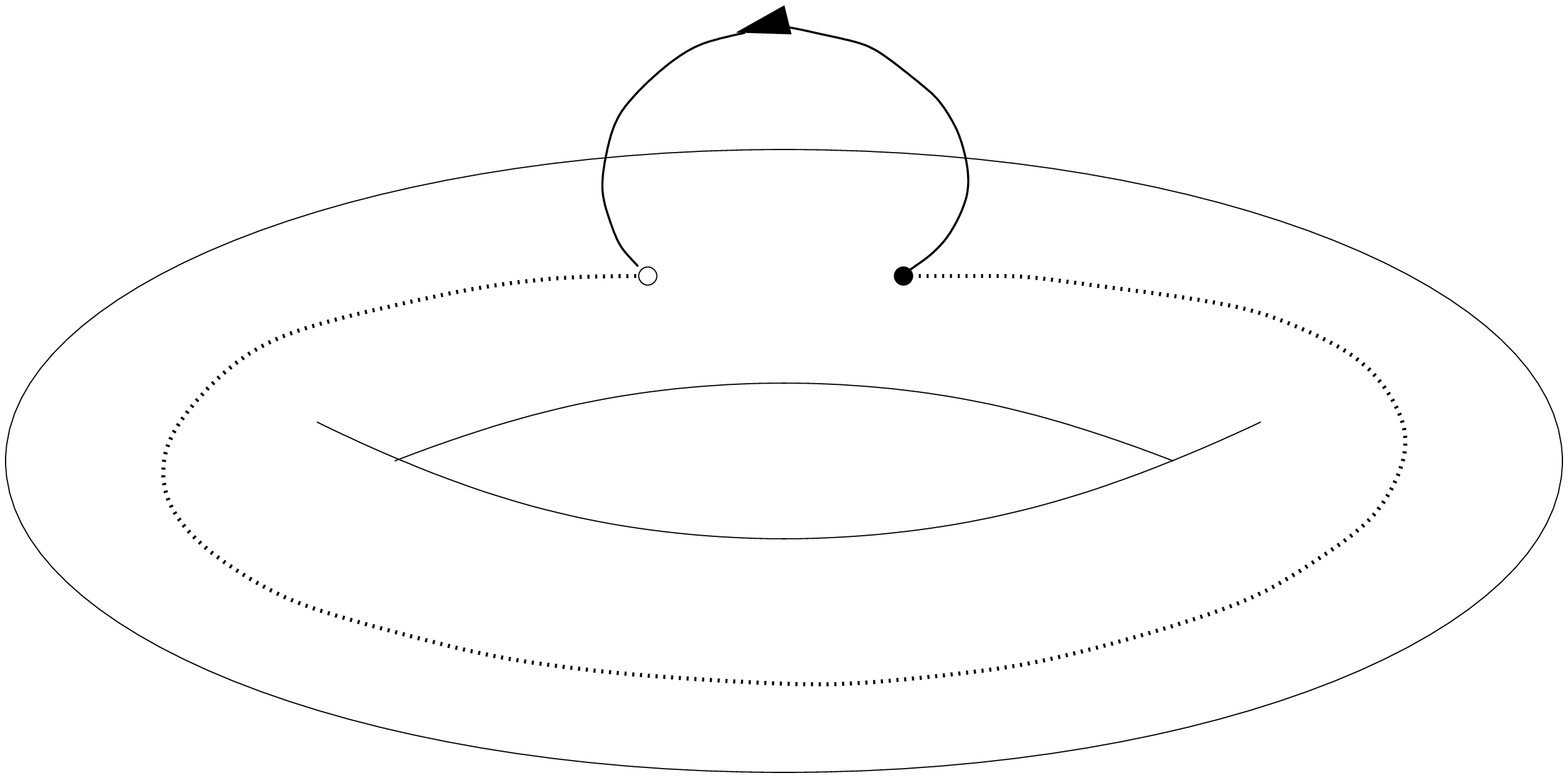}}}
\end{minipage}
\begin{minipage}[b]{5cm} \hspace{1cm}
\rotatebox{270}{\resizebox{!}{6cm}{\includegraphics{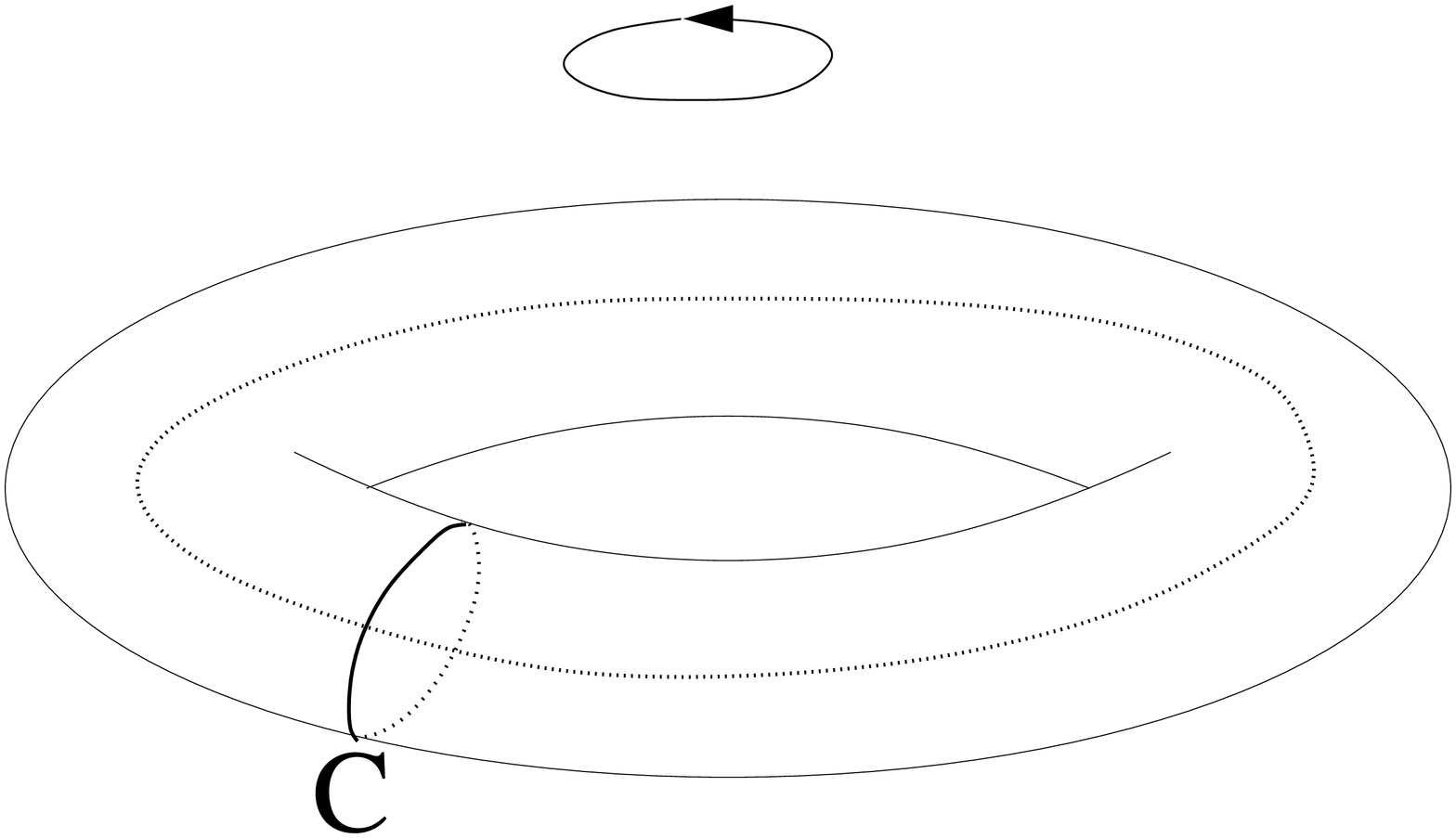}}}
\end{minipage}
\caption{A vortex tunneling process inserting a unit of magnetic flux inside the
torus. In this visualization it also leaves a flux loop outside. But as electrons on the surface on the torus cannot wind around this external flux, we can neglect it in the further discussion. This tunneling process connects ground states 
labelled by opposite values of the Wilson loop $e^{\oint_C \vec{dl}\vec a}\equiv 
e^{ie\Phi_M}$ where $\Phi_M$ is the magnetic flux threading C. }
\end{figure}

For each gauge field you can define two corresponding Wilson loops given by:
$A_j=e^{i\bar{a}_j}$ and $B_j=e^{i\bar{b}_j}$. Acting, for example, 
with $\mathcal{A}_i$ can either be seen as measuring the a-flux in the
j-direction or as inserting a b-flux in the other direction. 
Each set of eigenvalues for the Wilson loops $A_x$ and $A_y$ defines a sector. 
They can be connected
by tunneling processes: a vortex antivortex pair is created and recombines 
after having moved along eg. the non-contractible loop in x-direction. This
process creates a flux quanta in the y-direction and moves the system from one 
sector to another.

In the fully gapped case in an s-wave superconductor both types of quasiparticles 
are massive. Thus, tunneling processes are suppressed and the different sectors are 
separated by an energy barrier. Each sector has a unique ground state, so 
there is a four-fold degeneracy of the ground state on the torus \cite{Hans}.

\subsection{Casimir effect}

The Casimir effect is a manifestation of the zero point fluctuations of the vacuum. The simplest example of is two neutral, conducting plates hanging parallell to each other at a small distance. Classically, there is no interaction between the plates, but quantum-mechanically, the electromagnetic fields oscillate even at zero temperature and cause an effective, small, but measurable, attractive force between them. The size of this "Casimir force" was first calculated by H. Casimir. The total zero point energy (per unit volume) of the electromagnetic field is infinite, simply because there is one harmonic oscillator for each point in space and their zero point energies are all summed up. Nevertheless,  the {\em difference} in zero point energy for different positions of the plates is finite.\footnote{This is true for an electromagnetic field in the presence of perfect conductors, but in a more general context, the Casimir energy does not have to be finite.}It is this finite energy difference that is manifested in the Casimir force. 

In a more general setting, the Casimir force can be thought of as the finite response of an (interacting) field theory to varying boundary conditions. Also, 
there might be couplings to external potentials. In our case,  2d fermions are coupled to a (topological) gauge field in the presence of a point like impurity potential. Since the theory is defined on a torus $(L_1,L_2)$, the relevant quantity is the energy density, ${\cal E}(L_1,L_2)$.

In order to calculate the Casimir energy, the ultraviolet (UV) divergences in the sum of the energy levels in the Dirac sea, (or in the case of bosons, the sum over the zero point fluctuations) must be regularized, and the interactions renormalized, in order to extract a finite and physically meaningful result. The actual procedure will be briefly outlined in the next subsection, and described in detail in section 4.4 below. Since the full description is rather technical, however, I will first give a non-technical summary of the logic:
\begin{enumerate}
\item Since the gauge field is topological, it does not have any local degrees of freedom. More precisely, there are no fluctuations in the topological gauge field which could contribute to the Casimir energy. 
\item I will neglect the coupling between the fermions and the gauge field, except for the coupling to the external fluxes threading the torus. Technically this  means that I will only consider diagrams with a single fermion loop. 
\item I will assume that the impurity potential is weak, and can be treated in perturbation theory. The leading order term thus correspond to a free fermi theory coupled to external gauge fluxes defining the various sectors described in the previous sections.
\item There are several ways to calculate the Casimir energy of the noninteracting theory, but I will use a method that will generalize to the interacting case.  I relate $\cal E$ to the full interacting fermion Greens function $\hat{G}(\vec x, \vec y , \omega)$ on the torus, and the free field case amounts simply to use the free Greens function $G_0(\vec x, \vec y , \omega)$.  
\item
In order to isolate and subtract the leading free field theory divergence, I write the full torus propagator as $\hat{G}(\vec x, \vec y , \omega)= G_0(\vec x, \vec y , \omega)+ \sum_{\vec m\neq 0} G_0 (\vec x+\vec m, \vec y, \omega)$, where the sum is over windings (in multiples of $\vec L/(2\pi)$) around the cycles of the torus. This representation isolates the short distance singularity in the first term (corresponding to the free case) which can be directly subtracted as described in section 4.2.
\item Interactions with the impurity potential introduces new singularities, and in order to cancel these, the impurity interaction has to be renormalized - the procedure is described in section 4.5 .
\item The respresentation of the full propagator as a sum over windings is valid only when the sum actually converges. As is shown in section 4.5 this is not the case for $\omega = 0$ in one of the flux sectors. This introduces a spurious infrared divergence, and the associated technicalities are explaind in section 4.5.1
\end{enumerate}

\subsection{Regularization and renormalization of the impurity potential}

It is a well-known problem in QFT that point like interactions lead to divergent integrals. Those have to be renormalized in order to extract physical quantities. In the second part of this work, I consider a magnetic point impurity, which is a singular potential and causes similar problems. The quasiparticles scatter at the impurity, which shifts the energy levels. We are interested in the sum of all these shifts. 
However, as discussed above, this sum is in general infinite. To define the expressions we must first regularize and then renormalize the interaction. 
The divergencies first arise in second order perturbation theory, where the particles scatter two times at the impurity. If one considers this process in free space, it is well-known that it gives a divergent contribution which must be isolated and subtracted in order to calculate the renormalized Casimir energy. 
 The cause of the divergence is the singular nature of the impurity. It scatters into all momentum states with the same strength. When the particle scatters at the same point, the potential does not provide a natural momentum cutoff for the integral over the internal momentum. This is nothing but a loop in QFT and we can use the same techniques to regularize it. As soon as the particle travels along a non-contractible path between the scattering events, the minimal length of this path provides a cut-off and makes the expressions convergent. 

There are various ways to provide a momentum cutoff $\Lambda$.  The detailed form of the regularization does not effect physical results as long as it does not break any symmetry of the theory. Here, the regularization is achieved by point-splitting, {\em i.e.} by considering the interaction to happen at two different points $\vec x_0$ and $\vec x_0+\delta \vec x$. As long as $\delta \vec x$ is considered finite, though very small, the expressions are finite. Taking the limit, $\delta\vec x\rightarrow 0$, restores the original theory and gives rise to divergencies. In order to cancel the divergencies, I add local counterterms in the Lagrangian.  As we are interested in topological features of the model, it is important that we don't have to introduce nonlocal counterterms in order to regularize the expressions. 

Consider first the case where the quasiparticles are massive. In order to distinguish between the different topological sectors, the quasiparticles have to travel along non-contractible paths. On a torus, these are the paths around the two handles.  The different topological sectors are given by the flux quanta through the holes of the torus. When the particles are massive\footnote{In this work, I use both the notion of mass from particle physics and the notion of gapped particles from condensed matter interchangeably.}, each path around a hole is suppressed by a factor $e^{-m \vec n\vec L}$, where m is the mass of the particle, $L=(L_x,L_y)$ the torus size and $\vec n$ the winding number. Therefore, windings are suppressed in the limit of a large torus. As a result, the ground states in the different sectors are degenerate.  If a local impurity is inserted, the quasiparticles scatter on this impurity. However, as long as the windings are suppressed, a local impurity can not change the degeneracy of the ground states or, more generally, any topological feature. However, by inserting a non-local perturbation, windings could be facilitated and topological features could be destroyed. The same is true for nonlocal counterterms. 

In presence of gapless excitations, the  situation becomes more involved. The exponential factor vanishes for zero mass and windings around the handles are no longer suppressed. As a consequence, the ground state degeneracy is lifted.   In contrast to the case where all quasiparticles are massive, even a local impurity  can induce different energy shifts for the ground states  in the different sectors.  The issue addressed here is whether the energy shift shows some regularity. Also in the case of gapless excitations it is important to just consider local perturbations and to regularize them with local counterterms.

\newpage
\section{Dirac Theory}

\subsection{Quantization of Dirac Theory}
In the superconducting phase the electrons form Cooper pairs. To break these
Cooper pairs, energy is needed. In normal superconductors this energy does not depend on the direction in $\vec k$-space: the gap function has s-wave symmetry. This is different in high-$T_c$ superconductors. There the gap function has d-wave symmetry
$$ \Delta(\vec k) = v_\Delta(k_x^2 - k_y^2)$$
So along two lines in $\vec k$ space, $k_x = \pm k_y$ the gap function will be 
identically zero. Thus there will be four points on the Fermi surface where the 
excitations are massless. I will denote these with $k_i$ , $i=1,.., 4$. 
\begin{displaymath}
\vec k_1= \frac{k_F}{2}(1,1)=-\vec k_3 \quad \mbox{and} \quad
\vec k_2=\frac{k_F}{2}(-1,1)=-\vec k_4
\end{displaymath}
 
\begin{figure}[htp]
\centering
\rotatebox{270}{\resizebox{!}{8cm}{{\includegraphics{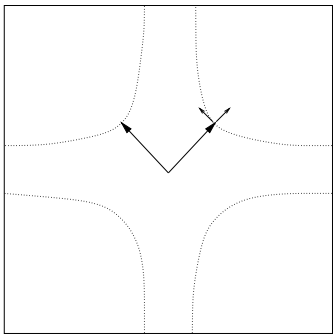}}}}
\caption{fermi surface in d-wave superconductors}
\label{fig:nodes}
\end{figure}
First the quasiparticle fields are expanded around around the nodes to obtain 
the low-energy description. Because of the d-wave symmetry of the gap function, 
the obtained model will be linear in both the time and the spatial derivatives. 
This indicates that Dirac theory is the natural language to describe 
these excitations. The quasiparticle fields can be seen as massless fermions, 
which are coupled to a gauge field $A_\mu$.   
Therefore I will give a short review of two-dimensional massless
Dirac theory. I start with the massive case and then discuss the massless case 
in detail. 

Normally two dimensional $\gamma$-matrices are sufficient to describe this 
theory. The most convenient choice is to use the three Pauli matrices as 
Dirac matrices. However, I will need to use four-dimensional 
matrices. 

After the expansion, there are two independent fields at each 
node, for spin-up and spin-down respectively. If the particles at every 
node act independently of the other nodes, then a two-dimensional 
description would be possible, consisting only of spin-up and spin-down
particle at a given node.  However, the nature of the interaction is to 
couple particles with opposite momentum and opposite spin. Thus, the spin-up 
particle at node 1 can not be treated independently of the spin-down 
particle at node 3. Due to this interaction, you need not consider an 
eight-dimensional analysis.
The quasiparticles are created by destroying Cooper pairs. In the Hamiltonian 
there are two types of terms:
$c^\dagger_\sigma(\veq)c_\sigma(\veq)$ and $c^{(\dagger)}_\sigma(\veq)
c^{(\dagger)}_{-\sigma}(-\veq)$. It is easy to see, that e.g. quasiparticles at
node 1 don't interact with those at node 2 and node 4. So, the theory can be 
restricted to one pair of nodes. 

In BCS theory of superconductors, the conserved quantity is spin, not charge. 
The goal, therefore, is to find a formulation, where the spin current can 
be identified with the conserved Dirac current. The naive choice for the 
spinors is to put all fermion fields into the spinor $\Psi$ and their 
adjoint into $\bar\Psi$. However, it is quite easy to see, that this is not
an appropriate choice. First of all, because it does not even enable you 
to write it as an Dirac equation. But if one wants to interpret spin as 
the conserved charge, the spinor needs to be an eigenstate of the spin 
operator. Therefore, you combine spin-up operators with the adjoint 
spin-down operators in the spinor, see \cite{Herbut}.  
Although I use a four-dimensional matrix representation, the theory itself 
is two-dimensional. Therefore, only three of the four Dirac matrices are needed. 
The third, $\gamma_3$ is redundant. It is still useful, though, to define it. 
It will appear later in the discussion as one of generators of the chiral 
symmetry of the two-dimensional system. 
 
Throughout the whole discussion I use the
metric $g^{\mu\nu} = diag(1,-1,-1)$. The Lagrangian for the spinor
fields can be written in following form: 

\be{Dirac equation}
\mathcal{L}(x) & = & \overline \Psi((x)
                          \gamma^\mu(i\partial_\mu-A_\mu)-m)\Psi(x) 
\ee 
using natural units where $c=\hbar=1$. 
It is helpful to expand the spinor fields in eigenfunctions of the Dirac 
equation:
\be {Spinorexpansion}
\Psi(x) &=& \int \frac{d^3p}{(2\pi)^3}\frac{1}{\sqrt{2E_p}}\sum_s
a^s_pu^s(\vec p) e^{-ipx} + b^{s\dagger}_pv^s(\vec p) e^{ipx}
\ee
$a^s$ and $b^s$ are operator valued coefficients. 
 $u^s$ and $v^s$ are the eigenfunctions of the Dirac equation for
positive and negative energy and $E=p_0$ is chosen to be positive. The factor 
$\frac{1}{\sqrt{2E_p}}$ was introduced as the spinors are normalized to 
$\sqrt{2E_p}$ instead of 1. 
The Hamiltonian is diagonal in the operators $a^s$ and $b^s$:
\be{Hamdiag}
H&=&\int \frac{d^3p}{(2\pi)^3}\sum_s
E_p(a^{s\dagger}_pa^s_p + b^{s\dagger}_pb^s_p)
\ee
$a^{s\dagger}$ creates a particle with momentum $\vec P$ , $b^{s\dagger}$ 
destroys one with momentum $-\vec p$. This can also be interpreted as 
creating a hole in the Dirac sea with positive momentum.

\subsection{Massless Fermions}
In the absence of a mass term, an additional symmetry arises, chirality. 
In 3+1 space dimension, one can define a matrix, $\gamma_5$,  which is
self-adjoint and anticommutes with the Dirac matrices:
\be {Gamma5} 
\gamma_5 & = & i\gamma_0\gamma_1\gamma_2\gamma_3 
\ee
  For every spinor $\Psi$, which is a solution of the Dirac equation, 
$\gamma_5\Psi$ is also a solution. As $\gamma_5^2=1$, its eigenvalues 
are $\pm 1$. Thus, the eigenstates of the Lagrangian can be chosen to have
a definite chirality, they are called right-handed for chirality $=1$, 
left-handed for chirality $-1$. 

In 2+1 dimensions, there are only three Dirac matrices instead of four. 
Analogous to the three dimensional case, one can define a $\gamma_3$
by claiming that it has to fulfill the Clifford algebra. So there
are two matrices, $\gamma_3$ and $\gamma_5$, which anticommute with
the Dirac matrices and one, which commutes, $\gamma_3\gamma_5$. These
matrices don't commute with each other. Instead they form an SU(2)
algebra like a spin. So the chirality in 2+1 dimensions has the 
same structure as the spin symmetry. Each matrix can be linked to a
direction and only one can be chosen to be diagonal.

\subsection{The Effective Lagrangian}
 First I will give a short review on BCS theory. I start with a rather general
ansatz for the Hamiltonian:
\be{ansatz}
\mathcal{H}&=&\mathcal{H}_0+\mathcal{H}_{int} \nonumber\\
&=& \sum_{\veq,\sigma}(\xi\ofq-\mu)c^\dagger_\sigma\ofq c_\sigma\ofq\nonumber \\
&+&\sum_{\veq, \vec q',\vec k, \sigma}
v_k(\vec q, \vec q')c^\dagger_\sigma(\veq +\vec k)c^\dagger_{-\sigma}(-\veq +\vec k)
c_{-\sigma}(-\vec q'+\vec k) c_\sigma(\vec q' + \vec k)
\ee
$c_\sigma$ and $c^\dagger_\sigma$ are the usual fermion operators and obey 
canonical anti-commutation relations:
$\{a^\dagger_\sigma(\veq),a_\rho(\vec{q}')\}=\delta_{\sigma,\rho}\delta_{\veq,\vec{q}'}$.
The interaction term, $\mathcal{H}_{int}$ is invariant under global charge U(1) and 
spin SU(2) symmetries and conserves both, momentum and spin. In BCS theory two electrons
with opposite momentum and opposite spin form a Cooper pair. You expect, therefore, a 
non-zero expectation value for terms like the pair operator: 
$c_\sigma(\veq)c_{-\sigma}(-\veq)$. You also assume, that the fluctuations in the pair 
operator are small, so you are allowed to make a mean-field approximation by introducing 
the superconducting order parameter (or gapfunction), $\Delta$:
\be{orderparameter}
\Delta_{\vec k}(\veq) &=& \sum_{\vec{q}'}v_k(\veq, \vec{q}')\langle
c_\sigma(\vec{q}'+\vec k) c_{-\sigma}(-\vec{q}'+\vec k)\rangle
\ee
You can now identify $\vec{k}$ with the center of mass momentum of a Cooper pair, and also 
define a spatially varying gap function by a Fourier transformation to real space:

\be{spatvargap}
\Delta_{\vec R}(\veq) = \sum_{\vec k} e^{i2\vec k \vec R}\Delta_{k}(\veq)
\ee
If the electrons form a pair with zero total momentum, the gap function is constant:
$\Delta_{\vec R}(\veq)=\Delta(\veq)$. With magnetic flux tubes present, Cooper pairs 
have non-zero momentum and the order parameter becomes a function of their center 
or mass coordinates, $\vec R$. 

The Hamiltonian for our system is given by:
\be{startingpoint}
\mathcal{H} & = & \sum_{\veq,\sigma}
c^\dagger_\sigma\ofq(\xi\ofq -\mu)c_\sigma\ofq \nonumber\\
&+& \frac{\sigma}{2} c^\dagger_{\sigma}\ofq \Delta_{\vec{R}}\ofq
c^\dagger_{-\sigma}(-\veq) +h.c.
\ee
where the dependence on the center of mass is coded in $\Delta_{\vec R}$. 
Furthermore, I will take the gap function to have d-wave symmetry. 
As I am interested in the low energy physics, I will expand the creation 
and annihilation operators in terms of continuous fields $\psi\ofq$ for $\veq$ 
values near the nodes and integrate out the high energy degrees of freedom:

\be{k-exp}
c_\uparrow(\vec k_j+\veq)&=&\psiu j\ofq\nonumber\\
c_\downarrow(\vec k_j-\veq)&=& \psid j\ofq
\ee
As we have restricted the momenta to regions near the nodes, it is legitimate to 
linearize both the energy and the gapfunction.

\be{linearization}
\xi(\vec k_j +\veq)-\mu &=& 2\vec k_j\cdot\veq +\mathcal{O}(q^2) 
\equiv v_Fq_j \nonumber\\ 
\Delta(k_j +\veq) &=& \Delta(\vec R) [(k_{j,x}q_x)^2-(k_{j,y}q_y)^2]
\nonumber\\
&=&2\Delta(\vec R)[k_{j,x}q_x-k_{j,y}q_y]
\ee
where $q_j$ is the momentum parallel to $\vec k_j$, 
$q_j = \hat k_j\cdot\veq$, and  $v_F$ denotes the Fermi velocity.   
Inserting this relation into (\ref{ansatz}) gives:

\be{inbetween}
\mathcal{H}&=& \int\frac{d^2q}{(2\pi)^2}\,\sum_{j=1}^{4}
\psiudag j\ofq \xi(\vec k_j+\veq)\psiu j\ofq +
\psiddag j\ofq \xi(\vec k_j-\veq)\psid j\ofq \nonumber\\
&+&\sum_{j=1}^{2}
\psiudag j \ofq\Delta_{\vec R}(\vec k_j+\veq)\psiddag {j+2}\ofq\nonumber\\
&&-\psiddag j\ofq\Delta_{\vec R}(\vec k_j-\veq)\psiudag {j+2}\ofq
+\mbox{h.c.} \\
&=&\int\frac{d^2q}{(2\pi)^2} \, \sum_{j=1}^{4}
\psiudag j\ofq v_F q_j\psiu j\ofq
-\psiddag j\ofq v_Fq_j\psid j\nonumber \\
&+&\sum_{j=1}^2
\psiudag j\ofq2\Delta(\vec R)[k_{j,x}q_x-k_{j,y}q_y]\psiddag{j+2}\ofq
\nonumber\\
&+&\psiddag j\ofq2\Delta(\vec R)[k_{j,x}q_x-k_{j,y}q_y]\psiudag {j+2}\ofq 
+\mbox{h.c.}
\nonumber
\ee
One can simplify this expression by introducing two 4-spinors
\be{spinor}
\Psi_j^\dagger = (\psiudag j, \psid {j+2}, \psiudag {j+2}, \psid j)
\ee
and writing it as a matrix equation. 
Furthermore I will write the gap function as 
$\Delta(\vec R) =\frac{v_\Delta}{k_F} e^{i\varphi}$. For simplicity 
I assume that the only space dependence lies in the phase $i\varphi$, since 
amplitude fluctuations are costly in energy.

\be{beforegauge}
\mathcal{H} & = & \sum_{j=1}^2 \int\frac{d^2q}{(2\pi)^2} \,
\Psi^\dagger_j\ofq v_Fq_j\left(
\begin{array}{cccc}1&&&\\&-1&&\\&&-1&\\&&&1\end{array}\right)\Psi_j\ofq
\nonumber\\
&+&\Psi^\dagger_1\ofq v_\Delta q_2 \left(\begin{array}{cccc}
0&-e^{i\varphi}&&\\-e^{-i\varphi}&0&&\\&&0&e^{i\varphi}\\
&&e^{-i\varphi}&0\end{array}\right)\Psi_1\ofq \nonumber\\
&+& \Psi_2^\dagger\ofq v_\Delta q_1 \left(\begin{array}{cccc}
0&-e^{i\varphi}&&\\-e^{-i\varphi}&0&&\\&&0&e^{i\varphi}\\
&&e ^{-i\varphi}&0\end{array}\right)\Psi_2\ofq
\ee
Now do a Fourier transformation back to real space. 

\be{fourier}
\psii j \uparrow\ofr&=&\int\frac{d^2q}{(2\pi)^2}e^{i\veq\ver}\psii j \uparrow
\ofq \nonumber\\
\psii j \downarrow\ofr&=&\int\frac{d^2q}{(2\pi)^2}e^{-i\veq\ver}
\psii j \downarrow\ofq
\ee
Note the "-" sign in the second line of (\ref{fourier}). This convention ensures 
that the spinors Fourier transform in a simple way:

\be{psiofx}
\Psi_j\ofr &=& \int\frac{d^2q}{(2\pi)^2}\,
e^{i\veq\ver}\Psi_j\ofq
\ee
We would have obtained the same low energy fields, if we had expanded the fermion 
fields in $\vec r$-space:

\be{lowenerx}
\psi_\sigma\ofr&=&\sum_{j=1}^4e^{i\vec k_j\ver}\psii j \sigma\ofr
\ee
In general, the phase of the order parameter will be space  dependent. In the Dirac 
theory the matrices are constants. So in order to rewrite (\ref{inbetween}) as a 
Dirac equation, I will redefine the spinors and, thus, remove the phase from the 
matrices. If the phase $\phi$ is well-defined at any point on the torus, there is no
problem since it can then be removed by a regular gauge transformation. 
The phase is divided equally between the spin-up and spin-down fields and the 
transformed spinor fields are still single-valued. However, $\phi$ is an angular 
variable, so generally it is not single valued. Consider a magnetic flux through one 
of the holes in the torus. The magnetic field on the torus will be zero, but the 
electromagnetic gauge potential is not. Instead, it will wind around the magnetic 
flux. By going around the hole including the flux tube, the phase does not return to 
its original value. As the magnetic flux in superconductors is quantized, the phase 
picks up an additional $2\pi$. If now the phase is divided equally between the two 
fields, they are not single-valued anymore. Thus, the naive choice how to transform 
the spinors does not work. I will denote that part of the phase as the singular part. 

There are different approaches to deal with that problem.  In \cite{Herbut} the 
author considers vortices in a superconductor. Those vortices are divided into two 
groups, labeled A and B. The spin-up fields are transformed with the singular part 
of the phase, which comes from the vortices in group A, spin-down with those in group 
B. Finally, you average over all possibilities to distribute a given number of 
vortices into the two groups to get rid of the anisotropy you introduced. In this paper, 
a similar but much simpler method is applied. Here for simplicity, no vortices are 
present. The singular behaviour of the phase arises only of the fluxes in the holes 
of the torus. In general, the phase has both a regular and a singular piece. 
The regular part of the phase will be divided symmetrically to both 
fields, the singular part, however, only to spin-up particles.  Thereby, I also create
an anisotropy between spin-up and spin-down particles. This will become manifest at 
the analysis of the ground state degeneracy. I include also an electromagnetic gauge 
potential to include external magnetic fields in the analysis. The transformation of 
the spinor fields is then given by:

\be{gaugetrans}
\psii j \sigma\ofr &\rightarrow&e^{i\varphi_\sigma}\psii j \sigma\ofr
\ee
where

\be{gaupha}
\varphi_\uparrow&=&\varphi_s+\varphi_r/2+\int^{\ver}d\vec r'\vec A \nonumber \\
\varphi_\downarrow &=&\varphi_r/2+\int^{\ver}d\vec r'\vec A
\ee
this is equivalent to transforming the spinors by a unitary matrix U:

\be{unimat}
\Psi_j\ofr&\rightarrow&U\ofr\Psi_j\ofr\nonumber\\
U\ofr&=&diag(e^{i\varphi_\uparrow},e^{-i\varphi_\downarrow},
e^{i\varphi_\uparrow},e^{-i\varphi_\downarrow})
\ee
As a result of performing this transformation there will appear additional
terms in the effective Hamiltonian. As in the Chern-Simons theory you can either choose to 
have multi-valued wave functions or you can keep them single-valued by introducing a 
Chern-Simons gauge field $a_\mu$. I introduce two matrices:

\be {matrices}
\alpha_1&=&
\left(\begin{array}{cc}\sigma_3&0\\0&-\sigma_3\end{array}\right)
\nonumber \\
\alpha_2&=&
\left(\begin{array}{cc}-\sigma_1&0\\0&\sigma_1\end{array}\right)
\ee
They correspond to the $\alpha$ matrices in the conventional Dirac 
theory. 
The additional terms which appear in (\ref{beforegauge}) after doing the gauge 
transformation, can be rearranged into two current terms coupling to different 
gauge fields. The Chern Simons gauge field: 
\be{def}
a_j &=& -\half\partial_j \varphi_s
\ee 
couples to the conserved Noether current, which in our model is not the em 
current but the spin current. The total phase, that is both the Chern-Simons field 
and the electromagnetic gauge field, couples to the em current. The spin current is 
denoted by $\hat\jmath_i$ while the electromagnetic current is $j_{em}^i$. A more 
detailed derivation of the spin current, is given in Appendix A.  

\be{currents}
\hat \jmath_x&=&-\Psi_1^\dagger v_F\alpha_1 \Psi_1
-\Psi_2 v_\Delta \alpha_2\Psi_2 \nonumber\\
\hat{\jmath}_y&=&-\Psi_1^\dagger v_\Delta \alpha_2\Psi_1
-\Psi_2 v_F \alpha_1\Psi_2 \nonumber\\
j_{em,j}&=&
\Psi_j^\dagger v_F \left(\begin{array}{cc}I&0\\0&-I\end{array}\right)\Psi_j
\ee
With this, we can define the covariant derivative $D_j=\partial_j-ia_j$ and
write the Hamiltonian in Dirac form,

\be{final Hamiltonian}
\mathcal{H}&=& 
-\Psi_1^\dagger (iv_FD_x\alpha_1+iv_\Delta D_y \alpha_2)\Psi_1
-\Psi_2^\dagger( iv_\Delta D_x\alpha_2 +iv_F D_y \alpha_1)\Psi_2
\nonumber \\ &+&\half(\partial_j\varphi+2A_i)j_{em}^j
\ee
The Hamiltonian and the spin current depend on two velocities, one being just 
the normal Fermi velocity, $v_F$. The other, $v_\Delta$, characterizes 
excitations parallel to the Fermi surface. In cuprates, these velocities are 
not equal to each other. This anisotropy has consequences on the symmetries 
of the system. A more detailed analysis can be found in the discussion of 
the symmetries of the model.

\newpage
\section{The Effective Lagrangian for a d-wave superconductor}

To analyze our model, it is convenient to use a Lagrangian
description, where the time dependence is included. For later convenience I 
will also transform the effective Lagrangian to Euclidean space.  
I will also include a dynamical term for the $\vec A$ field,
$-\frac{1}{4}F_{\mu\nu}^2$, and introduce a vortex current, parameterized 
by a gauge potential $b_\mu$:
\be{vortexcurrent}
\tilde \jmath^\mu&=&\frac{i}{\pi}\epsilon^{\mu\nu\lambda}
\partial_\nu \varphi a_\lambda
\ee
I account for the phase fluctuations by also introducing a mass term:

$$ \frac{\lambda_L^{-2}}{2}|\partial_\mu+A_\mu|^2$$
The mass $\lambda_L$ will depend on the size of the torus and go to zero
in the limit of $|\vec L|\rightarrow\infty$
I will also introduce $\gamma_0$. As we are in Euclidean space, the gamma
matrices have to fulfill the Clifford algebra for Euclidean space:
\be{Clifford}
\{\gamma_\mu,\gamma_\nu\}&=&2\delta_{\mu,\nu}
\ee
which leads to the following definition of $\gamma_\mu$
\be{gammamatrices}
\gamma_0&=&\left(\begin{array}{cc}0&I\\I&0\end{array}\right)\nonumber\\
\gamma_1&=&\gamma_0\cdot\tilde \gamma_1\nonumber\\
\gamma_2 &=&\gamma_0\cdot\tilde\gamma_2
\ee
The effective low energy Lagrangian is then given by:

\be{lowenact} 
{\cal L} &=& \frac 1 4 F_{\mu\nu}^2 + \frac {\lambda_L^{-2}} {2} 
(A_\mu - a_\mu)^2 - 
b_\mu \tilde \jmath^\mu + \frac i \pi \varepsilon^{\mu\nu\lambda}
b_\mu\partial_\nu a_\lambda \\ 
& + & \overline\Psi \gamma_E^\mu(\partial_\mu -ia_\mu) \Psi - (A_\mu +
\half \partial\varphi_r -a_\mu)J^\mu_{em}\nonumber
\ee 
I introduced the abbreviation:
\be{abbrev}
\bar\Psi\gamma_E^\mu\partial_\mu\Psi&\equiv&
+\bar\Psi_1 v_F\gamma_1\partial_1+v_\Delta\gamma_2\partial_2\Psi_1
+\bar\Psi_2 v_F\gamma_1\partial_2+v_\Delta\gamma_2\partial_1\Psi_2
\ee
The conserved Noether current in our model is not the electromagnetic 
current but the spin current given by:
\be{spincurrent}
\hat \jmath^\mu&\equiv&\bar\Psi\gamma_E^\mu\bar\Psi\nonumber\\
\ee
Which in detail reads:
\be{spincurr}
\hat \jmath^0 &=& \bar\Psi_1\gamma^0\Psi_1+\bar\Psi_2\gamma^0\Psi_2
\nonumber\\
\hat \jmath^1&=& \bar\Psi_1v_F\gamma_1\Psi_1+\bar\Psi_2v_\Delta
\gamma_2\Psi_2 \nonumber\\
\hat \jmath^2&=& \bar\Psi_1v_\Delta\gamma_2\Psi_1+\bar\Psi_2v_F
\gamma_1\Psi_2\nonumber\\
\ee
For a more detailed discussion of the spin current see Appendix A.

\subsection{Derivation of the effective BF action}
First I note that the regular phase can be absorbed by a regular
gauge transformation and then I identify $A_\mu + a_\mu = \delta
a_\mu$ where $\delta a_\mu$ is a "small" quantity. In the obtained
equation 

\be {before integrating} 
{\cal L} & = & \frac 1 4
F_{\mu\nu}^2 + \frac{\lambda_L^{-2}}{2} \delta a_\mu^2 - b_\mu \tilde
\jmath^\mu + \frac i \pi \varepsilon^{\mu\nu\lambda}b_\mu\partial_\nu
(\delta a_\lambda - A_\lambda) \\ &+& \overline\Psi
\gamma_E^\mu(i\partial_\mu - A_\mu + \delta a_\mu) \Psi - \delta a_\mu
J^\mu_{em}\nonumber \ee the $\delta a_\mu$ field can be integrated
out. In the final equation we identify again the $a_\mu$ field as the
singular piece of the gauge field $A_\mu$ and set 
$A_\mu =a_\mu$.   

\be{after integrating} 
{\cal L} & = & \frac 1 4
{f_{\mu\nu}^{(a)}}^2 -b_\mu \tilde \jmath^\mu + \frac i \pi
\varepsilon^{\mu\nu\lambda}b_\mu\partial_\nu a_\lambda \nonumber\\ &+&
\overline \Psi\gamma_E^\mu(i\partial_\mu-a_\mu)\Psi-\half
\lambda_L^2(\frac i \pi \varepsilon^{\mu\nu\lambda}\partial_\mu b_\nu
+\hat j^\lambda - J_{em}^\lambda)^2 \nonumber\\ & = & \frac 1 4
{f_{\mu\nu}^{(a)}}^2 +\frac {\lambda_L^2}
{4\pi^2}{f_{\mu\nu}^{(b)}}^2-b_\mu\tilde j^\mu +\frac i \pi
\varepsilon^{\mu\nu\lambda}b_\mu\partial_\nu a_\lambda
+\overline\Psi\gamma_E^\mu(i\partial_\mu+a_\mu)\Psi \nonumber\\ &+&
i\frac{\lambda_L^2}{\pi}\varepsilon^{\mu\nu\lambda}(\partial_\mu
b_\nu)(\hat j_\mu-J_{em,\mu})-\frac{\lambda_L^2}{2}(\hat
j_\mu-J_{em,\mu})^2 
\ee

In order to simplify the analysis, I will assume that the vortex current
vanishes: $\tilde \jmath_\mu = 0 $
and I will keep only lowest order terms

\be{lowest}
\mathcal{L}&=& \frac{i}{\pi}\epsilon^{\mu\nu\lambda}b_\mu\partial_\nu
 a_\lambda +\overline \Psi\gamma_E^\mu D_\mu\Psi
\ee

\subsection{BF-action on a torus}
First I will discuss the BF part in Minkowski space. It is illuminating
to rearrange the terms in the BF-Lagrangian:
\be{Chern}
\mathcal{L}&=&\frac{1}{\pi}\epsilon^{\mu\nu\lambda}b_\mu\partial_\nu 
a_\lambda + a_\mu\hat \jmath_\mu\\
&=& a_0(\hat \jmath^0+\frac{1}{\pi}\epsilon^{ij}\partial_ib_j)
+\frac{1}{\pi} b_0\epsilon^{ij}\partial_ia_j +
\frac{1}{\pi}\epsilon^{ij}\dot a_ib_j
-a_i\hat \jmath^i
\ee

There are no time derivatives acting on $a_0$ and $b_0$, so they can be 
regarded as Lagrange multipliers. As described in the introduction, they 
enforce constraints on the spatial gauge fields. Although in this case, 
the constraints look more complicated:
                       
\be{constraints}
\epsilon^{ij}\partial_ib_j&=&-\pi\hat \jmath_0\nonumber\\
\epsilon^{ij}\partial_ia_j&=& 0
\ee
The second constraint is easily solved using Hodge's theorem
\be{asol}
a_i&=& \partial_i\Lambda+\frac{\bar a_i}{L_i}
\ee

For $b_i$ there will be an additional term in the solution.
\be{bsol}
b_i&=&\partial_i\Omega+\frac{\bar b_i}{L_i}-\pi\epsilon^{ij}
\frac{\partial_j}{\partial^2}\hat \jmath^0
\ee
where $\Lambda$ and $\Omega$  are periodic functions on the torus 
and $\bar a_i$, as well as $\bar b_i$, is spatially constant. So far
the analysis followed the standard BF-theory, described in the 
introduction. The new feature is the inhomogeneous part in the 
differential equation for $\bar b_i$. In general, you can always 
solve these equations with help of Green's function, in this case the 
Green's function for a torus, denoted by: $\frac{1}{\partial^2}$. 
The interesting (topological) behavior is given by $\bar a_i$ and 
$\bar b_i$. As $e^{i\varphi_s}$ is invariant under the transformation
\be{periodicity}
\ver &\rightarrow &\ver + \vec L\nonumber \\
\vec L = (L_x, 0) \quad &\mbox{and} &\quad \vec L = (0,L_y)
\ee

\noi
$\varphi_s$ is only allowed to change up to an additional $2\pi$.
With the definition
of $a_i=\half\partial_i\varphi_s$ 
the Wilson loops 
$A_i=e^{i\oint \ai i}$ and $B_i=e^{i\oint\bi i}$ can be 1 or $-1$. If 
a quasiparticle encircles a flux quantum, it picks up an additional 
minus sign. Otherwise, it's wave function remains unchanged under a 
transformation along a closed path.  

After inserting this into (\ref{Chern}), the Lagrangian is simplified
considerably:
\be{easyLag}
\mathcal{L}&=& \varepsilon^{ij}\frac{\dot{\bar{a}_i}}{L_i} \bi j - 
\Lambda(\pi\partial_t \hat \jmath^0-\partial_i \hat \jmath^i)
-\ai i \hat\jmath^i
\ee
Also $\Lambda$ can be regarded as a Lagrange multiplier which 
ensures the current conservation. $\ai i$ and $\bi j$ form a 
canonical conjugated pair
\be{comm}
[\ai 1, \bi 2]&=&i
\ee

\noi
In contrast to the s-wave case, the BF-Hamiltonian is not identically
zero. Instead, the conserved current, in this case the spin current, 
couples to a (spatially) constant gauge field. 
\be{CSH}
\mathcal{H}&=& \ai i\hat\jmath^i
\ee
So the Wilson loops, set up by the $\ai i$, commute with the
Hamiltonian for both i, while the ones for the $\bi i$ do not.
$A_i = \exp(i\oint \ai i)$ measures the magnetic flux. In other words, 
the number operator of the magnetic flux can be simultaneously 
diagonalized with the energy. Thus, the
ground state is labeled by the number of fluxes. Creating
quasiparticles does not require a finite energy, therefore their
particle number is not fixed in the ground state. The total electric
flux is measured by $B_i = \exp(i\oint \bi i$. Due to the commutation 
relations of the $\bi i$ with the Hamiltonian, the ground states 
can not be labelled by the number of electric fluxes. 
Note, however, that the Wilson loops cannot distinguish a positive 
from a negative flux quantum. $A_i$ takes the same value for 
$\bar a_i=\pi$ and $\bar a_i=-\pi$. In addition, you can't count 
flux quanta with help of the Wilson loops. Two flux quanta give rise 
to a phase of $2\pi$ and are, therefore, not distinguishable by any other, 
even number of vortices. Each Wilson loop has the eigenvalues 1 and -1, for 
even respectively odd number of flux quanta present.

\subsection{Dirac Hamiltonian}
I use (\ref{CSH}) to write down the final result for the Hamiltonian. 
Note that $\bar{a}_i$ is spatially constant. Thus, the Hamiltonian 
can be treated as in the field-free case. The topological phase 
only shifts the momentum by a constant amount. The Dirac Hamiltonian 
is given by:

\be{Dirac Hamiltonian}
\mathcal{H}& = & \left( \begin{array}{cc}
-iv_F\sigma_3D_x+iv_\Delta\sigma_1D_y & 0 \\ 
0 & iv_F\sigma_3D_x-iv_\Delta\sigma_1D_y
\end{array}\right)
\ee 
It is block diagonal and, therefore, splits into two two-component
equations. The two equations differ only by an overall minus sign, so I will 
concentrate on one.  
The eigenfunctions of the Hamiltonian can be expanded in momentum space by
writing
$$\Psi_j\ofr =\int \frac{d^2k}{(2\pi)^2}\frac{1}{\sqrt{2E_k}}\sum_{s=1,2} 
(a_s(\vec k) u^s_k + b^{\dagger}_s(\vec k) v^s_k) e^{i\vec k\ver}$$
where $u^s_k$ are solutions of the Dirac equation for positive
resp. $v^s_k$ for negative energies. Here E will always denote a positive quantity.  
There are nontrivial solutions for $(\mathcal H\mp E)\Psi_j=0$ if the 
determinant of the matrix equation vanishes:

\be{det} 
\left| \begin{array}{cc}v_F(k_x-\ai x) 
\mp E & -v_\Delta(k_y - \ai y) \\ -v_\Delta( k_y - \ai y) & -v_F( k_x -
\ai x)\mp E 
\end{array} \right|
& =& 0 
\ee

\be{energy relation}
\Rightarrow E^2 & = & v_F^2(k_x - \ai x)^2+v_\Delta^2(k_y-\ai y)^2 
\ee
for node 1 and 3, I denote this pair by (1/3). 
For the other pair of nodes, (2/4), you obtain the same expression but
with x and y interchanged. So

\be{energy}
E_{(1/3)}&=&\sqrt{v_F^2(k_x - \ai x)^2 + v_\Delta^2( k_y - \ai y)^2}\nonumber\\
E_{(2/4)}&=&\sqrt{v_F^2( k_y - \ai y)^2 + v_\Delta^2( k_x - \ai x)^2}
\ee
The total energy is given by the sum of the partial energies. 

As we're on the torus, we need to specify boundary conditions for the spinor 
fields. On the torus, the points $\vec x=(x_1, x_2)$ and 
$\vec x'=(x_1+nL_x, x_2+mL_y)$ are identified.  
In general, the spinor fields can differ by a phase $\alpha$ at both points. 
The phase of the adjoint fields is then given by $-\alpha$. Both combined 
give the relation of the spinors at both points:      

\be{spinortransformation}
\Psi_j(\vec x') &=& U\Psi_j(\vec x)
\ee
With U given by:
$U=\mbox{diag}(e^{i\alpha},e^{-i\alpha},e^{i\alpha},e^{-i\alpha})$
The Lagrangian has to be the same for $\vec x $ and $\vec x'$. 
Therefore, the matrices have to fulfill following constraints:
\be{boundary}
U^\dagger\tilde\gamma_1U&=&\tilde\gamma_1\nonumber\\
U^\dagger\tilde\gamma_2U&=&\tilde\gamma_2
\ee
The first one is trivially satisfied, while the second one gives a 
non-trivial constraint on $\alpha$:
$$e^{2i\alpha}=1 \quad \mbox{or} \quad \alpha = n\pi$$ 

This determines the boundary conditions to be either periodic or
antiperiodic. Which one is chosen is not important as long as I also include 
all the flux sectors. Inserting a magnetic flux
changes the boundary conditions from periodic to antiperiodic ones and
vice versa and will only change the labelling of the different sectors but not the 
overall structure. Here I chose the antiperiodic one which sets $\vec k$ to:
$$ k_i = \frac{2n+1}{L_i}\pi$$ 
This choice makes sure that the ground state is unique for 
$\ai x = \ai y =0$.

I introduce $\pi_i = k_i-\ai i$ in order to shorten the notation and to 
make it uniform in the different sectors. The solutions for the spinors 
$u^s_k$ and $v^s_k$ depend on the value of $\vec k$ and the Chern-Simons 
potentials. The cases, 
where one or both of the $\pi_i$'s are zero, have to be treated separately. 
However, it is still possible to write down a closed expression for the spinor 
fields with momentum-dependent coefficients. 
 
In the sector $A_x=A_y=1$, and also  for an excited state in any other sector,
both $\pi_x$ and $\pi_y$ are nonzero.
The additional "nodon" symmetries, see Appendix B, allow me to treat the two 
pairs of nodes separately. I will therefore only solve the equations for 
the pair (1/3). Solving for the other pairs is completely analogous. As the 
theory is not relativistic I chose to normalize the spinors to 1 instead of
the more common $\sqrt{2E(\vec k)}$.
\be{1spinor}
u^{1\dagger}_{k}&=& c(-v_\Delta\pi_y,E-v_F\pi_x,0,0)\nonumber\\
v^{1\dagger}_{k}&=& c(E-v_F\pi_x,v_\Delta \pi_y,0,0)\nonumber\\ 
u^{2\dagger}_{k}&=& c(0,0,E-v_F\pi_x, v_\Delta\pi_y)\nonumber\\
v^{2\dagger}_{k}&=& c(0,0,-v_\Delta\pi_y,E-v_F\pi_x)
\ee
with
\be{const}
c &=&\frac{1}{\sqrt{2E(E-v_F\pi_x)}}\nonumber\\
\ee 

If  $\pi_x=0$ and $\pi_y\ne 0$, the energy will be given solely by the
canonical momentum in the y-direction. In this case, the quasiparticles 
will be an equal mixture of particles and holes. 

\be{2spinor}
u^{1\dagger}_{k}&=& \frac{1}{\sqrt 2 E}(-\pi_y,E,0,0)\nonumber\\
v^{1\dagger}_{k}&=& \frac{1}{\sqrt 2 E}(E,\pi_y,0,0)\nonumber\\ 
u^{2\dagger}_{k}&=& \frac{1}{\sqrt 2 E}(0,0,E,\pi_y)\nonumber\\
v^{2\dagger}_{k}&=& \frac{1}{\sqrt 2 E}(0,0,-\pi_y,E)
\ee
In the case where $\pi_x\ne0$ and $\pi_y=0$ the gap function
does not have any effect. The Hamiltonian is already diagonal. However, I 
choose a representation of the eigenvectors, which simplifies later 
computations.  

\be{3spinor}
u^{1\dagger}_{k}&=& \frac{1}{2E}(-(E+\pi_x),E-\pi_x,0,0)\nonumber\\
v^{1\dagger}_{k}&=& \frac{1}{2E}(E-\pi_x,E+\pi_x,0,0)\nonumber\\ 
u^{2\dagger}_{k}&=& \frac{1}{2E}(0,0,E-\pi_x,E+\pi_x)\nonumber\\
v^{2\dagger}_{k}&=& \frac{1}{2E}(0,0,-(E+\pi_x),E-\pi_x)
\ee
Finally consider the zero mode:  $\pi_x=0=\pi_y$.
In this case, the Hamiltonian for the ground state is equal to zero. 
Thus, the eigenspinors can be chosen freely. The simplest choice is 
most convenient in this case, as it gives the most convenient representation 
for the spin operators. 
\be{4spinor}
u^{1\dagger}_{k}&=& (1,0,0,0)\nonumber\\
v^{1\dagger}_{k}&=& (0,1,0,0)\nonumber\\ 
u^{2\dagger}_{k}&=& (0,0,1,0)\nonumber\\
v^{2\dagger}_{k}&=& (0,0,0,1)
\ee

In general, 
the new operators, $a_i$ and $b_i$, are a mixture of particles and holes, similar 
to the s-wave case. In order to express the Hamiltonian in the new creation and 
annihilation, you have to rotate the spinors by a (momentum-dependent) unitary
matrix U($\vec k)$. You have to do this in each sector separately. After this 
rotation, the Hamiltonian is diagonal and can be written as:

\be{Haminab}
\mathcal{H}&=& \sum_{\vec k,s} E_{\vec k}\{ a_s^+(\vec k)a_s(\vec k) 
+ b_s^+(\vec k)b_s(\vec k)\}
\ee

In all the sectors, apart from the one where $A_x=A_y=-1$, the energy is
nonzero for all momenta. Thus, by filling up all the states with negative 
energy and leaving all the other ones empty, one obtains a unique
ground state. In the forth case, the ground state has zero energy
and at first it seems arbitrary, whether to fill it or not. 
For each node, the zero mode can be empty, completely filled or occupied by
either a spin-up or a spin-down particle. For the node pair (1-3) this means 
a 16-fold degeneracy of the ground state.  In order to label the ground states 
one needs to find a set of commuting operators. The eigenvalues of those 
operators don't change under adiabatically switching on a magnetic flux. Later, 
we will see that only two of the 16 ground states can be connected to the ground 
states of the other sectors by adiabatic processes.

In a Cooper pair, the electrons have opposite spin and momentum. Breaking up
a pair and, thus, creating a pair of quasiparticles does not conserve the 
charge but it does conserve spin and momentum. However, the spinors are 
defined in a way that spin-up particles have positive momenta, while spin-down
particles have negative ones. So momentum does not provide additional information 
for the states and I will ignore it. 
An additional quantum number is needed to give a complete characterization of 
the states. As mentioned earlier, the chirality is an appropriate choice. 
In the following section I will derive how to obtain the quantum numbers and 
discuss their physical meaning.

\subsection{Symmetries}

The original Hamiltonian is invariant under SU(2) spin rotations, so the spin 
current is conserved and the spin in the z-direction, together with the total 
spin, are  good quantum
numbers. However, when I introduced the spinors, I obscured this symmetry. For
a detailed analysis of how to obtain the spin in the different directions, see
Appendix A.  
\be {spinoperator}
\hat S_3&=& \half\hat \jmath^0\nonumber \\
        &=& \half\int d^2r \, \Psi^\dagger\ofr\Psi\ofr\nonumber\\
        &=& \half \sum_{s,q}(a^{\dagger}_s(\veq)a_s(\veq)-b^{\dagger}_s(\veq)
b_s(\veq))
\ee
The momentum is also diagonal in the new set of operators. 
\be{momentumoperator}
\hat P&=& \int d^2r \, \Psi^\dagger\ofr(-i\grad)\Psi\ofr \nonumber \\
      &=& \sum_{s,q}(a^{\dagger}_s(\veq)a_s(\veq)-b^{\dagger}_s(\veq)b_s(\veq))
\ee
However, spin-up is coupled to positive momenta and spin-down to
negative. Therefore, the momentum does not provide any new information and I
will omit it. 

Analogous to the 3+1 dimensional case, one can construct a fourth 
$\gamma$-matrix, $\gamma_3$, 
\be{gamma3}
\gamma_3&=& \left(\begin{array}{cc}
            0&\sigma_2\\ -\sigma_2&0\end{array}\right),
\ee 
which fulfills the Clifford algebra. 
In the 3+1 dimensional case, the chirality is defined by a symmetry transformation, 
which leaves the massless Dirac equation invariant. Its (hermitian) operator is 
given by the product of all $\gamma$-matrices: 

\be{gamma5}
\gamma_5&=& \gamma_0\gamma_1\gamma_2\gamma_3 \nonumber \\
&=& \left(\begin{array}{cc}
                  I&0\\0&-I\end{array}\right)
\ee
In 2+1 dimensions, $\gamma_3$ is redundant for the Dirac equation and can, therefore,
be combined with $\gamma_5$ to give a bigger internal symmetry group. 
Both matrices anticommute with the other $\gamma$-matrices. Therefore, there
is a set of three hermitian matrices, which anticommute with each other but commute
with the matrices $\alpha_1$ and $\alpha_2$. Explicitly, they are given by 
 $i\gamma_3$, $\gamma_5$ and $\gamma_{35}=\gamma_3\gamma_5$. One of these
can be included into the set of commuting operators. I will use $\gamma_5$ as
it is already diagonal. Using another notation, 
\be {chiralityalgebra}
\Gamma_1 &=& \half\gamma_3\gamma_5 \nonumber\\
\Gamma_2 &=& \frac{i}{2}\gamma_3 \nonumber\\
\Gamma_3 &=& \half\gamma_5
\ee
it is easy to see that they fulfill the spin algebra:
\be{spinalgebra}
[\Gamma_1,\Gamma_2] &=& i\Gamma_3
\ee
The eigenvalue of $\Gamma_3$, I call chiral charge, in analogy to spin and
electric charge. 
\be{chiralityoperator}
\Gamma_3&=& \half\int d^2r \, \Psi^\dagger\ofr\gamma_5\Psi\ofr \nonumber\\
        &=& \half\sum_k \,  a^{\dagger}_1(\vec k)a_1(\vec k)-a^{\dagger}_2(\vec k)
a_2(\vec k)
         - b^{\dagger}_1(\vec k)b_1(\vec k)+b^{\dagger}_2(\vec k)b_2(\vec k)
\ee
To gain further insight into the chirality, I rewrite it in the old variables. 
\be{chiralityoperatorold}
\Gamma_3 &=& \half\sum_k \, 
  \psiudag 1 \psiu 1 - \psiddag 3 \psid 3 - 
  \psiudag 3 \psiu 3 + \psiddag 1 \psid 1 
\ee
From (\ref{chiralityoperatorold}) it is apparent that 
$2\Gamma_3$ does nothing else than count the difference of the number of
particles at the node pair (1-3). In contrast, the ordinary number operator $\hat N$
is not a good quantum number and does not commute with the Hamiltonian. This
is what you would expect, since the quasiparticles are massless around the
nodes. Thus, their number is not conserved. 
However, these quasiparticles are created by destroying a Cooper pair. Creating a 
quasiparticle at eg. node 1 is always connected with creating one at node 3. The 
difference of the number of particles at node 1 and node 3 is unchanged by this 
process.  

In the isotropic case, $v_F=v_\Delta$, the chirality forms an U(4) symmetry 
group. For a more detailed analysis of this symmetry see \cite{Herbut}.  
By letting both velocities differ from each other, U(4) breaks down to 
U(2)xU(2). For each pair of nodes, the chirality behaves as the well-known 
spin algebra. Measurements on cuprates show, that the Fermi velocity is 
approximately one order of magnitude bigger than the velocity tangential to the 
Fermi surface, so the U(2)xU(2) description is appropriate. Thus, it is legitimate 
to regard the chirality as an isospin variable.

\subsection{Ground State Degeneracy}
I will now discuss the ground state degeneracy. The only non-trivial case is 
the fourth sector, where both $a_x$ and $a_y$ are non-zero. 
In the other sectors, the energy gap is always finite. The ground state is determined
by filling all the negative energy states and leaving the positive ones empty.
However, in the fourth sector there are states with zero energy and there is an 
arbitrariness, whether to fill these states or not.  

Let $\vec k_0$ be the momentum of this state, fulfilling
$$ \vec k_0 = \left(\begin{array}{c}\bar a_x\\\bar a_y\end{array}\right) $$
and let $\zero$ be the state, where the zero mode is empty, 
$$a_s(k_0)\zero = b_s(k_0)\zero = 0$$ 
Creating particles with momentum $\vec k_0$ does not require energy. Therefore, 
acting on $\zero$ with 
$a_s^\dagger(\vec k_0)$, $b_s^\dagger (\vec k_0)$ or a product of these 
operators, will also generate a ground state. As these operators fulfill 
anti-commutation relations, you can construct 16 
linearly independent ground states and label these states with spin and 
chirality. Both symmetries form one triplet, four doublets and five singlets. 
For a detailed analysis, see Appendix C. 

The ground states of the different sectors can be connected by tunneling 
processes as described in \cite{Hans}. Consider creating a 
vortex-antivortex pair. Then move  the vortex along a non-contractible, 
closed path and let them recombine afterward. Moving the vortex around a 
hole in the torus, creates a magnetic flux in that hole and changes the singular
part of the corresponding gauge field, $\bar{a}_x$ or $\bar{a}_y$, by $\pi$. 
Thus, doing this adiabatically, connects the different sectors.
This adiabatic process corresponds to slowly switching on the flux, $a_x$ 
respectively $a_y$. 

For each value of the flux, the spin in z-direction $S_z$, the total spin $S^2$, 
 the chirality $\Gamma_3$, which describes the difference of numbers of particles 
 and $\Gamma^2$ 
commute with the Hamiltonian. During an adiabatic process, the states remain 
eigenstates of these operators. 

So first I want to 
compute the eigenvalues of spin and chirality for the unique vacua in the 
other sectors. As $a_s\zero=b_s\zero =0$, it is straightforward to see 
that the unique vacua are eigenstates to all the operators with eigenvalue
zero.
In the degenerate sector, there are only two states which also fulfill these 
requirements: the first is a linear combination of the completely empty and the 
completely filled state, the other one is a combination of 2 excitations each:  
\be{4sectorgs}
\phi_1&=& \frac{1}{\sqrt{2}} 
\left(1+\frac{1}{L_x^2L_y^2}a_1^\dagger(k_0)a_2^\dagger(k_0)b_1^\dagger(k_0)
b_2^\dagger(k_0)\right)\zero \nonumber\\
\phi_2&=& \frac{1}{\sqrt{2}L_x L_y}\left(a_1^\dagger(k_0)b_1^\dagger(\vec k_0) + 
a_2^\dagger(\vec k_0) b_1^\dagger(\vec k_0)\right)\zero
\ee

Thus, if you consider the system under adiabatic processes, the huge degeneracy is
reduced from 16 to two. Depending on the order in which the fluxes were turned on, one of those two is obtained. If the flux in the x-direction is switched on first, one obtains $\phi_1$ at the nodal pair (1,3) and $\phi_2$ at nodal pair (2,4). Taking both nodes into account, there are four states which are singlets in spin and chirality.  However, only two of those can be reached by the tunneling processes I considered.

\subsection{Summary of the first part}

In the first part of my thesis, I derived the effective low-energy theory of d-wave superconductors. As it could be expected, the appropriate description is a massless Dirac Lagrangian, given by equation (\ref{after integrating}). Furthermore, we saw that the conserved current is nothing else than the spin current which couples to a topological gauge field $a_\mu = -\half \partial_\mu\varphi_s$. We then proceeded by analyzing the topological theory on the torus\footnote{including magnetic flux tubes through the holes} and solving for the energy. The results we found, differ mainly in two aspects from the ones you expect in a conventional topological theory (with gapped quasiparticles): the ground states in the four different sectors are not degenerate, in contrast to the gapped case. In addition, in one sector the ground state is not unique but there is instead a rather huge degeneracy, in total given by 256 states. Finally, we argued why you could focus on only two of these ground states. These were the ones, which could be connected to the ground states of the other sectors by tunneling processes.  The structure we obtained for d-wave superconductor differs from the one for s-wave superconductors. It does not show topological features, such as the ground state degeneracy on the torus. The second part of this thesis deals with the question whether there are other regularities in the theory due to the topology.

\newpage

\section{Effects of a local perturbation}

\subsection{general remarks}

In the first part, I derived a topological description for the low-energy 
excitations and solved for the energy spectrum on a torus in the different flux 
sectors.  I proceeded by discussing the novel features that arise because of the 
gapless excitations, emphasizing  that there are several important differences to 
the fully gapped case.  The most significant being, that gapless modes destroy 
the (topological) ground state degeneracy and make the system sensitive to 
local perturbations.  However, I will argue that  perhaps there are topological features left, even in the presence of gapless particles. 

For s-wave superconductors, one obtained a four-fold ground state degeneracy on the torus, which was solely determined by the topology \cite{Hans}.  The ground states of the different sectors can be connected by adiabatic tunneling processes.  The energy gap of the ground states in the different sectors is proportional to the exponential factor $e^{-c\epsilon n L}$, where $\epsilon$ is the energy of the quasiparticle, n the winding number, L the size of the torus and c a normalization constant .  Hence, the ground states become degenerate in the limit of a large torus\footnote{In this limit, tunneling processes are suppressed as the action of the quasiparticle $S=\frac{\epsilon}{v_F} n L$ gives an exponential fall-off in Euclidean space}. 
This argument does not hold if there are gapless excitations. The 
quasiparticles are massless and free to tunnel, even if the torus becomes very large. Therefore, it is to be expected that the degeneracy between the different sectors is lifted. The gapless modes also cause one of the sectors to behave exceptionally as they give rise to a zero mode and, thus,  to a large ground state degeneracy.  I argued that only two of the states can be connected to the other sectors via adiabatic processes. In the following discussion I focus on these two states and their behavior in relation to the unique ground states in the other sectors. 

I have already argued that some characteristics of topological orders (such as the topological degeneracy on the torus) are destroyed by the gapless excitations. However, the superconductor is still described by a topological theory and even in the presence of gapless modes, there may be topological features left. In order to analyze this possibility, I consider local perturbations and study their effects on the ground state energies. In particular, we look for quantities which show some regularities in the different sectors. The idea is, to include a magnetic point defect at  point $\vec x_0 $ with strength $V_0$: $\hat{V}= V_0\, \delta(\vec x-\vec x_0)$. The coordinates of the impurity are arbitrary. They do not affect physical results as the the torus is translational invariant. In the fully gapped case, such an impurity does not effect the ground state degeneracy. In order to distinguish between the different sectors, the quasiparticles have to travel around large cycles and measure the flux through the holes. Those paths are not suppressed any longer in the gapless case and the local perturbation will shift the energy levels. However, it is not obvious, whether these shifts are distinct for each sector or whether some regularity reflects the topological degeneracy of the gapped system. 

The perturbed Hamiltonian is given by:
\be{perturbedHamilt}\mathcal{H}&=&\mathcal{H}_0 + \hat{V}\nonumber \\ &=& \int d^2x \bar{\Psi}_1(\vec x)\left(-iD_xv_f\gamma_1-iD_yv_\Delta\gamma_2\right)\Psi_1(\vec x) \nonumber \\& & +  \bar{\Psi}_2(\vec x)\left(-iD_yv_f\gamma_1-iD_xv_\Delta\gamma_2\right)\Psi_2(\vec x) \nonumber \\& & + V_0\left(\Psi^\dagger_1\delta(\vec x-\vec x_0)\Psi_1(\vec x)    + \Psi^\dagger_2\delta(\vec x-\vec x_0)\Psi_2(\vec x)\right)\ee

The characteristic quantity for the vacuum is given by the Casimir energy.  It is distinct for each flux sector and is, in general, shifted by the perturbation. The shift, however, could be a quantity which is independent of the sectors and only  determined by the topology. In the following sections I calculate the energy shift in the Casimir energy in all four sectors and compare the results. 

The most convenient method to calculate the Casimir energy is, as mentioned earlier, by using Green's functions. In section 4.2, I derive how the Green's function is related to the vacuum energy:
\be{vacuum energy}
E &=& \int d^2x \int \frac{d\omega}{2\pi} \, i\omega\Tr(\hat{G}(\vec x,\vec x,\omega)\gamma_0)
\ee
where $\hat{G}(\vec x,\vec x,\omega)$ is the full propagator. 

The reader may note that special care is needed, when zero modes are concerned. Whether this mode is filled or not,  affects the shift of all the energy levels. This contribution can be computed in two ways.  One possible way is to include this mode in the Green's function. An occupied zero mode simply shifts the Green's function by a constant. Another option is to exclude the zero mode explicitly and calculate its contribution separately. The Green's function representation I use to compute the Casimir energy,  is not suitable when there is a zero mode present.  To first order, both approaches are equivalent and I will compute the shift in both ways. However, to compute the shift to second order you have to exclude the zero mode from the Green's function as it gives raise to a (fake) infrared divergence. 

At this point some general remarks on the interaction potential are in order. Choosing a point defect is certainly the simplest perturbation, one can consider. However, there are some well-known difficulties with a singular interaction. It scatters into all momentum states with the same strength. Therefore, it does not provide a cutoff for high momenta and leads to ultraviolet divergencies. Normally, this divergence occurs already in the integrals for the first order shift. In our case, however,  only scattering at the impurity can change momentum, whereas the insertion of $i\omega\gamma_0$ at $\vec x$ is momentum conserving. Therefore, to first order only scattering processes contribute, where the interaction does not change the momentum eigenvalue of the quasiparticles. 

The Casimir energy is obtained by regularizing the vacuum energy on the torus as well as in free space. When the vacuum energy on the torus is expressed as a sum over windings, the free space vacuum energy is subtracted by omitting the term with no windings.  As soon as windings are included, the size of the torus yields a natural length scale and provides a cutoff for high momenta. This is sufficient to ensure a finite expression for the energy up to the first order shift, but fails for second order. When the particle scatters two times at the impurity, it may have any momentum between the two scattering events. This scattering process yields an infinite result, even after the subtraction of the free space vacuum energy. The energy shift can be regularized by splitting the point of interaction and assuming a finite distance between both scattering events. If the quasiparticle does not wind around one of the holes of the torus, then one encounters a short distance divergence in the limit where both points coincide. However, windings between the scattering events are not sufficient to render the integrals finite.  The integration over the torus surface diverges also when the interaction points coincide.  In addition to subtracting the vacuum energy of the free space, one must also renormalize the interaction strength by adding local counterterms. 

As mentioned earlier, the delta potential does not conserve momentum. More specifically, it introduces an interaction between particles at arbitrary momentum. Thus, the decoupling of the node pairs is not valid any more and particles get scattered between different nodal pairs. To first order, this process is irrelevant as $i\omega\gamma_0$ is momentum independent and the trace ensures an overall momentum conservation. There the different nodal pairs can still be regarded as independent and the total shift is given by the sum of two independent contributions. In the second order approximation, the nodal pairs (1,3) and (2,4) can mix.  As there are two scattering events, the overall momentum conservation does not prohibit the particle to be scattered to another node pair. The way to implement such processes is straightforward.  The divergent behaviour and the corresponding regularization techniques are the same for both nodal pairs. They differ only in the coupling of the x- and y-direction to the characteristic velocities and $\gamma$ matrices: $v_F\gamma_1$ and $v_\Delta\gamma_2$. Therefore, the only difference is  given in the dependence on the parameters $L_x$,$L_y$, $v_F$ and $v_\Delta$.

\subsection{Casimir energy}
In this section, I first give a general derivation of how the Green's function (in Euclidean space) can be related to the vacuum energy density. We obtain the Casimir energy density by subtracting the free space energy density from the one on the torus. I also give a derivation how the Green's function looks like in our specific case of 2d massless fermions on the torus. For each sector without a zero mode, there will be a unique Green's function which will, in general, lead to different energy densities in the different sectors. I also derive, how the zero mode affects the Green's function and specify it for the degenerate ground states in that sector.  

In general, the fermionic Green's function is written as the time-ordered product of the spinor fields:
\be{euclgreen}
\hat{G}_{\alpha,\beta}(\vec x,\vec y,\tau- \tau')&=&\zerod T\{\Psi_\alpha(\vec x,\tau)
\bar{\Psi}_\beta(\vec y,\tau')\}\zero
\ee
expressed in imaginary time $\tau = it$. It is connected to the real-time Green's function by 
$\hat{G}_{\alpha,\beta}(\vec x,\vec y, \tau)\big|_{\tau=it} = i \hat{G}_{\alpha, \beta}(\vec x,\vec y, t)$.
The spinors can be expanded in the (positive and negative) energy eigensolutions $\Psi^\pm_n$ of the total Hamiltonian:
\be{eigenspinor}
\Psi_\alpha(\vec x,\tau)& =&\sum_n \Psi^+_{\alpha,n}(\vec x)e^{(-\epsilon_n^+ \tau)} +
\Psi^-_{\alpha,n}(\vec x)e^{(-\epsilon_n^-\tau)}\nonumber\\
&=& \frac{1}{\sqrt{V}}\sum_{\vec p, s} a_s(\vec p)u_s(\vec p)e^{-i\vec p\vec x}+b^\dagger_s(\vec p)v_s(\vec p)e^{i\vec p\vec x}
\ee
with the convention that
$$ a_s(\vec p)\zero = b_s(\vec p)\zero=0 $$
$$ \zerod a_s^\dagger(\vec p) = \zerod b_s^\dagger(\vec p) = 0$$

To calculate the vacuum energy, all the negative energies $\epsilon_n^-$ have to be summed up. For that purpose, I assume\footnote{to choose $\tau-\tau' <0$ ensures, that the sum runs over the filled levels in the Fermi sea, for $\tau>\tau$ one sums instead over all empty states} $\tau<\tau'$ and take the derivative of the Green's 
function with respect to $\tau$:
\be{derivgreen}
\partial_\tau \hat{G}_{\alpha,\beta}
(\vec x,\vec y, \tau-\tau')\Big |_{\tau<\tau'}&=&
\sum_{n,m}\epsilon_n^-\zerod\{\bar{\Psi}_{\beta,n}^+(\vec y)
\Psi_{\alpha,m}^-(\vec x)\}\zero
e^{(-\epsilon_n^-\tau+\epsilon_m^-\tau')}
\ee
By using the anticommutation relations it is straightforward to see that
\be{greenenergy}
\partial_\tau \int d\vec x \, tr(\gamma_0 \hat{G}(\vec x,\vec x,\tau-\tau'))\big|_{\tau'>\tau}
&=&\sum_n \epsilon_n^-\,e^{\epsilon_n^-(\tau'-\tau)}
\ee
This sum\footnote{In the formula above, you can already see  that the Green's function formalism may be problematic in presence of the zero modes as it is ambigous whether you include the zero mode (and its shift) in the sum or not. } is still regulated by $\tau'-\tau$ and finite as long as $\tau'-\tau>0$. In the limit 
$\tau \rightarrow \tau'$, the vacuum energy is obtained:  
\be{vacuum energy1}
E&=& \int d^2x \, \Tr\left(\partial\tau \hat{G}(\vec x,\vec x,\tau)
\gamma_0\right)\big|_{\tau\rightarrow 0^-}\nonumber\\
&=&\int d^2x \int\frac{d\omega}{2\pi}\, i\omega \Tr\left(
\hat{G}(\vec x,\vec x,\omega)\gamma_0\right)
\ee
This integral diverges in the limit $\tau\rightarrow 0^-$ and it must be regularized. In addition, one has to subtract the part which corresponds to free space. For that purpose, it is helpful to rewrite the propagator. A formal derivation of the Poisson resummation formula can be found in Appendix D. However, you can also give an argument on how to rewrite the propagator by looking at the path integral formalism.  

Consider the probability amplitude $K( \vec x_a, \vec x_b)$ for a particle to go from some initial point $\vec x_a$ to a final point $\vec x_b$. 
\be{amplitude}
K( \vec x_a, \vec x_b)&=& \sum_{\mbox{paths p}}e^{iS[p]}
\ee
As you can only measure absolute values of the probability amplitude, you can also add a constant phase $i\phi$ to the action $S[p]$.  On the torus, all paths  can be divided into classes. Two paths are in the same class, if they can be continuously transformed into each other. As paths around the handles of the torus can not be contracted to a point, it is clear that the classes can be labeled by the winding numbers. Thus, the sum over all paths can be broken into two sums, where the one over windings encodes all the information of the topology. The remaining sum yields nothing else but the probability amplitude in free space.  In each of the classes, the phase $i\phi$ has to be constant, but as the paths of different classes cannot be transformed continuously into each other, they may be different constants in different classes. In this example, the phase can be related to the magnetic flux quanta enclosed by the path by $\phi=i\int d^2x \, \vec a$. 

In my case, there is also a further contribution originating from the boundary conditions. In order to obtain a unique ground state in the non-flux sector, I chose the boundary conditions on the spinor fields to be antiperiodic. This boundary conditions has the same effect as inserting one flux quantum into each one of the holes. More specifically, by inserting flux quanta, one can switch between periodic and antiperiodic boundary conditions.  Windings around a hole with no flux quanta (more generally, an even number of flux quanta) have an alternating factor. Adding one flux quanta, consequently, eliminates that factor.

Now we can insert the (total) Green's function into (\ref{vacuum energy1}) and compute the Casimir energy and its shift in presence of a perturbing potential. 
The interaction will be regarded as small and we switch to the interaction picture:
\be{HInt}
\mathcal{H}_I(\tau)&=& \int d^2x \,e^{H_0\tau}\hat{V}(\vec x)e^{-H_0\tau}\ee 

Then, the Green's function in expression (\ref{greenenergy}) is expanded with regard to the interaction strength $V_0$:

\be{greenexpt}
\hat{G}(\vec x, \vec x, \tau,\tau')&=& \sum_{i=1,2}\zerod T\{\Psi_{i,\alpha}(\vec x,\tau)\bar{\Psi}_{i,\beta}(\vec x, \tau')e^{-\int d\tau_0 H_I(\tau_0)}\}\zero\nonumber \\
&=&  \sum_{i=1,2}\zerod T\{\Psi_{i,\alpha}(\vec x,\tau)\bar{\Psi}_{i,\beta}(\vec x, \tau')\}\zero\nonumber \\
&-& \sum_{i=1,2}\int d\tau_0\,\zerod T\{\Psi_{i,\alpha}(\vec x,\tau)\bar{\Psi}_{i,\beta}(\vec x, \tau')H_I(\tau_0)\}\zero\nonumber \\
&+& \sum_{i=1,2}\int d\tau_1\int d\tau_2 \,\zerod T\{\Psi_{i,\alpha}(\vec x,\tau)\bar{\Psi}_{i,\beta}(\vec x, \tau')H_I(\tau_1)H_I(\tau_2)\}\zero\nonumber\\
&+& \mathcal{O}(V_0^3)
\ee
By using Wick's theorem, one can write the time ordered product as the normal ordered product and the sum over all contractions. The contractions are nothing but the non-interacting propagators. In the sum, I neglect all contributions which correspond to disconnected Feynman diagrams. Note that the non-interacting propagator is always "located" at one node pair\footnote{The nodal pairs (1,3) and (2,4) are independent without the perturbing potential}. I denote the pair (1,3) with a superscript (I) and (2,4) with a superscript (II) respectively. 

The impurity is taken to be rigid. It can transfer momentum but not energy to the scattering particles. All Green's functions are, therefore, evaluated at the same frequency $\omega$. This can also be shown formally, by doing a Fourier transform of each Green's function. The time integrals then enforce energy conservation. 

\be{greenexpw}
\hat{G}(\vec x,\vec x,\omega)&=& \sum_{i=I,II}\{G^{(i)}(\vec x,\vec y,\omega)\nonumber\\
&-&V_0G^{(i)}(\vec x,\vec x_0,\omega)\gamma_0G^{(i)}(\vec x_0,\vec x, \omega)\nonumber \\
&+&V_0^2\sum_jG^{(i)}(\vec x,\vec x_0,\omega)\gamma_0G^{(j)}(\vec x_0, \vec x_0, \omega)\gamma_0G^{(i)}(\vec x_0,\vec x,\omega)\}\nonumber\\
&+& \mathcal{O}(V_0^3)
\ee
With $G^{(i)}(\vec x,\vec x_0,\omega)$ I denote the unperturbed propagator on the torus, whereas $\hat{G}(\vec x,\vec y,\omega)$ is the full interacting propagator. For simplicity, I chose to omit the index for the different sectors. However, I keep the index for the nodal pairs explicit as it will be needed in the calculations for the second order shift.

To keep track of the various terms in the expansion, it is convenient to introduce Feynman diagrams to label the different summands. The notation originates from the standard notation for the expansion of a propagator going from an initial point $\vec x$ to a final point $\vec y$. All the propagators are assumed to be at the same frequency $\omega$.

\begin{figure}[ht]
\centering
\scalebox{0.3}{\includegraphics{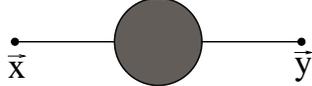}}
\caption{standard representation for an interacting  greens function, with distinct initial and final  point}
\label{fig:green}
\end{figure}

\begin{figure}
\centering
\scalebox{0.2}{\includegraphics{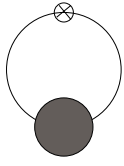}}
\caption{\label{fig:torusgreen} diagrammatic representation of the vacuum, or Casimir energy as an interacting Green's function, where initial and final point coincide}
\end{figure}

According to  (\ref{vacuum energy1}), the initial and final point are identified with each other. Thus, each diagram is closed into a loop. Moreover, the operator $i \omega \gamma_0$ is inserted at point $\vec x$. The trace and both integrations are implicit in all expressions. For the Feynman diagrams of the total Casimir energy, see figure \ref{fig:torusgreen}

Each propagator $G^{(i)}(\vec x,\vec y,\omega)$ can be written as a sum over windings of the free propagators $G^{(i)}_0(\vec x +\vec m,\vec y,\omega)$. The free propagator itself can be separated into two pieces, which differ in their short distance behaviour.  One couples to $\gamma_0$, the other to $(\vec x -\vec y)\vec{\gamma}$.  As the commutation relations for the Dirac matrices give constraints on the diagrams, it is convenient to label each propagator by the Dirac matrix which is picked up. In addition, the winding number can be specified.  The index for the node pair is, however,  kept implicit in all diagrams. The free propagator at the different nodes differ only in its dependence on the parameters $L_x$, $L_y$, $v_F$ and $v_\Delta$:

\be{besselgreenI+II}
G_0^{(i)}(\vec{x},\vec{y},\omega)&=&\frac{1}{2\pi v_Fv_\Delta}\{-i\omega K_0^{(i)}(|\omega|
|\vec{x}'-\vec{y}'|)\nonumber \\
&+&|\omega|
\frac{\vec{\gamma}(\vec{x}'-\vec{y}')}
{|\vec{x}'+-\vec{y}'|}K_1^{(i)}(|\omega||\vec{x}'-\vec{y}'|)\}
\ee
where $x_1'= \frac{x_1}{v_F}$ and $x_2'=\frac{x_2}{v_\Delta}$ for the nodal pair (1,3). For the other nodal pair, the velocities have to be switched. The full propagator is written as a sum over windings of the free propagators. 
\be{fullbessel}
G^{(i)}(\vec{x},\vec{y},\omega)&=&\sum_{\vec n}\hat{f}_i(\vec n)G_0^{(i)}(\vec x+\vec n, \vec y, \omega)\nonumber \\
&=&
\sum_{\vec n}\hat{f}_i(\vec n)\frac{1}{2\pi v_Fv_\Delta}\{-i\omega K_0^{(i)}(|\omega|
|\vec{x}'+\vec n'-\vec{y}'|)\nonumber \\
&+&|\omega|
\frac{\vec{\gamma}(\vec{x}'+\vec n'-\vec{y}')}
{|\vec{x}'+-\vec{y}'|}K_1^{(i)}(|\omega||\vec{x}'+\vec n'-\vec{y}'|)\}
\ee
where the winding numbers are given in multiples of $\frac{\vec L}{2\pi}$. In this thesis, I will mostly  write $\vec x$ instead of $\vec x'$ and keep the dependence on the nodes implicit in the expressions. 

Special care is needed in the zero mode sector. In general, the time ordered product of operators is given by the normal ordered product and the contractions. When there is a zero mode, this mode can be partly filled. This will of course change the notion of normal ordering. When one writes all creation operators to the right of the destruction operators, this order will, in general, not destroy the ground state if this state is not empty. The zero mode is special in that respect, that it does not depend on time or the spatial variables. Thus, the Green's functions for the different ground states differ only by a constant. 
\be{zeromodeGreen}
\tilde{G}^{(i)}_{\alpha\beta}(\vec x,\vec y,\tau)&=&\langle\tilde 0|T\{\Psi_{i\alpha}(\vec x, \tau)\bar{\Psi}_{i\beta}(\vec y,0)\}|\tilde0\rangle\nonumber\\
&=&\hat{G}^{(i)}_{\alpha\beta}(\vec x,\vec y,\tau)\nonumber \\
&+& \langle\tilde0|\Psi^-_{i\alpha}(\vec x,\tau)\bar{\Psi}^+_{i\beta}(\vec y,0)-\bar{\Psi}^-_{i\beta}(\vec y)\Psi^+_{i\alpha}(\vec x,\tau)|\tilde 0 \rangle \nonumber\\
&=& \hat{G}^{(i)}_{\alpha\beta}(\vec x,\vec y,\tau)+M^{(i)}_{\alpha\beta}
\ee

The constant contribution becomes simple, when we look at $\hat{G}\gamma_0$ instead as $M^{(i)}\gamma_0$ is diagonal. Using the anticummutation relations, it is easy to see, which matrices you pick up when the zero mode is occupied by a quasiparticle: 

\be{zeromodeContr}
b_1(\vec k_0) &\rightarrow& E_{22}\nonumber \\
b_2(\vec k_0)&\rightarrow& E_{44}\nonumber \\
a_1(\vec k_0)&\rightarrow& -E_{11}\nonumber \\
a_2(\vec k_0)&\rightarrow& -E_{33}
\ee
Analogously, for states with several quasiparticles the total contribution is obtained by the sum of the single ones. For example,  the ground states $\phi_1$ and $\phi_2$ give the same contribution:
\be{zeromodesinglett}
\phi_1&\rightarrow& \frac{1}{\sqrt{2}V}\left(\begin{array}{cc} \sigma_3&0\\0&\sigma_3\end{array}\right)\nonumber\\
\phi_2&\rightarrow& \frac{1}{\sqrt{2}V}\{\left(\begin{array}{cc} \sigma_3&0\\0&0\end{array}\right)+\left(\begin{array}{cc} 0&0\\0&\sigma_3\end{array}\right)\}=\left(\begin{array}{cc} \sigma_3&0\\0&\sigma_3\end{array}\right)
\ee

The different notion of normal ordering does not give any contribution for the non-interacting case, as the zero mode contribution is cancelled by the time derivative and does not give any contribution to the Casimir energy. However, we will have to take it into account in other orders of perturbation theory.

\subsection{Numerical results in zeroth order}

The Casimir energy in absence of a perturbing potential can be written in a closed form, given by:

\be{casimirenergy}
E^{(0)}_j &=& {\sum_{i,\vec{n}}}^{'} \hat{f}_j(\vec{n}) 
\int d^2x \, \int_{-\infty}^\infty \frac{d\omega}{2\pi} \, 
\Tr( i\omega\gamma_0G_0^{(i)}(\vec x+\vec{n},\vec x,\omega))\nonumber \\
\ee
where the winding numbers $n_i$ are given in multiples of $\frac{L_i}{2\pi}$
and $\hat{f}_j(\vec n)$ denotes the alternating factors which depend on the sector j.  

After inserting the definition of the Green's function and using the trace properties of the $\gamma$-matrices, it is straightforward to see that only the  term coupled to $\gamma_0$ survives. Both integrals can then be performed. The one over space is trivial and gives just the surface of the torus. The sum is regularized by omitting the term without windings and a finite energy is obtained in all four sectors. 

\begin{figure}[tb]
\begin{minipage}{12cm}
\centering
\scalebox{0.2}{\includegraphics{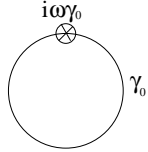}} 
\label{fig:torusgreen0g}
\caption{lowest order graph corresponding to figure (\ref{fig:torusgreen})}
\end{minipage}
\end{figure}

\be{casimirfinal}
E^{(0)}_j&=&{\sum_{n,m}}^{'}\hat{f}_j(n,m)
8\pi\sqrt{\frac{v_fv_\Delta}{L_1 L_2}}((
n^2\frac{v_\Delta L_1}{v_F L_2}+m^2\frac{v_F L_2}{v_\Delta L_1})^{-3/2}\nonumber \\
& &+ (
n^2\frac{v_F L_1}{v_\Delta L_2}+m^2\frac{v_\Delta L_2}{v_F L_1})^{-3/2})
\ee

First, one can try to draw some conclusions from the analytic expression. For example, one can compare the sectors, where there is one flux tube through one of the holes. The energy shift of these sectors differs only in the dependence on the torus length. 
You would expect the ground states in these two sectors to behave similarly because of symmetries in the original d-wave descripton. In fact, if the length scales are the same $L_x=L_y$, their energy shift is exactly the same. However, the more the two length scales differ, the more differs also the Casimir energy in both sectors. This is not a coincidence, but originates in the way we quantized the momentum in the first place. After expanding around the nodes, we rotated the coordinate system and chose to quantize the theory in terms of the new coordinates. Thereby, we assigned the different length scales to the node pairs in an arbitrary manner. If we had chosen to quantize in the old set of coordinates, the energy dependence on the length scale would most probably have been symmetric at the two nodal pairs and, thus, also their behaviour for varying this scales. Of course the way to quantize the momentum should not make a difference in the final result.  However, the description would be more involved when quantized in terms of the old coordinate system and it is not straightforward to see that the argumentation can be generalized to different boundary conditions in a simple way. 

For the numerical calculation, I assume the two lengths of the torus to be equal to each other: $ L_x=L_y=L$. Using that choice ensures that the two nodal pairs act in the same way. In addition, the velocities $v_F$ and $v_\Delta$ differ approximately by one order of magnitude in cuprates,\cite{}. Therefore, I assume $ \frac{v_F}{v_\Delta}=10$ in the calculations.  After inserting these values, the Casimir energy is given by:
\be{finalCasimir0}
E&=& 800\pi\frac{v_F}{L}\sum_{n,m}\,'(-A_x)^n(-A_y)^m\cdot\nonumber\\
&&(\frac{1}{\sqrt{n^2+(10m)^2}^3}+\frac{1}{\sqrt{(10n)^2+m^2}^3})
\ee
where $A_x$ and $A_y$ are the Wilson loops introduced in section 3.2. The Casimir energy can easily be computed in units of $v_F/L$ in the four sectors:
\begin{displaymath}
\begin{array}{c|c|c|c|c}
(A_x,A_y)&(1,1)&(1,-1)&(-1,1)&(-1,-1)\\\hline
E&-99063.3&1427.87&1427.87&12415.1
\end{array}
\end{displaymath}

One finds that the the Casimir energy of the first sectors \footnote{with no magnetic fluxes present} is lowest,  while the sector with the degenerate ground states has the highest Casimir energy. The other two sectors with only one flux quantum have the same energy \footnote{as both lengths coincide} which lies approximately in the middle of the other two. The splitting of the ground state energies is algebraic in L,  $\Delta E\sim 1/L$ instead of the exponential splitting in the s-wave case.

\subsection{Perturbation theory to first order}

The first order shift in the Casimir energy corresponds to the physical process were a 
quasiparticle scatters once at the impurity. It is calculated by inserting the first 
order pertubation of (\ref{greenexpw}) into (\ref{vacuum energy1}). Using  the trace 
formulas of the Dirac matrices, only two terms survive:
\begin{eqnarray*}
\begin{minipage}{5cm}
\centering
\scalebox{0.2}{\includegraphics{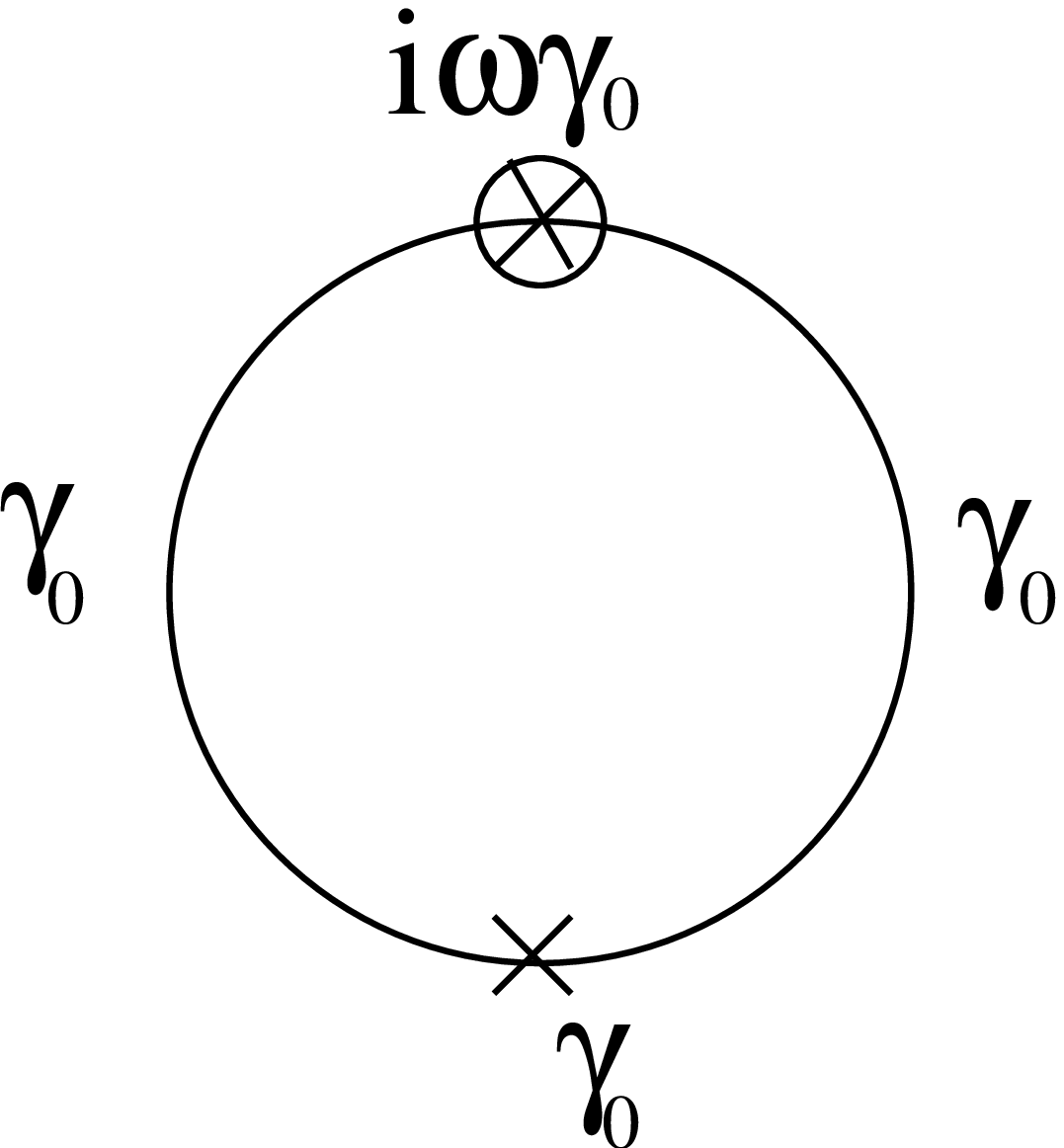}}\\
\label{fig:torusgreen10} 
\end{minipage}
&&
\begin{minipage}{5cm}
\centering
\scalebox{0.2}{\includegraphics{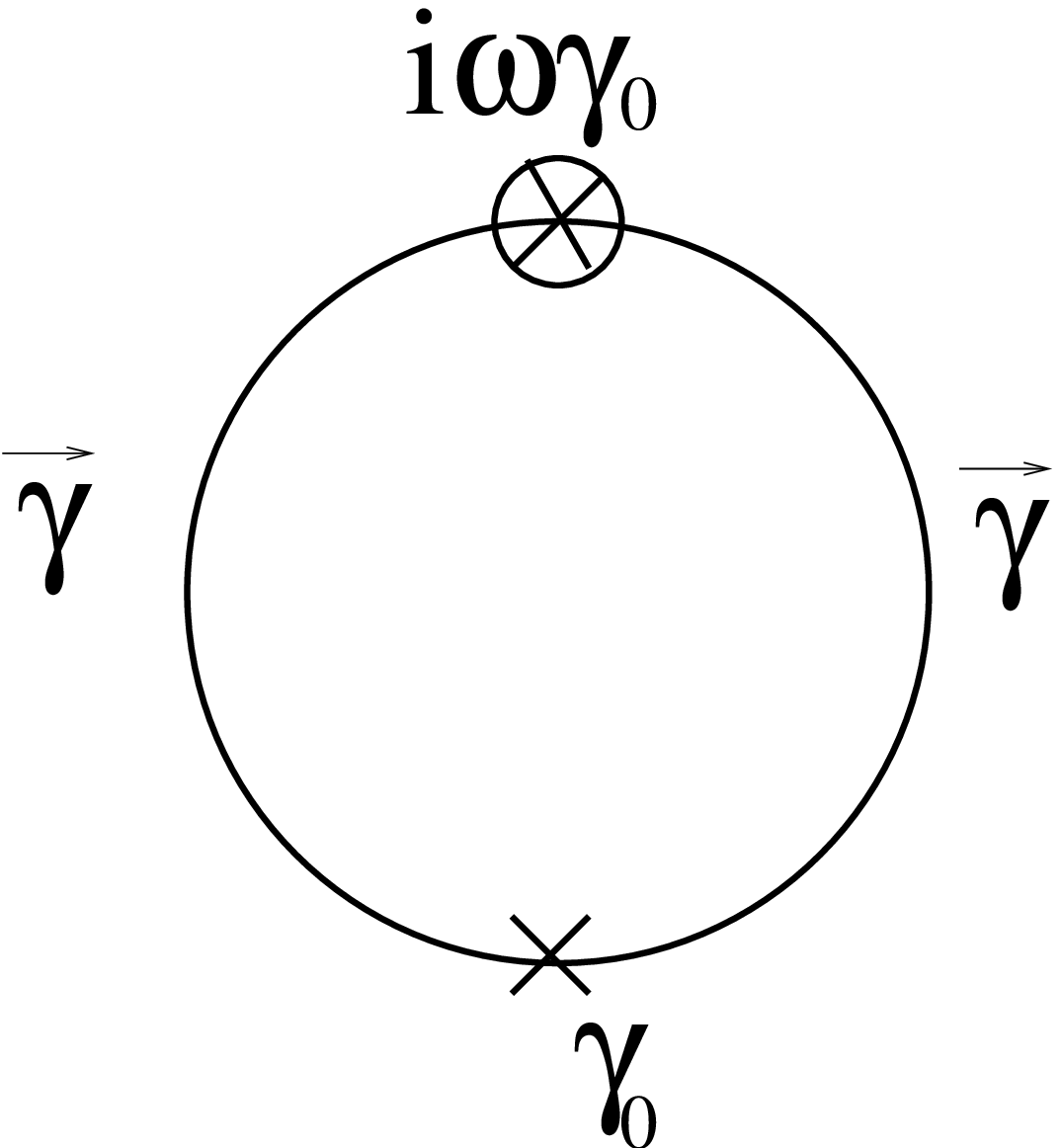}} 
\label{fig:torusgreen11} 
\end{minipage}
\end{eqnarray*}
\be{Casfirst}
\Delta E^{(1)}&=&- V_0\sum_{\vec{n},\vec{m}} \int d^2x\int \frac{d\omega}{2\pi}\, 4 i\omega(
(\frac{-i\omega}{2\pi})^2 K_0(|\omega||\vec x+\vec n-\vec x_0|)
K_0(|\omega||\vec x_0-\vec m-\vec x|)\nonumber \\
&+ &(\frac{|\omega|}{2\pi})^2 
\frac{(\vec x+\vec{n}-\vec{x}_0)(\vec x_0-\vec{x}-\vec{m})}
{|\vec x+\vec{n}-\vec{x}_0||\vec x_0-\vec{x}-\vec{m}|}
K_1(|\omega||\vec x+\vec{n}-\vec{x}_0|)
K_1(|\omega||\vec x_0-\vec{x}-\vec{m}|))\nonumber\\
\ee

However, both expressions are odd in $\omega$ while the integrals extends over all values from $-\infty$ to $\infty$. Thus, every contribution from a positive 
frequency is cancelled by an equally big one for the negative counterpart and the integral vanishes. Therefore, the energy shift to first order vanishes in the sectors without the zero mode. 

Note, that this argument does not apply for $\omega=0$, so the contribution of the zero mode must be computed in an alternative way. As it is just one mode, one can use standard degenerate perturbation theory. The matrix elements of the perturbing potential are given by:
\be{potDef}
\langle \vec k|\hat{V}|\vec q\rangle&=&\sum_{\vec{k},\vec{q},s}
a^\dagger_s(\vec{k})a_s(\vec q)u^\dagger_s(\vec k)u_s(\vec q)
-b^\dagger_s(\vec{q})b_s(\vec{k})v^\dagger_s(\vec k) v_s(\vec q)
\nonumber\\
& &+a^\dagger_s(\vec{k})b^\dagger_s(\vec{q})u^\dagger_s(\vec k)v_s(\vec q)
+b_s(\vec k)a_s(\vec q)v^\dagger_s(\vec k) u_s(\vec q)
\ee
To first order perturbation theory, the potential does change neither the energy nor the momentum of the scattering quasiparticle. This can be seen from the Feynman diagrams as explained above, but it is also a straightforward consequence of degenerate perturbation theory. In first order, the perturbation scatters only to other state in the degenerate manifold. However, when restricted to $\vec k_0$, the interaction becomes diagonal in the quasiparticle operators and is given by:
\be{potDefspecial}
\hat{V}&=& V_0\sum_{s}a^\dagger_s(\vec k_0)a_s(\vec k_0)-b^\dagger_s(\vec k_0)b_s(\vec k_0)
\ee

Generally, the perturbing potential lifts the degeneracy. In this case, the 
degeneracy is only partly lifted. The kets are chosen as the spin and chirality eigenstates which are defined in appendix B. In this representation the impurity interaction is diagonal and the energy shift is proportional to the difference in the number of particles and antiparticles. 
\be{shift}
\Delta E^{(i)} &=& \langle i|\hat{V}|i \rangle \nonumber\\
&=& V_0 \sum_s  \langle i|\hat{N}_{a,s}-\hat{N}_{b,s}|i \rangle
\ee
As particles have a "charge" $+\frac{1}{2}$ and antiparticles $-\frac{1}{2}$, the energy shift is proportional to the total charge, i.e. the spin eigenvalue of the state. Consequently, the ground states which have $S_3$-eigenvalue 0 are not shifted and behave as the ground states in the other three sectors. The two states which can be connected to the ground states in the other sectors are spin singletts, and therefore, their energy shift (to first order) is the same as for the other sectors. 

However, first order perturbation does not distinguish whether the states are 
singlets in chirality. So the energy levels are unperturbed also for other 
states than the ones, which can be connected to the other three sectors by adiabatic tunneling processes. Thus, it is of interest to consider also the shift in second order perturbation theory and see whether this shift distinguishes between the chirality eigenstates.

The same calculation can be done using Green's functions. In first order perturbation theory, one needs to consider 
\be{zeromode1}
\Delta E^{(1)}&=&-\sum_i\int d^2x\int d\tau_0 \partial_\tau (\gamma_0)_{\beta\alpha}\nonumber \\
&& \langle\tilde 0|T\{\Psi_{i\alpha}^\dagger(\vec x, \tau)\bar{\Psi}_{i\beta}(\vec y,0)V_0 \Psi_{i\sigma}^\dagger(\vec x_0, \tau_0) \Psi_{i\sigma}(\vec x_0,\tau_0)\}|\tilde0\rangle\nonumber\\
&=&-V_0\Tr\sum_{\vec n,i} \int d^2 x \int d\tau_0 \int \frac{d\omega}{2\pi} i\omega e^{i\omega(\tau-\tau_0)}\gamma_0  G^{(i)}_0(\vec x+\vec n, \vec x_0, \omega) \gamma_0\nonumber \\
&& \langle \tilde 0| N [ \Psi_{i\sigma}(\vec x_0, \tau_0) \bar{\Psi}_{i\beta}(\vec y,0)]|\tilde{0}\rangle\nonumber\\
&=& -V_0\Tr \sum_{\vec{n}i} \int d^2x \int \frac{d\omega}{2\pi} i\omega 2\pi\delta(\omega)\gamma_0G^{(i)}_0(\vec x+\vec n, \vec x_0, \omega)\gamma_0 M^{(i)}
\ee
The normal ordered product I abbreviate by $M$: it depends only on the state $|\tilde 0 \rangle$. The sum combined with the integral over the torus gives an integral over all space which can be performed easily:
\be{spaceints}
\int d^2x K_0(|\omega| |\vec x|) &=& \frac{2\pi}{|\omega|^2}\nonumber\\
\int d^2x \frac{\vec\gamma\vec x}{|\vec x|}K_1(|\omega| |\vec x|)&=& 0
\ee

\be{zeromode2}
\Delta E^{(1)}&=& -V_0\Tr\int  \int \frac{d\omega}{2\pi} i \omega 2\pi\delta(\omega) e^{-i\omega\tau_0} \frac{-i\omega}{|\omega|^2}\gamma_0M\nonumber \\
&=& -V_0\Tr(\gamma_0M)
\ee
The matrix $M$ can be read off from (\ref{zeromodeContr}). If  the ground state $|\tilde0 \rangle$ has the same amount of particles as antiparticles, the trace vanishes and the energy shift is zero. This method gives exactly the same result as when we calculated the energy shift with help of degenerate perturbation theory.

\subsection{Perturbation theory to second order}

The Dirac delta function yields infinite energy shifts to second order in perturbation theory.  As the theory is two-dimensional and its energy relation is linear, one expects the divergence to be linear. The special form of the Casimir energy takes out the linear divergence leaving a logarithmic one.  
All the divergencies I discussed so far arise in every sector and can be cancelled by local counterterms. In the three sectors without a zero mode, a finite expression is obtained after introducing the counterterms. The summation over the windings leads to a well-defined expression for the Green's function and the integral over the frequency $\omega $ can be performed. This is not the case, when there is a zero mode. 

The energy shift involves now the product of three Green's functions: 

\be{secondorder}
E&=&V_0^2 \sum_{i,j=I,II}\int d^2x \int d\omega\, i\omega\nonumber \\
&&\Tr\left(\gamma_0 \hat{G}^{(i)}(\vec x,\vec x_0,\omega)\gamma_0\hat{G}^{(j)}(\vec x_0, \vec x_0, \omega)\gamma_0 \hat{G}^{(i)}(\vec x_0,\vec x,\omega)\right)
\ee
\begin{figure}[hb]
\centering
\scalebox{0.2}{\includegraphics{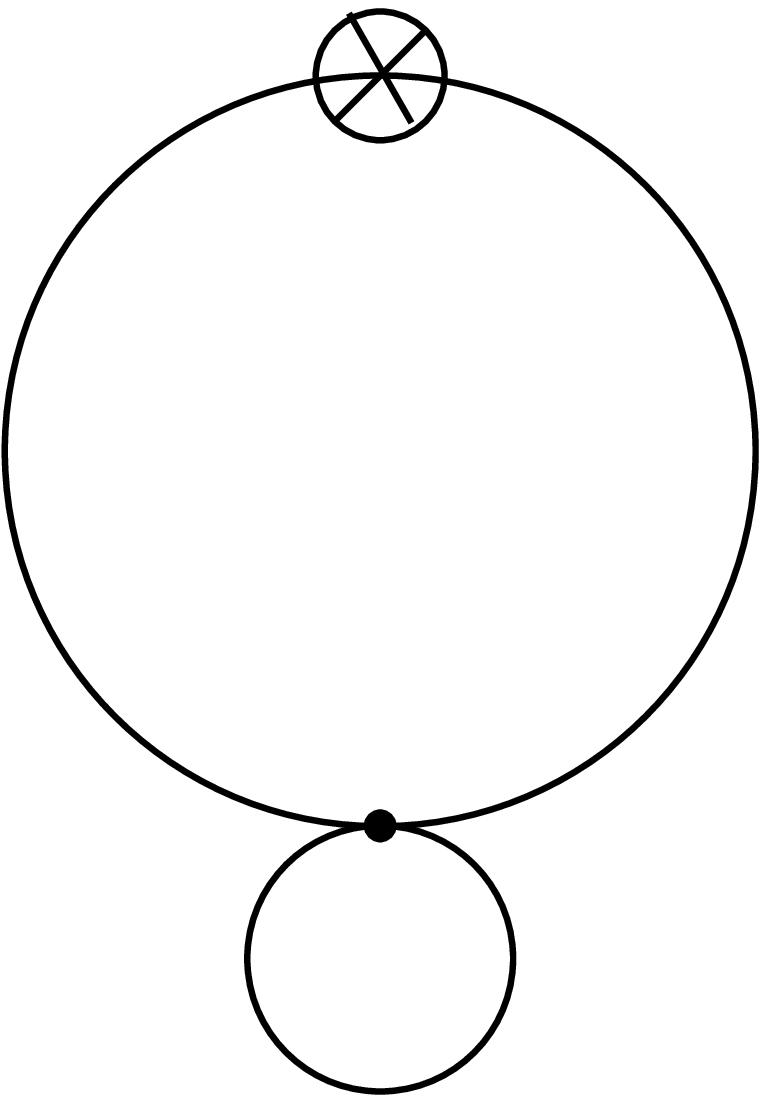}}
\caption{\label{torusgreen2}}
\end{figure}

Even though this does not cause any fundamental difficulties, it makes the numerical calculations much more involved. Due to the commutation relations of the Dirac matrices, only certain terms in (\ref{secondorder}) survive after performing the trace. All the remaining terms are even functions in $\omega$ and the integration over the frequency gives a finite result. So, second order perturbation theory gives the first non-vanishing contribution to the Casimir energy shift in all sectors. In fact, there are only four kinds of diagrams which have to be computed as all the other terms are cancelled by the trace properties. Only two of them give a finite contribution to the Casimir energy after the regularization, see figure (\ref{00contribution}):

\begin{figure}[ht]
\begin{minipage}{6cm}
\centering
\scalebox{0.2}{\includegraphics{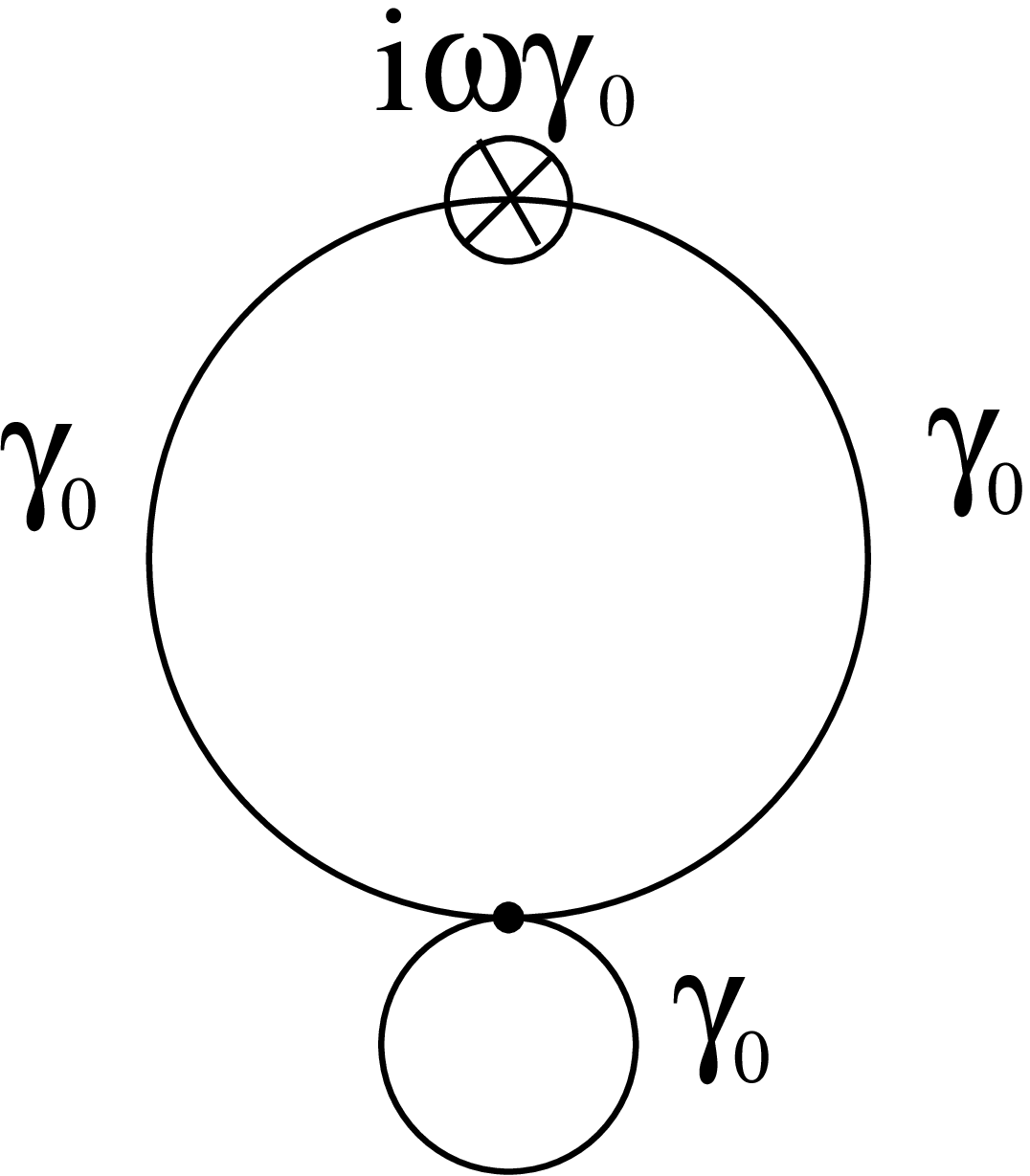}}
\caption{\label{00contribution} (1)}
\end{minipage}
\begin{minipage}{6cm}
\centering
\scalebox{0.2}{\includegraphics{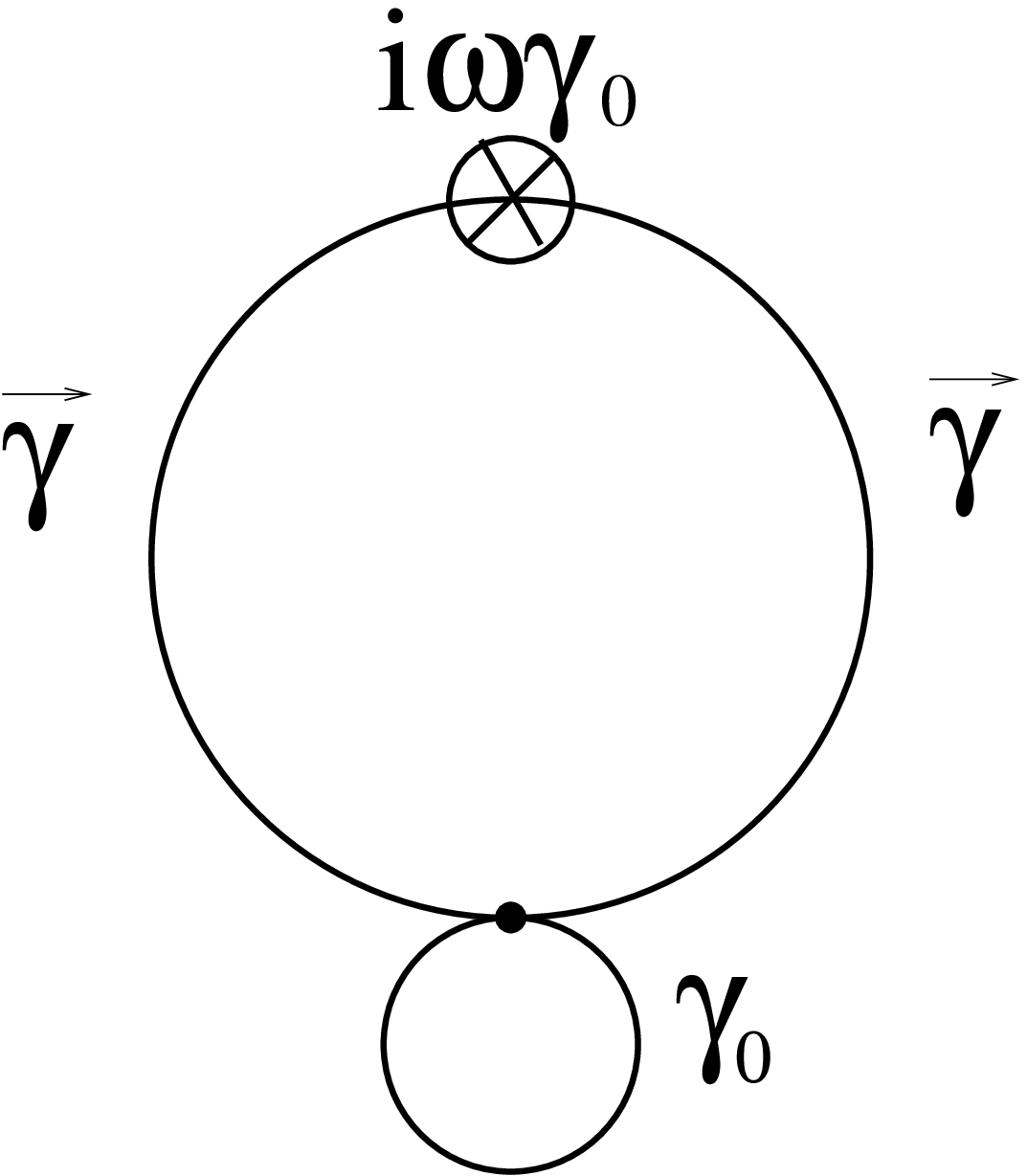}}
\caption{\label{11contribution} (2)}
\end{minipage}
\end{figure}

The most obvious divergency arises, when the quasiparticles scatter two times without winding around the torus in between. The Green's function becomes singular when the two points of interaction coincide.

Another divergence is given by the spatial integral, which may look surprising at first sight. It is instructive to take a closer look on that particular divergence in the fourth sector, where the spatial integral (together with one of the sums) can be performed easily and gives nothing but a delta function in momentum space. 
\be{k_1div}
E&=& \Tr(\sum_{\vec{n},\vec{m}}\int\frac{d^2p}{(2\pi)^2}\int\frac{d^2q}{(2\pi)^2} \, i\omega\gamma_0 e^{i\vec{p}\vec{n}}e^{i\vec{q}\vec{m}}(G(\vec p,\omega)\gamma_0)^2G(\vec q,\omega))
\ee

The regularization must be done for both momentum integrals. Windings in between the scattering events regularize only the loop at $\vec x_0$.  

There are also two asymmetric diagrams which could give a contribution to the energy shift.  It is easiest to look at the sum of both integrals:
\begin{figure}[ht]
\be{asymdiag}
\begin{minipage}{3cm}
\centering
\scalebox{0.2}{\includegraphics{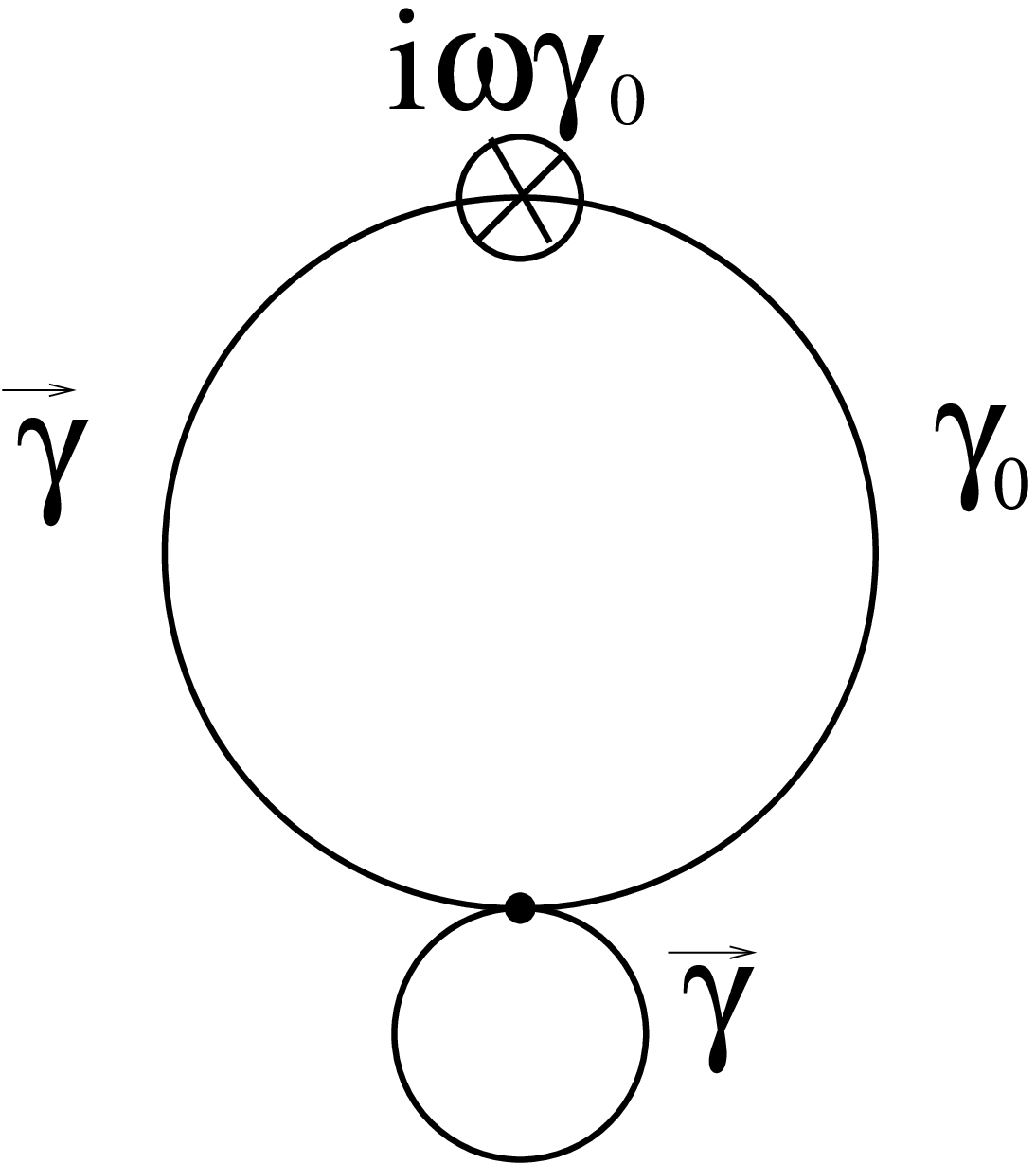}}
\end{minipage} &+&
\begin{minipage}{3cm}
\centering
\scalebox{0.2}{\includegraphics{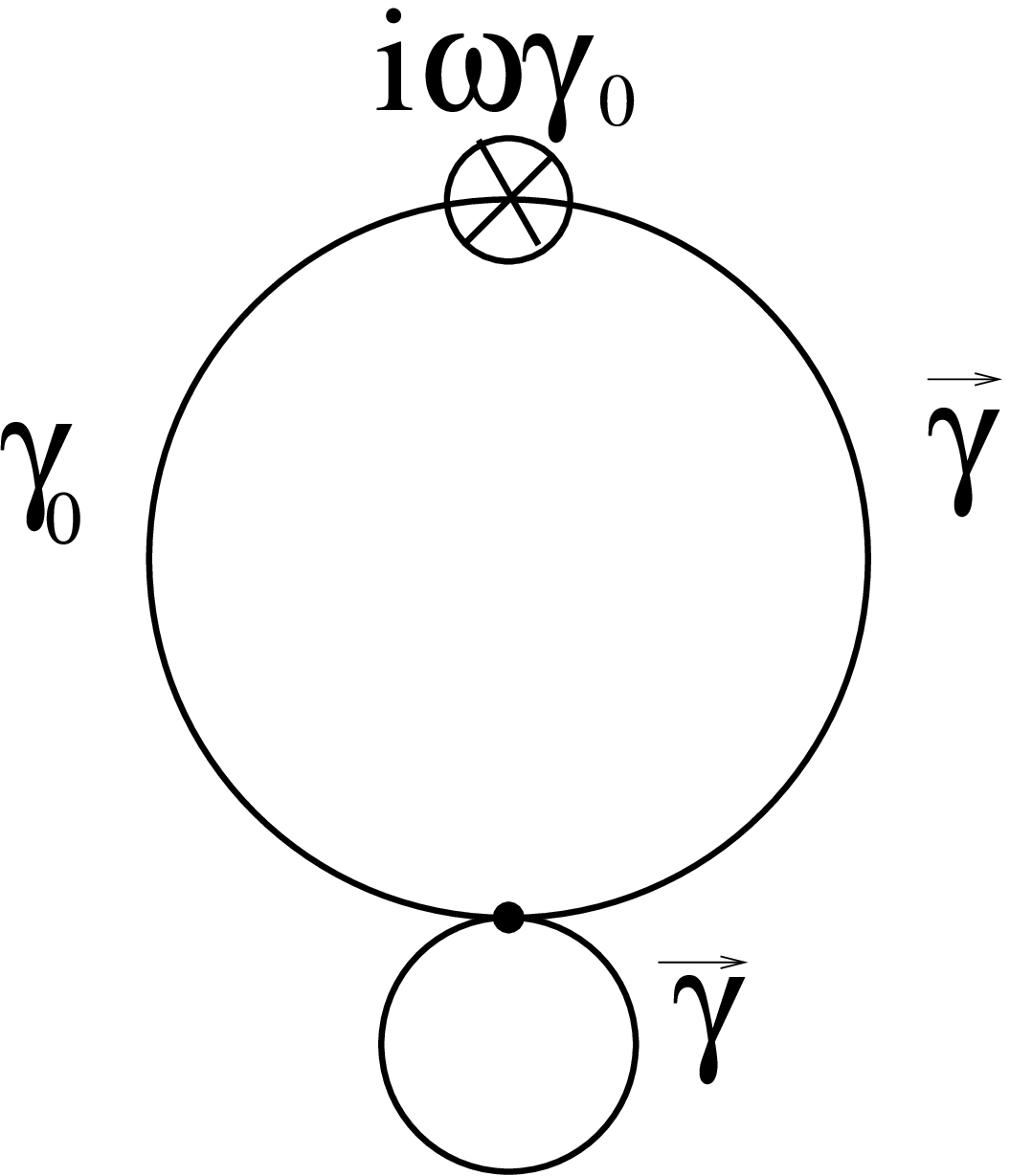}}
\end{minipage}
\ee
\end{figure}

\noindent
The two diagrams can be seen as the same one, but with different directions for the $\vec x-$integration. In first order perturbation theory, I argued that such diagrams should not contribute to the energy shift. I will show that the contribution of the sum of those two, indeed, vanishes. 

All the divergencies I discussed so far arise in every sector and can be cancelled by local counterterms. In the three sectors without a zero mode, a finite expression is obtained after introducing the counterterms. The summation over the windings leads to a well-defined expression for the Green's function and the integral over the frequency $\omega $ can be performed. This is not the case, when there is a zero mode.  The zero mode sector is special in that respect that the spatial integral can be performed, yielding a delta function in momentum space. 
\be{4energygreen}
E=\sum_{\vec{n}}\sum_{\vec{m}}\int\frac{d^2q}{(2\pi)^2}\int\frac{d^2k}{(2pi)^2}\int\frac{d\omega}{2\pi}\Tr(i\omega\gamma_0(\hat{G}(\vec k,\omega)\gamma_0)^2\hat{G}(\vec q,\omega))e^{i\veq\vec{n}}e^{i\vec k \vec m}
\ee
When you consider only the $\vec{p}=0$ part in the propagator, you can replace each propagator by $\frac{-i}{\omega}\gamma_0$. The expression you obtain is ill-defined and diverges at the lower limit of the $\omega$ integral: 
$\int d\omega \frac{1}{\omega^2}$. 
Thus, in the fourth sector you need a low-energy cut-off for this integral. Such a cut-off is provided by excluding the zero mode. Once it is excluded, the integral over the frequency becomes finite. The contribution of the zero mode has to be calculated in an alternative way. As was the case for the first order shift, the zero mode is insensitive to the symmetries in the theory. Therefore, its divergence is linear, instead of logarithmic. 

In the Green's function representation as a sum over windings, this divergence appears to be an infrared divergence. However, there should not be an infrared divergence on the torus as it has a finite volume and, therefore, only a finite number of zero modes. The contradiction is solved by realizing that the representation of the Green's function as a sum over windings is a good one, if the summation converges. This is not the case for $\omega=0$ and therefore, we need to use an alternative way to calculate its contribution to the energy shift. When the energy shift of the zero mode is expressed in terms of single particle perturbation theory, also an interpretation of the divergence is provided: it is then easy to see that the divergence occurs for the scattering into high momenta and is not an infrared, but an ultraviolet divergence.

\subsubsection{Proof of finiteness}

Even though the divergent integrals originate from the same short distance divergence, I am going to use two different approaches to deal with them. During this whole paragraph, I will keep the expressions as general as possible, so that the discussion is valid for both nodal pairs and in all sectors. 

First, I consider the divergence which arises from the zero mode. As said earlier, in order to provide a low-energy cut-off for the integral over the frequency $\omega$, I exclude the zero mode from two of the Green's functions. In the remaining Green's function I assume that the quasiparticle winds at least once around the torus. The case without windings is considered later as it is completely analogous to the other three sectors.  To see that the divergence is removed, it is sufficient to look at the low-energy behaviour of the regularized Green function:
\be{zeromodesubtr}
\tilde{G}(\vec x,\vec x_0,\omega)&=& \sum_{\vec n}\{G_0(\vec x+\vec n,\vec x_0,\omega)-\frac{1}{L_xL_y}\int d^2y \,G_0(\vec y+\vec n,\vec x_0,\omega)\}
\ee
It has to be finite for $\omega=0$ otherwise the integral over the frequency is divergent. 

As the sum over windings is responsible for the $\omega\rightarrow0$ divergence, I neglect the first terms and focus on winding numbers where $|\vec m|\gg|\vec x|$. For large $\vec m$ the difference whether you integrate over the torus or whether you choose a specific point on the torus becomes negligible and the two terms cancel. By doing a Taylor expansion around $\vec m$, I show that the subtraction of the zero mode effectively regularizes the Green's function. In fact, it is sufficient to show that $K_0(\omega|\vec n|)$ is regularized. The properties for $K_1(\omega|\vec n|)$ follow from it.  

The first non-vanishing contribution from the Taylor expansion is second order in $\vec x$. The first derivative, $\vec{\nabla} K_0(\omega|\vec{y}+\vec n|)$ as well as the mixed term in the second order $\partial_1\partial_2 K_0(\omega|\vec{y}+\vec n|)$ vanish because of symmetry reasons. Therefore, the first contribution comes from 
\be{secondordertaylor}
&&(\partial_1^2+\partial_2^2)K_0(\omega|\vec y+\vec n|)\nonumber \\
&=& \{-\frac{\omega}{|\vec n|}K_1(\omega|\vec n|) + \omega^2K_0(\omega|\vec n|)\}\frac{L_x^3+L_y^3}{L_xL_y}
\ee
Both terms are finite in the limit where $\omega\rightarrow0$. Once the zero mode is excluded from one of the Green's functions, the integration over $\omega$ becomes finite. 

The missing contribution of the zero-mode is computed with help of degenerate perturbation theory. There it is well-known that the Dirac delta function gives rise to a (in our case linear) divergence and regularization techniques have been developed, \cite{Jackiw}.  I excluded the zero mode from the Green's function when there were windings between the scattering. The term with no windings must be omitted in the calculation. Thus, the energy shift for the scattering in free space must be subtracted:
\be{secondzero}
\Delta E^{(2)}&=& -\frac{1}{V}\sum_{\vec k}\frac{|\langle\vec k|\hat{V}|\tilde0\rangle|^2}{\epsilon_k}+
\int \frac{d^2k}{(2\pi)^2} \frac{|\langle\vec k|\hat{V}|\tilde0\rangle|^2}{\epsilon_k}
\ee
where the sum runs over all states with finite energy $\epsilon_k$. To show the convergence of (\ref{secondzero}), it is most convenient to expand the integrand in a Taylor expansion. As the divergence is linear, there could be  a subdivergent term,  which is logarithmic in $\vec q$. By powercounting, the logarithmic divergence must sit in the first derivative in the expansion. However, this term vanishes because of the integration. The first non-vanishing term is proportional to $\frac{1}{\epsilon_p^3}$ and is therefore finite.

Now I want to consider the term when there are no windings between two scattering events: 
\be{00}
\begin{minipage}{2.5cm}
\scalebox{0.2}{\includegraphics{torusgreen2g}}
\end{minipage}
&=&-4 \int d^2x\int_{-\infty}^{\infty}d\omega \,  (\frac{\omega}{2\pi})^4 \sum_{\vec n,\vec m,\vec l} f(\vec n,\vec m,\vec l)  K_0^{(j)}(|\omega|\sqrt{\vec m^2})  \cdot \nonumber \\ 
&& K_0^{(i)}(|\omega|\sqrt{(\vec x-\vec x_0+\vec n)^2}) K_0^{(i)}(|\omega|\sqrt{(\vec x_0-\vec x-\vec l)^2})
\ee
where the sectors are encoded in $f(\vec n,\vec m,\vec l)$. 
Setting $\vec m$ to zero gives an undefined expression. When the second scattering site is displaced by a vector $\delta\vec x$, the expression becomes finite for $\vec m=0$:
\be{00splitted}
\begin{minipage}{2.5cm}
\scalebox{0.2}{\includegraphics{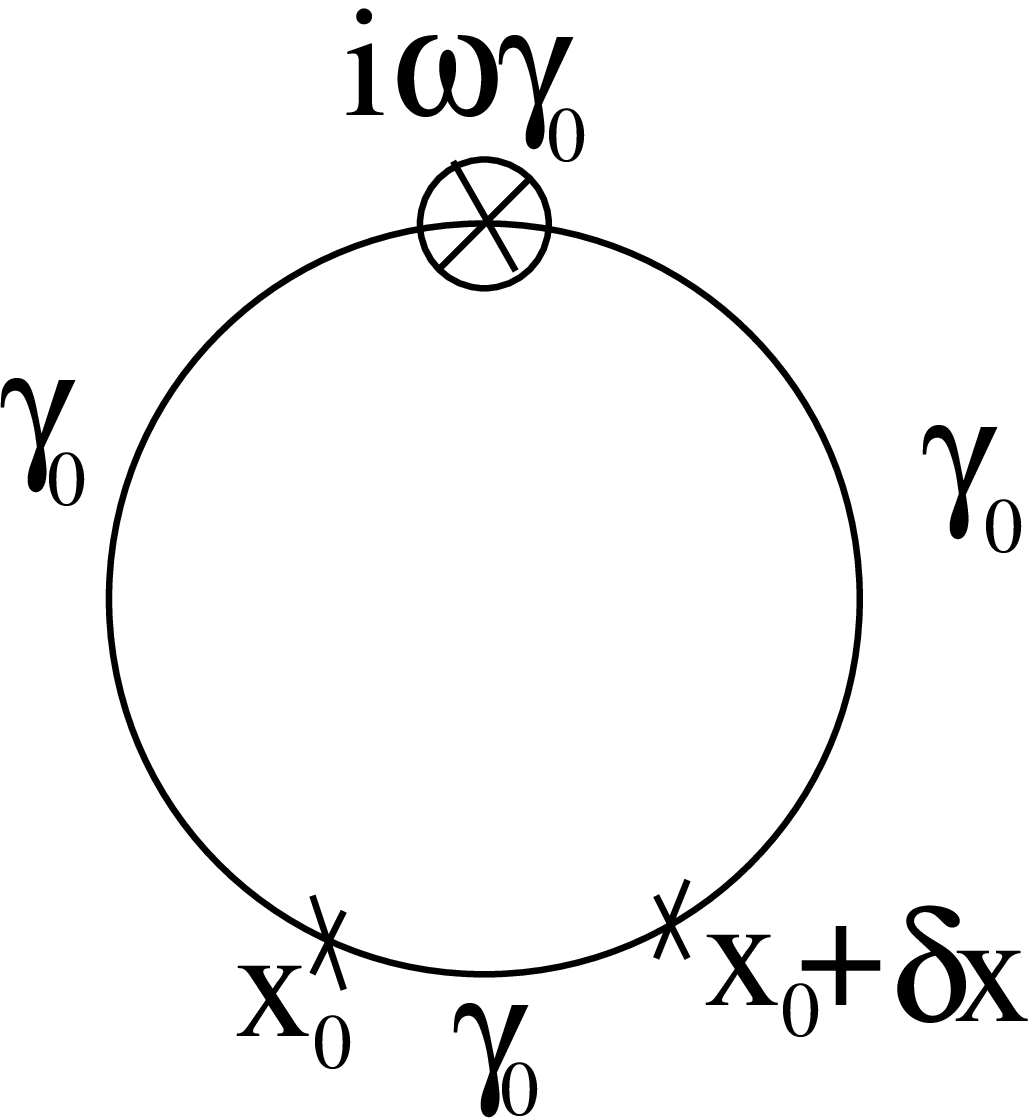}}
\end{minipage}
&-&4\int d^2x\int_{-\infty}^{\infty}d\omega \,   (\frac{\omega}{2\pi})^4 \sum_{\vec n,\vec l} f(\vec n,\vec l)K_0^{(j)}(|\omega|\sqrt{(\delta\vec x)^2})\cdot \nonumber \\ && K_0^{(i)}(|\omega|\sqrt{(\vec x-\vec x_0+\vec n)^2})  
 K_0^{(i)}(|\omega|\sqrt{(\vec x_0+\delta\vec x-\vec x-\vec l)^2})\nonumber\\
\ee
For small separations, the Besselfunction can be replaced by its asymptotic behaviour:
$$ K_0^{(j)}(|\omega\sqrt{(\delta\vec x)^2}|)\rightarrow -\ln(|\omega'|)+\ln(2)+\Psi(1) -\ln|\delta\vec x|^{(j)} \hspace{1cm}\mbox{ for } \delta\vec x\rightarrow0 $$
where $\Psi(1)$ is the Euler constant and $|\delta\vec x|^{(j)}$ is labeled by the nodal pair. $\omega'$ is given by $\omega'=\omega\frac{L_xL_y}{v_Fv_\Delta}$. 
The first three terms give finite contributions for the case where there are no windings. The last can be cancelled by a local counterterm. 
\be{countertermalpha}
H_\alpha&=& V_0^2\delta^2(\vec x-\vec x_0)\Psi^\dagger_{i\sigma}(\vec x,\tau_0) \alpha \partial_{\tau_0} \Psi_{i\sigma}(\vec x, \tau)
\ee
which gives  a non vanishing contribution in first order perturbation theory:
\be{countalpha}
&\alpha V_0^2\Tr(\int d^2x\int \frac{d\omega}{2\pi} \,  i\omega\gamma_0G^{(i)}(\vec x, \vec x_0, \omega)\omega \gamma_0 G^{(i)}(\vec x_0,\vec x,\omega))&
\ee
The expression above can be separated into two terms. First consider the term where both propagators contribute a $\gamma_0$ to the trace. That term cancels exactly with (\ref{00splitted}) if $\alpha $ is chosen to be:
$$ \alpha = -\frac{1}{2\pi}( \ln|\delta \vec x|^{(I)}+\ln|\delta \vec x|^{(II)})$$
The term, where both propagator contribute with a $(\vec x-\vec x_0)\vec \gamma$ is needed for the second diverging diagram: 

\be{11}
\begin{minipage}{2.5cm}
\scalebox{0.2}{\includegraphics{torusgreen2g2}}
\end{minipage}
&=&4\int d^2x\int_{-\infty}^{\infty}d\omega \, (\frac{|\omega|}{2\pi})^4\sum_{\vec n,\vec m,\vec l} f(\vec n,\vec m,\vec l) 
\nonumber\\
&&
K_0^{(i)}(|\omega|\sqrt{\vec m^2})  \frac{(\vec x-\vec x_0+\vec n)(\vec x-\vec x_0+\vec l)}{|\vec x-\vec x_0+\vec n||\vec x-\vec x_0+\vec l|}
 \nonumber \\
&&  K_1^{(j)}(|\omega|\sqrt{(\vec x-\vec x_0+\vec n)^2})K_1^{(j)}(|\omega|\sqrt{(\vec x_0-\vec x-\vec l)^2})\nonumber \\
\ee
 As in the previous diagram, this diagram has a divergence when the quasiparticles do not wind between the scattering events. We split the point of interaction in the same way as above:
\be{11splitted}
\begin{minipage}{2.5cm}
\scalebox{0.2}{\includegraphics{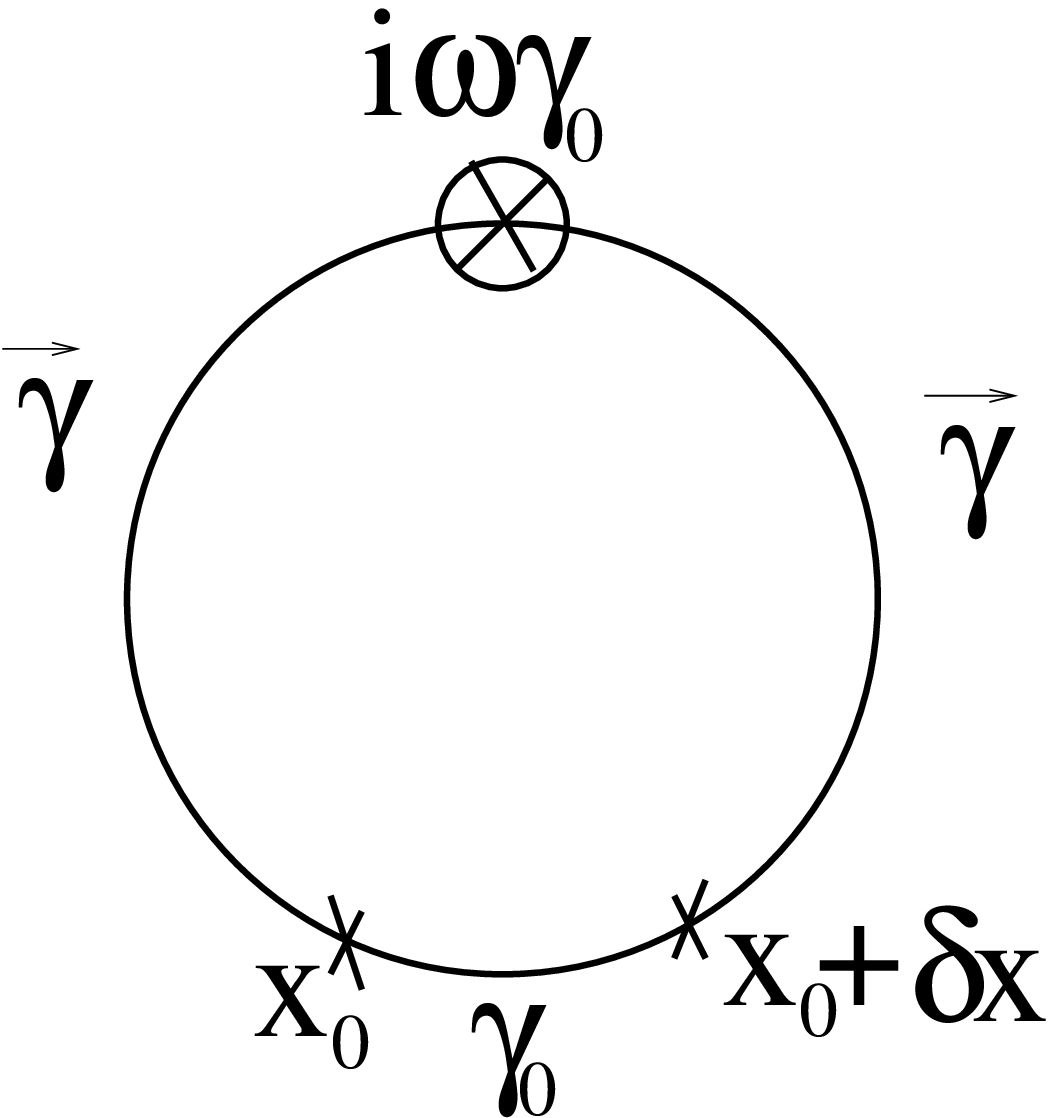}}
\end{minipage}
&=&4\int d^2x\int_{-\infty}^{\infty} d\omega \,   (\frac{\omega}{2\pi})^4 \sum_{\vec n,\vec l,\vec m} f(\vec n,\vec m,\vec l)  K_0^{(i)}(|\omega|\sqrt{(\delta\vec x+\vec m)^2}) \cdot \nonumber \\
&&\frac{(\vec x-\vec x_0+\vec n)}{|\vec x-\vec x_0+\vec n|}\frac{(\vec x-\vec x_0+\vec l)}{|\vec x-\vec x_0+\vec l|} K_1^{(j)}(|\omega|\sqrt{(\vec x-\vec x_0+\vec n)^2})\nonumber\\
&& K_1^{(j)}(|\omega|\sqrt{(\vec x_0+\delta\vec x-\vec x-\vec l)^2})
\ee
The term where $\vec m=0$ is treated in the same way as above. The Besselfunction is replaced by its asymptotic behaviour at the origin and the singular piece is cancelled by $H_\alpha$. 

The other divergence arises for $\vec m\neq 0$ but $\vec n = \vec l =0$. 
In order to obtain the same divergent behaviour in a first order diagram, this counterterm needs to couple to $\vec \gamma \vec \nabla$, see figure (\ref{fig:countertermb}):
\begin{figure}[tb]
\centering
\scalebox{0.2}{\includegraphics{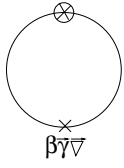}}
\caption{\label{fig:countertermb} Feynman diagram corresponding to $H_\beta$}\end{figure}
\be{countertermbeta}
H_\beta&=& V_0^2\delta^2(\vec x-\vec x_0)\Psi^\dagger_{i\sigma}(\vec x,\tau_0) \beta \gamma_0 \vec\gamma \vec \nabla_0  \Psi_{i\sigma}(\vec x, \tau_0)
\ee
This counterterm includes also finite contributions to the energy shift:
\be{countbeta}
&&\beta V_0^2\Tr(\int d^2x\int \frac{d\omega}{2\pi} \,  i\omega\gamma_0G^{(i)}(\vec x, \vec x_0, \omega)\gamma_0\vec \gamma \vec \nabla_0  G^{(i)}(\vec x_0,\vec x,\omega)) \\
&=&4\beta\int d^2x \int d\omega \,  \frac{|\omega|}{(2\pi)^3}\sum_{\vec n, \vec m}\nonumber \\
 && K_0^{(i)}(|\omega|\sqrt{(\vec x-\vec x_0+\vec n)^2})\frac{1}{|\vec x_0-\vec x +\vec m|} K_1^{(i)}(|\omega|\sqrt{(\vec x_0-\vec x+\vec m)^2})\nonumber\\
&-& K_0^{(i)}(|\omega|\sqrt{(\vec x-\vec x_0+\vec n)^2})|\omega| K_0^{(i)}(|\omega|\sqrt{(\vec x_0-\vec x+\vec m)^2})\nonumber\\
&+& K_1^{(i)}(|\omega|\sqrt{(\vec x-\vec x_0+\vec n)^2})|\omega| K_1^{(i)}(|\omega|\sqrt{(\vec x_0-\vec x+\vec m)^2})\frac{(\vec x -\vec x_0+\vec n)(\vec x -\vec x_0 +\vec m)}{|\vec x -\vec x_0 +\vec n||\vec x -\vec x_0 +\vec m|}\nonumber 
\ee
Only the first term is logarithmic divergent when $\vec m=0$. When $\beta$ is set to $\beta = -\frac{2}{\pi}$ it cancels the divergence in (\ref{11splitted}) in the limit where $\delta\vec x\rightarrow0$. The spatial integration over the sum of both divergent terms becomes finite:
$$\int d^2x 
(K_0(|\omega(\vec x-\vec x_0+\vec n)|)\frac{K_1(|\omega(\vec x-\vec x_0)|)}{|\omega(\vec x-\vec x_0)|}-K_0(|\omega\vec n|)K_1(|\omega(\vec x-\vec x_0)|)^2<\infty$$

The divergence arises for small $|\vec x-\vec x_0|$, so by reducing the area of integration such that $K_0(|\omega(\vec x-\vec x_0+\vec n)|)\approx K_0(|\omega\vec n|$ the leading divergence cancels. Note, that it is not important at which nodal pair the Green's function is located. Their contribution differs by a finite amount, while the divergent part is exactly the same for both node pairs. 

Finally,  I consider the sum of the asymmetric diagrams:
\be{3divstart}
&&
\begin{minipage}{3cm}
\centering
\scalebox{0.2}{\includegraphics{torusgreen2g10}}
\end{minipage} +
\begin{minipage}{3cm}
\centering
\scalebox{0.2}{\includegraphics{torusgreen2g01}}
\end{minipage}
\nonumber\\
\nonumber\\
&=&4\sum_{\vec n,\vec m,\vec l}\int d^2x \int d\omega \, (\frac{\omega}{(2\pi)})^4  \gamma_0^2 \frac{(\vec x-\vec x_0+\vec n)\vec m}{|\vec x-\vec x_0+\vec n||\vec m|}K_1^{(j)}(|\omega\vec m|)\{K_1^{(i)}(|\omega(\vec x-\vec x_0+\vec n)|)\nonumber \\
&&K_0^{(i)}(|\omega(\vec x-\vec x_0+\vec m)|)-K_0^{(i)}(|\omega(\vec x-\vec x_0+\vec n)|)K_1^{(i)}(|\omega(\vec x-\vec x_0+\vec n)|\}
\ee
For $\vec m\neq0$ the diagrams vanish because of symmetry reasons. For zero windings, the diagrams in equation (\ref{3divstart}) are not properly defined as one of the Besselfunctions diverges, while the expression in the parenthesis becomes zero. When the point of interaction is splitted, (\ref{3divstart}) becomes finite and one can study the limiting behaviour for $\delta\vec x\rightarrow 0$. For the nodal pair (1,3), $\delta\vec x$ is given by $(\delta x/v_F, \delta y/v_\Delta)$. For the other node pair, the velocities are switched. In the following discussion, I also include a subscript on $\delta\vec x$ to keep track of the different dependence of the velocities:

\be{asymsplitted}
&&\Tr \sum_{\vec n,\vec l}\int d^2x \int d\omega \, \frac{\omega^4}{(2\pi)^4}  \gamma_0^2K_1^{(j)}(|\omega\delta\vec x_j|)\nonumber \\
&&\{ \frac{(\vec x-\vec x_0+\vec n)\delta\vec x_j}{|\vec x-\vec x_0+\vec n||\delta\vec x_j|}K_1^{(i)}(|\omega(\vec x-\vec x_0+\vec n)|)K_0^{(i)}(|\omega(\vec x-\vec x_0-\delta\vec x_i+\vec l)|)\nonumber \\
&-& \frac{(\vec x-\vec x_0-\delta\vec x_i+\vec n)\delta\vec x_j}{|\vec x-\vec x_0-\delta\vec x_i+\vec n||\delta\vec x_j|}K_0^{(i)}(|\omega(\vec x-\vec x_0+\vec l)|)K_1^{(i)}(|\omega(\vec x-\vec x_0-\delta\vec x_i+\vec n)|\}\nonumber\\
\ee
The limiting behaviour for the Besselfunctions is known, each diagram for itself diverges linearly in $|\delta\vec x|$. The sum of both diagrams will take out the leading order in the divergence. But there may still be a divergence in subleading order left. Therefore, I  analyze how the expression in the parenthesis acts for small separations $|\delta\vec x|$. For derivating the Bessel functions I use:
\be{besselderivatives}
K_0'(x)&=& -K_1(x)\nonumber\\
K_1'(x)&=& -K_0(x)-\frac{1}{x}K_1(x)
\ee
so that the result can be written in terms of these two functions, only. The total contribution from the two asymmetric diagrams can be written as:
\be{asymtaylor}
&&\Tr\sum_{\vec n,\vec l}\int d^2 \int d\omega x\, (\frac{\omega}{2\pi})^4\frac{|\delta\vec x_i|}{|\delta\vec x_j|}\nonumber\\
&-&\frac{(\vec x-\vec x_0+\vec n)\hat\Delta_j}{|\vec x-\vec x_0+\vec n|}\frac{(\vec x-\vec x_0+\vec l)\hat\Delta_i}{|\vec x-\vec x_0+\vec l|}K_1^{(i)}(|\omega(\vec x-\vec x_0+\vec n)|)K_1^{(i)}(|\omega(\vec x-\vec x_0+\vec l)|)\nonumber \\
&+&\frac{((\vec x-\vec x_0+\vec n)\hat\Delta_i)(\vec x-\vec x_0+\vec n)\hat\Delta_j)}{|\vec x-\vec x_0+\vec n|^2}K_0^{(i)}(|\omega(\vec x-\vec x_0+\vec l)|)K_0^{(i)}(|\omega(\vec x-\vec x_0+\vec n)|)\nonumber\\
&+& 2 \frac{((\vec x-\vec x_0+\vec n)\hat\Delta_j)((\vec x-\vec x_0+\vec n)\hat\Delta_i)}{|\vec x-\vec x_0+\vec n|^2}K_0^{(i)}(|\omega(\vec x-\vec x_0+\vec l)|)\frac{K_1^{(i)}(|\omega(\vec x-\vec x_0+\vec n)|)}{|\omega(\vec x-\vec x_0+\vec n)|} \nonumber\\
&-&\hat\Delta_i\cdot\hat\Delta_j K_0^{(i)}(|\omega(\vec x-\vec x_0+\vec l)\frac{K_1^{(i)}(|\omega(\vec x-\vec x_0+\vec n)|)}{|\vec x-\vec x_0+\vec n|}
\ee
Where $\hat\Delta $ is a unit vector with direction $\delta\vec x$. The first two terms give finite contributions to the energy shift, but the others exhibit a subleading divergence which is logarithmic in $\vec x$.  
The counterterms which we have introduced so far are not able to cancel this divergence as it is direction dependent. Point splitting introduced a specific direction which was not present in the original graph. This can be remedied either by integrating over all directions or by a counterterm which is also direction dependent. The former is more intuitive. Moreover, it is quite straightforward to see that the divergent part vanishes, when integrated over all directions. However, it is much simpler to compute the finite contribution with help of counterterms  and, therefore, I use this method instead.  In the end, both methods give the same result, namely that the asymmetric diagrams do not contribute to the energy shift:
\be{countertermepsilon}
&&\int d^2x \int \frac{d\omega}{2\pi} i\omega \gamma_0\hat{G}^{(i)}(\vec x,\vec x_0,\omega) (\hat\epsilon_i\vec \gamma)(\hat\Delta_i\vec\nabla_0)\hat{G}^{(i)}(\vec x_0,\vec x,\omega)\nonumber\\
&=&\sum_{\vec n,\vec m}\int d^2x\int d\omega \frac{|\omega|^3}{(2\pi)^3}K_0^{(i)}(|\omega(\vec x-\vec x_0+\vec n|))(\hat\epsilon_i\vec\gamma) \{\nonumber \\
&&(\frac{\vec\gamma\hat\Delta_i}{|\vec x-\vec x_0+\vec m|}-\frac{(\vec\gamma(\vec x-\vec x_0+\vec m))}{|\vec x-\vec x_0+\vec m|}\frac{(\vec x-\vec x_0+\vec m)\hat\Delta_i}{|\vec x-\vec x_0+\vec m|^3})K_1^{(i)}(|\omega(\vec x-\vec x_0+\vec m)|)\nonumber \\
&+& \frac{(\vec x-\vec x_0+\vec m)\vec \gamma}{|\vec x-\vec x_0+\vec m|}( -K_0^{(i)}(|\omega|(\vec x-\vec x_0+\vec m)|)-\frac{K_1^{(i)}(|\omega(\vec x-\vec x_0+\vec m)|)}{|\omega(\vec x-\vec x_0+\vec m)|})\frac{|\omega|(\vec x-\vec x_0+\vec m)\hat\Delta_i}{|\vec x-\vec x_0+\vec m|} \}\nonumber \\
&+&\frac{\vec\gamma(\vec x-\vec x_0+\vec n)}{|\vec x-\vec x_0+\vec n|}K_1^{(i)}(|\omega(\vec x-\vec x_0+\vec n)|)\frac{|\omega|(\vec x-\vec x_0+\vec m)\hat\Delta}{\vec x-\vec x_0+\vec m} K_1^{(i)}(|\omega(\vec x-\vec x_0+\vec m)|)\nonumber \\
&=&\Tr\sum_{\vec n,\vec m}\int d^2x\int d\omega \frac{|\omega|^4}{(2\pi)^3}\gamma_0^2\nonumber \\
&-& \frac{((\vec x-\vec x_0+\vec m)\hat\Delta_i)((\vec x-\vec x_0+\vec m)\hat\epsilon_i)}{|\vec x-\vec x_0+\vec m|^2}K_0^{(i)}(|\omega(\vec x-\vec x_0+\vec n)|)K_0^{(i)}(|\omega(\vec x-\vec x_0+\vec m)|)\nonumber \\
&+&\frac{(\vec x-\vec x_0+\vec n)\hat\epsilon_i}{|\vec x-\vec x_0+\vec n|}\frac{(\vec x-\vec x_0+\vec m)\hat\Delta_i}{|\vec x-\vec x_0+\vec m|}K_1^{(i)}(|\omega(\vec x-\vec x_0+\vec n)|)K_1^{(i)}(|\omega(\vec x-\vec x_0+\vec m)|)\nonumber\\
&-&2\frac{((\vec x-\vec x_0+\vec m)\hat\Delta_i)((\vec x-\vec x_0+\vec m)\hat\epsilon_i)}{|\vec x-\vec x_0+\vec m|^2}K_0^{(i)}(|\omega(\vec x-\vec x_0+\vec n)|)\frac{K_1^{(i)}(|\omega(\vec x-\vec x_0+\vec m)|)}{|\omega(\vec x-\vec x_0+\vec m)|}\}\nonumber\\
&+&\hat\epsilon_i\hat\Delta_iK_0^{(i)}(|\omega(\vec x-\vec x_0+\vec n)|)\frac{K_1^{(i)}(|\omega(\vec x-\vec x_0+\vec m)|)}{|\omega(\vec x-\vec x_0+\vec m)|}\}
\ee
In order to cancel the divergence in (\ref{asymtaylor}), $\hat\epsilon_i$ must be chosen as $$\hat\epsilon_i=-\frac{1}{2\pi}\sum_{j=I,II} \frac{|\delta\vec x_i|}{|\delta\vec x_j|}\hat\Delta_j$$ In that case, the counterterm cancels not only the divergent term but also all the remaining, finite terms from (\ref{asymsplitted}).  A finite contribution from the asymmetric diagrams would have been inconsistent  with the underlying symmetry. Furthermore, there is no specific direction in the model. A local impurity on the torus cannot introduce a direction dependent energy shift.

\subsubsection{second order results}
 In this section, I am going to summarize the results for the energy shift to second order perturbation theory. The analytic expressions I obtained are finite and can be computed numerically. 
As there are a lot of terms to take into account, I organize the results by the diagrams which they originate from. The total energy shift to second order  in $V_0$ is most conveniently written as a sum of five diagrams:
\be{totalEnergyShift}
\Delta E^{(2)} &=& \Delta E_\alpha + \Delta E_\beta +\Delta E_r\nonumber\\
\Delta E_\alpha &=&
\begin{minipage}{3.5cm}
\centering
\scalebox{0.2}{\includegraphics{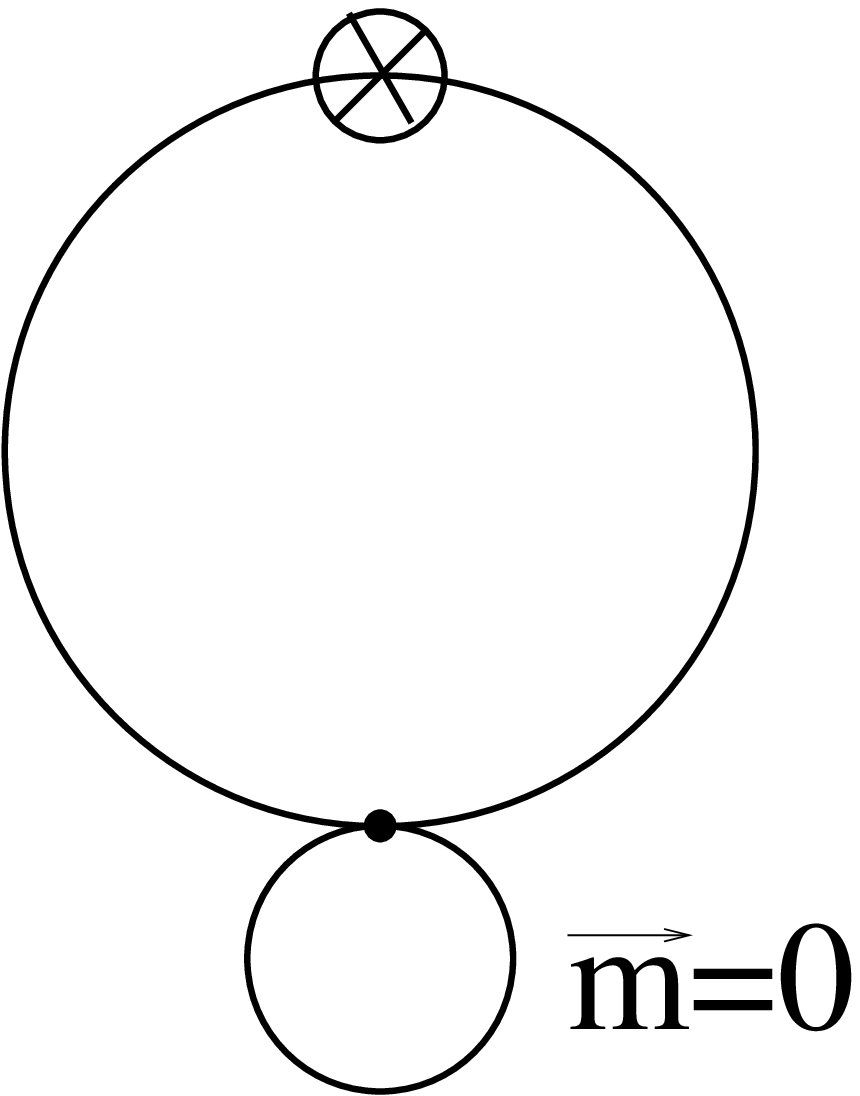}}
\end{minipage}+
\begin{minipage}{3.5cm}
\centering
\scalebox{0.2}{\includegraphics{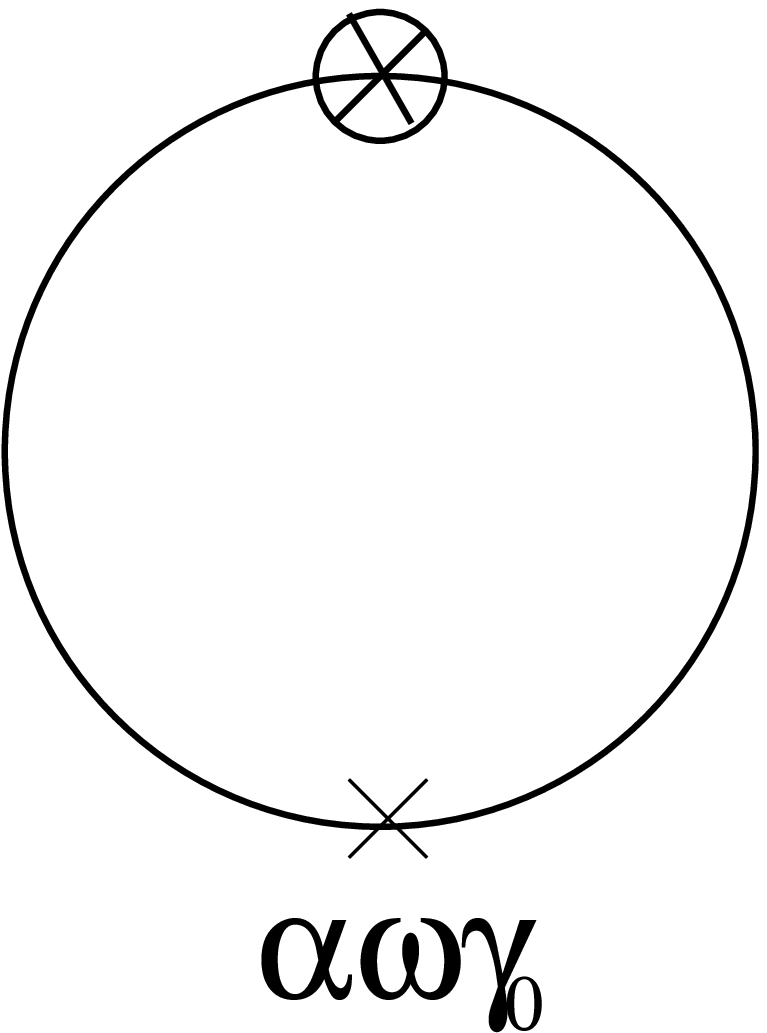}}
\end{minipage}\nonumber\\
\Delta E_\beta&=&
\begin{minipage}{3.5cm}
\centering
\scalebox{0.2}{\includegraphics{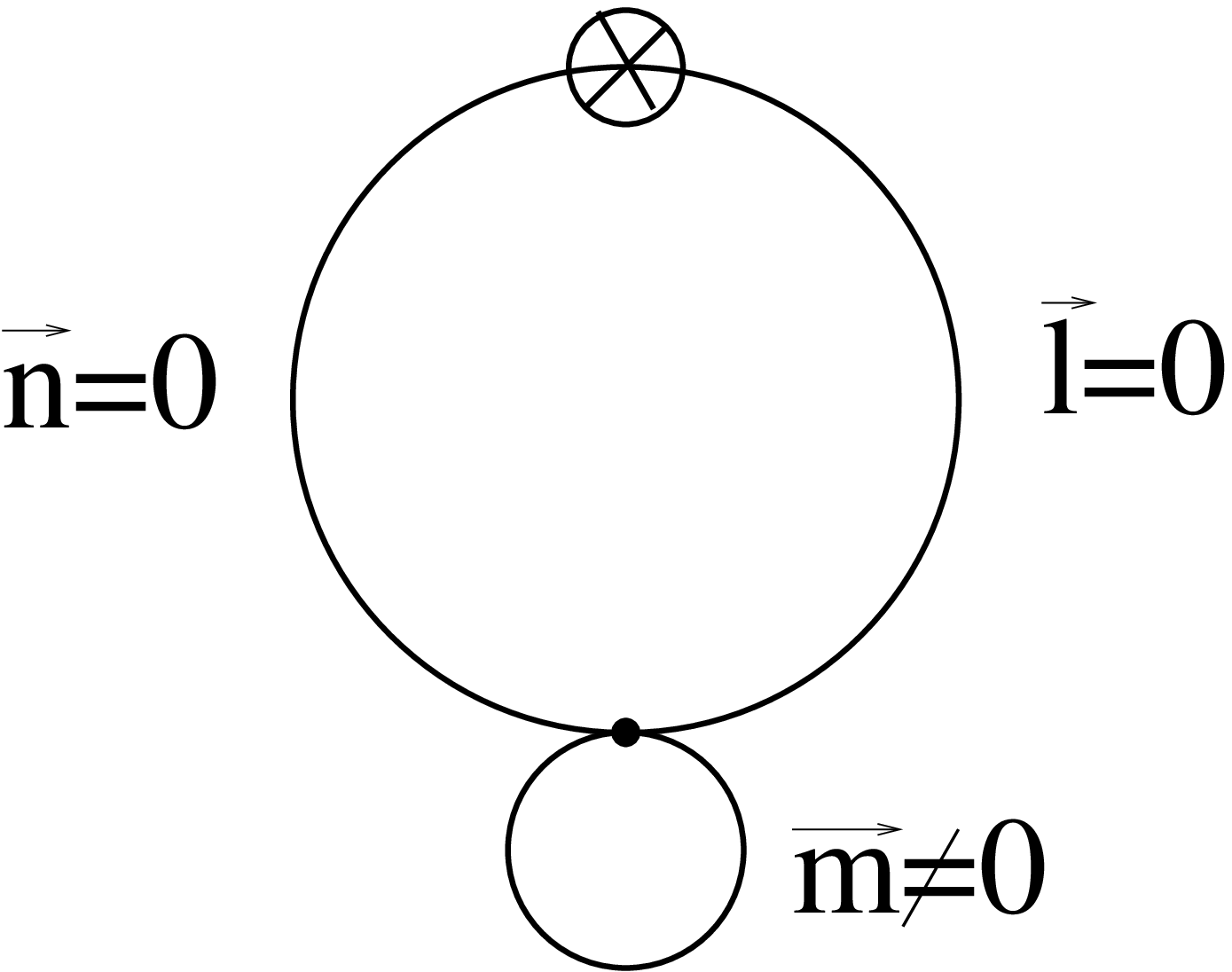}}
\end{minipage}+
\begin{minipage}{3.5cm}
\centering
\scalebox{0.2}{\includegraphics{counterb}}
\end{minipage}\nonumber\\
\Delta E_r&=& 
\begin{minipage}{3.5cm}
\centering
\scalebox{0.2}{\includegraphics{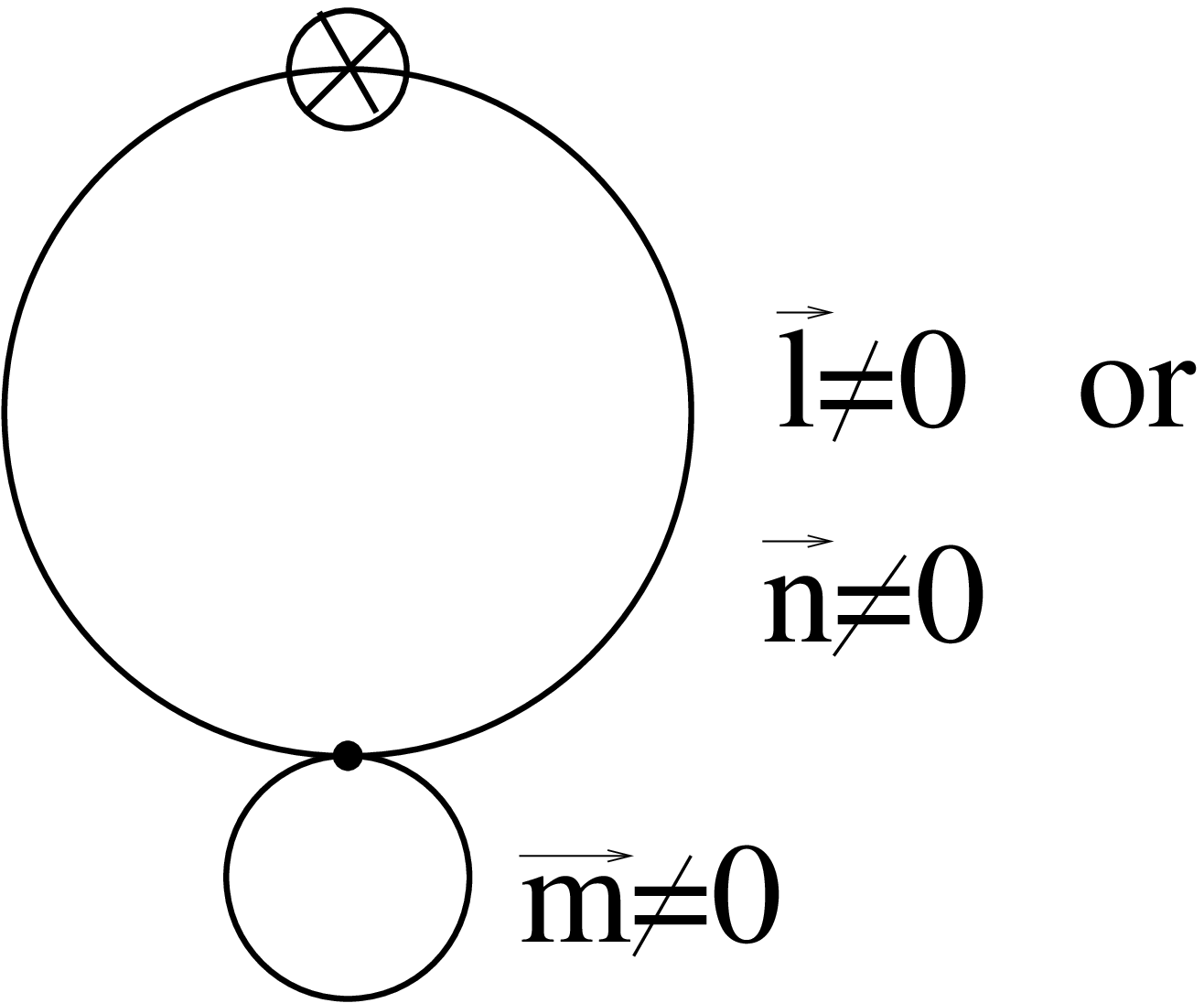}}
\end{minipage}
\ee
The first two contributions are analogous in all sectors. In the previous section I derived that both $\Delta E_\alpha$ and $\Delta E_\beta$ are finite in all sectors. In the three sectors without a zero mode, $\Delta E_r$ does not need regularization. In the zero mode sector, I excluded the zero mode from two of the Green's functions. In that sector, I need to compute the shift $\Delta E_r^0$,  which is due to the zero mode, separately. I will use the notion of the Wilson loops $A_x$ and $A_y$, introduced in section 3.2, to give an explicit dependence of the energy shifts on the sectors.

\be{secOrderalpha}
\Delta E_\alpha&=&8 \int d^2x\int d\omega \,(\frac{\omega}{2\pi})^4 \sum_{j}\sum_{\vec n,\vec l} \,' (-A_x)^{n_x+l_x}(-A_y)^{n_y+l_y} \nonumber\\
&& ( \ln(|\omega|\sqrt{V/v_Fv_\Delta})-\ln(2)-\Psi(1) ) \cdot \nonumber \\ 
&& \{K_0^{(j)}(|\omega(\vec x-\vec x_0+\vec n)|) K_0^{(j)}(|\omega(\vec x_0-\vec x-\vec l)|)\nonumber\\
&+& \frac{(\vec x-\vec x_0+\vec n)}{|\vec x-\vec x_0+\vec n|}\frac{(\vec x-\vec x_0+\vec l)}{|\vec x-\vec x_0+\vec l|}K_1^{(j)}(|\omega(\vec x-\vec x_0+\vec n)|) K_1^{(j)}(|\omega(\vec x_0-\vec x-\vec l)|)\}\nonumber\\
\ee

\be{secOrderbeta}
\Delta E_\beta&=&4\int d^2x\int d\omega \, (\frac{\omega}{2\pi})^4 \sum_{i} \sum_{\vec{m}}\,'  (-A_x)^{m_x}(-A_y)^{m_y} [ \nonumber \\
&&\sum_{j}K_0^{(i)}(|\omega\vec m|)K_0^{(j)}(|\omega(\vec x-\vec x_0)|) K_0^{(j)}(|\omega(\vec x_0-\vec x)|)\nonumber\\
&+&\{\sum_{j}K_0^{(i)}(|\omega\vec m|)   K_1^{(j)}(|\omega(\vec x-\vec x_0)|)\nonumber\\
&& K_1^{(j)}(|\omega(\vec x_0-\vec x)|)
\nonumber \\
&-&2 K_0^{(i)}(|\omega(\vec x-\vec x_0+\vec m)|)\frac{K_1^{(i)}(|\omega(\vec x_0-\vec x)|)}{|\omega(\vec x_0-\vec x)|} \}\nonumber\\
&-&2\sum_{i}\sum_{\vec n}\,'  (-A_x)^{n_x}(-A_y)^{n_y} (K_0^{(i)}(|\omega(\vec x-\vec x_0+\vec m)|)\frac{K_1^{(i)}(|\omega(\vec x_0-\vec x+\vec n)|)}{|\omega(\vec x_0-\vec x +\vec n)|} ]\nonumber\\
&+&2\sum_{i} \sum_{\vec n,\vec m}\,' (-A_x)^{n_x+m_x}(-A_y)^{n_y+m_y} [K_1^{(i)}(|\omega(\vec x-\vec x_0+\vec n)|) \nonumber\\
&&K_1^{(i)}(|\omega(\vec x_0-\vec x+\vec m)|)\frac{(\vec x -\vec x_0+\vec n)\cdot(\vec x -\vec x_0 +\vec m)}{|\vec x -\vec x_0 +\vec n| \cdot |\vec x -\vec x_0 +\vec m|}\nonumber \\
&-& K_0^{(i)}(|\omega(\vec x-\vec x_0+\vec n)|) K_0^{(i)}(|\omega(\vec x_0-\vec x+\vec m)|)\}]
\ee

\noindent
In the non-zero mode sectors, the contribution from the other summands is trivial to write down:
\be{secOrder2nonzero}
\Delta E_r &= & 4 \int d^2x\int d\omega \,  (\frac{\omega}{2\pi})^4\sum_{i,j} \sum_{\vec n,\vec l}\,' \sum_{\vec m}\,'(-A_x)^{n_x+m_x+l_x}(-A_y)^{n_y+m_y+l_y}   [ \nonumber \\ 
&&K_0^{(i)}(|\omega\vec m|)  K_0^{(j)}(|\omega(\vec x-\vec x_0+\vec n)|) K_0^{(j)}(|\omega(\vec x_0-\vec x-\vec l)|)\nonumber\\
&+&
K_0^{(i)}(|\omega\vec m^2|)  \frac{(\vec x-\vec x_0+\vec n)\cdot(\vec x-\vec x_0+\vec m)}{|\vec x-\vec x_0+\vec n|\cdot |\vec x-\vec x_0+\vec m|}
 \nonumber \\
&&  K_1^{(j)}(|\omega(\vec x-\vec x_0+\vec n)|)K_1^{(j)}(|\omega(\vec x_0-\vec x-\vec l)|)
\ee

\noindent
In the zero mode sector, I replace the non-interacting Green's function with the one, where the zero mode is excluded: 
\be{regGreen0}\tilde{G}(\vec x,\vec x_0,\omega)&=& \sum_{\vec n}\{G(\vec x+\vec n,\vec x_0,\omega)-\frac{1}{L_xL_y}\int d^2y \,G(\vec y+\vec n,\vec x_0,\omega)\}\nonumber\\
&=& \sum_{\vec n}\frac{1}{V} \int d^2 y \,[ -\frac{i\omega}{2\pi}\gamma_0(K_0(|\omega(\vec x-\vec x_0)|)-K_0(|\omega\vec y|))\nonumber\\
&+& \frac{|\omega|}{2\pi}\vec\gamma\{\frac{\vec x-\vec x_0}{|\vec x-\vec x_0|}K_1(|\omega(\vec x-\vec x_0)|)-\frac{\vec y}{|\vec y|}K_1(|\omega\vec y|)\}]
\ee
and insert the modified Green's functions in (\ref{totalEnergyShift})
\be{secOrder2zero}
\Delta E_r&=&\frac{4}{V}\int d^2x \int d^2y \int d\omega  \,  (\frac{\omega}{2\pi})^4 \sum_{\vec n,\vec l}\,' \sum_{\vec m}\,' K_0^{(i)}(|\omega\vec m|)[ \nonumber \\ 
&&  \{K_0^{(j)}(|\omega(\vec x-\vec x_0+\vec n)|)-K_0^{(j)}(|\omega(\vec y+\vec n)|)\} \cdot\nonumber\\
&& \{K_0^{(j)}(|\omega(\vec x_0-\vec x-\vec l)|)-K_0^{(j)}(|\omega(\vec y-\vec l)|)\}\nonumber\\
&+&\{\frac{\vec x-\vec x_0+\vec n}{|\vec x-\vec x_0+\vec n|}K_1^{(j)}(|\omega(\vec x-\vec x_0+\vec n)|)-\frac{\vec y+\vec n}{|\vec y+\vec n|}K_1(|\omega(\vec y+\vec n)|)\}\cdot\nonumber\\
&&\{\frac{\vec x_0-\vec x-\vec l}{|\vec x_0-\vec x-\vec l|}K_1^{(j)}(|\omega(\vec x_0-\vec x-\vec l)|)-\frac{\vec y-\vec l}{|\vec y-\vec l|}K_1^{(j)}(|\omega(\vec y-\vec l)|)\}]
\ee

Even though I replaced only two of the three Green's functions\footnote{The Green's function $\hat{G}(\vec x_0,\vec x_0,\omega)$ is unchanged and could, in principle, give different contributions for the different ground states.},  one finds that the energy shift of all the filled levels is {\it independent} of whether the zero mode is (partly) filled or not. However, the energy shift of the zero mode itself to the Casimir energy is, in general, distinct for the different ground states:

\be{secOrder3zeromode}
\Delta E^{0}_r(|\tilde{0}\rangle)&=& -\frac{1}{V}\sum_{\vec k}\frac{|\langle\vec k|\hat{V}|\tilde0\rangle|^2}{\epsilon_k}+
\int \frac{d^2k}{(2\pi)^2} \frac{|\langle\vec k|\hat{V}|\tilde0\rangle|^2}{\epsilon_k}
\ee

The expressions do not show a simple dependence on the windings. It is not possible to extract some significant features from the analytic results. In addition, the computation of the sums is very time-consuming. For low frequencies, the convergence of the sums is very slow. In addition, the expressions are given by modified Bessel functions. While this does not give any fundamental problems, it makes the numerical calculations very slow. With a computational program like mathematica or matlab, the expressions can not be computed in an acceptable time intervall. 

\subsection{Conclusions of the second part}
 In the second part of this work, we considered a local perturbation, namely a magnetic point defect, and studied its effects on the Casimir energy in the different sectors. The total Casimir energy can be  written in terms  of Green's functions and be calculated perturbatively. I showed that the Casimir energy (in case of no perturbing potential) is distinct for different sectors\footnote{as could be expected in presence of gapless particles} and looked for regularities in the energy shift induced by the magnetic defect. In the discussion I focused on the three unique ground states in the non-zero mode sectors and the two (degenerate) ground states in the zero-mode sector which were connected to the other ones by adiabatic tunneling processes. 
 
 A major result of this discussion is  that there are, indeed, some regularities for these ground states as their energy shift to first order perturbation theory vanishes. Note that the first order shift is not generically zero, but that there are certain ground states energies which are shifted by the perturbation.  
 
I also derived the analytic expressions for the energy shift in second order. As the singular perturbation introduced new UV divergencies, I renormalized the interaction and showed that the contributions are finite in all sectors. To write the programs which are necessary to calculate the second order shift is beyond the scope of this work and will be left for later work.

\newpage
\section{Final conclusions and Outlook}

The 2+1 Dirac theory proves an appropriate model for the d-wave superconductor. In general, a two-dimensional representation is sufficient to describe a two-dimensional Dirac theory. The four-dimensional representation has, however, several advantages, when it comes to the interpretation of the physical properties. In a two-dimensional representation, there are no closed expressions for the spin charge or the spin current. Combining the fermion fields of opposite nodes into a four dimensional spinor, provides a simple interpretation for the density as the spin and the conserved current as the spin current.

An equal copy of the representation arises for the other node pair. 
There are various versions of how to distribute the fermion operators 
into the spinors and the suitable choice depends strongly on the quantities you 
want to study and which symmetries you want to keep manifest in the model. 
As I wanted to study the coupling to magnetic flux, it was most convenient to 
take an approach which masks the spin rotational symmetry. On the other hand,
it allows us to identify $\Psi^\dagger\Psi$ with the spin density and to 
write down the spin current in a simple form. 

The most delicate part in the derivation of the Dirac theory, is the point 
where you have to eliminate the topological phase in order to obtain spatially 
independent matrices. As I already mentioned earlier, the naive choice to distribute the 
phase equally to spin-up and spin-down leads to multi-valued spinors. To 
distribute it solely to spin-up particles gives a reasonable result, but one 
has to keep in mind that this induces an artificial asymmetry in the behavior 
of the fermion operators. The spin current is only coupled to the topological 
phase, a result which remains true, even if you choose another way of attaching 
the topological phase to spin, see \cite{Herbut}. 

The effective theory you obtain, looks like an ordinary Dirac theory for massless spin 1/2 fermions. In order to analyze the theory for topological properties, it is convenient to consider a torus as the underlying manifold. Further, you neglect vortices to simplify the analysis. Magnetic fluxes are now implemented by flux tubes through the holes of the torus and can be measured by taking quasiparticles along non-contractible paths around these holes. The gauge field, that enters the effective theory, can be regarded as spatially constant and classifies four different sectors, one for each configuration of flux quanta. For an s-wave superconductor, the structure of each of the sectors is identical. The ground state can be labeled by either the magnetic or the electric flux as both, quasiparticles and vortices, are gapped. In the d-wave case, only magnetic fluxes can be used to describe the different topological sector. You find, that one of the sectors behaves exceptional. Which one it is, depends on the boundary conditions you impose on the fermion operators. The ground state is highly degenerate in this sector. It is 16-fold degenerate for each pair of nodes. The complete degeneracy is, thus, given by 256, as both pairs can be treated independently. 

In order to describe the degenerate ground state, the observables spin, momentum 
and charge are not sufficient. As stated earlier, there is also another quantity, besides spin and momentum, which is conserved. The chiral charge measures the difference in the number of quasiparticles at opposite nodes. It provides the missing tool to label each one of the degenerate states uniquely. However, by changing boundary conditions, you can change which of the sectors behaves exceptional. This suggests, that this (huge) degeneracy in one of the sectors might not be fundamental. As in the s-wave case, you can connect the different sectors by tunneling processes. Doing this adiabatically, in particular when going from a (unique) ground state to the degenerate sector, you pick out one particular ground state out of all degenerate ones. One finds, that the obtained state depends on the sector you started from. In the end, there are only two states which can be obtained by adiabatic tunneling processes. They look very similar to each other and can be translated into each other by replacing $x\leftrightarrow y$.  

As the quasiparticles are massless, the ground states in the different sectors are not degenerate and generally the different sectors will mix due to tunneling processes. It is an interesting question whether some regularity still reflects the topological order. In this thesis, we chose to examine the response of the Casimir energy  to a local impurity. For gapped quasiparticles, local impurities do not affect the ground state energy. We found that this is also true in case of massless particles to first order perturbation theory. The Casimir energy of the  two states which could be connected to the ground states in the other sectors, are not shifted and, thus, behave as the ground states of the other sectors. In second order perturbation theory, I derived the analytic expression for the energy shift in second order. The calculations become much more involved and the expressions become divergent. The divergencies can be canceled by local counterterms and the interaction strength $V_0$ is renormalized. The expressions show no simple dependence on the winding modes and there is no possibility to make statements on the second order energy shift without numerical calculation.   However, the numerical results require a lot of time and computational power and will not be presented in this thesis.

Some experiments indicate that cuprates do not have a pure d-wave pairing but rather a mixture of d- and s-wave pairing. As the cuprates are anisotropic, the relative strength of the order parameters depends on the direction and is, therefore, tunable. In appendix D I show, how the excitation spectrum changes, if a (small) s-wave contribution is added. It would also be interesting to see, what happens to the degenerate ground states if you (slowly) switch on this s-wave contribution. With an s-wave contribution, there is a unique ground state in all four sectors. By a perturbative calculation, one should be able to find out, which one is selected.

Another interesting feature is to discuss the effect of shifting the nodes away from the position $\vec{k} =(\pm k_F, \pm k_F)$. This certainly effects the energy spectrum. The simplest possibility is just to rotate the nodes with a small angle. This should not change the structure of the ground state in a qualitative way. Actually, by expanding the field operators around the nodes and changing to relative coordinates, namely changing from $(q_x,q_y)$ to $(q_1,q_2)$, the absolute position of the nodes becomes less important and is reflected only in the parameter $v_\Delta$. The only needed feature is the symmetric arrangement of the nodes. This should lead to a certain robustness of the structure against rotations. Deviations of the symmetric arrangement, however, should affect the ground state structure. 

In addition, I completely neglected vortices on the torus by setting the vortex current $\tilde \jmath^\mu = 0$. By allowing vortices only as thin fluxes through the holes, the theory remains regular on the surface of the torus. The next step would be to consider stationary thin flux tubes on the torus. If the core of the flux tube is much smaller than the magnetic length, it is still a good approximation to consider the absolute value of the order parameter to be constant.

\newpage

\begin{appendix}
\appendix

\renewcommand{\theequation}{A-\arabic{equation}}
\setcounter{equation}{0}

\section{Spin Current}
As mentioned in the main text, the spin current is the conserved Noether current 
of our model. However, our choice of spinors masks the invariance under the SU(2)
spin rotation group. In order to derive the Noether current, another choice of
the spinor representation \cite{Fisher} is more useful as it leaves the 
rotational invariance manifest:
\be{Fisherspinor}
\Psi_j^\dagger&=&(\psiudag j, \psiddag j, \psiu {j+2}, -\psid {j+2})
\ee
The minus sign for $\psid{j+2}$ was chosen so that $\Psi_j$ transforms as a
spinor. With this convention, spin transformation can be written as matrices, but 
the chiral symmetry is obscured. 

With this choice of the spinor, it is now easy to formulate 
the operator for spin rotations. It is obvious that the Lagrangian will
be invariant under this (global) transformation. 
\be{spin rotations}
U&=&\exp\left(i\vec\theta  \cdot\vec\Sigma_i\right)
\ee
where 
$$\Sigma_i=\half\left(\begin{array}{cc}\sigma_i&0\\0&\sigma_i\end{array}\right)$$
and $\sigma_i$ being the well-known Pauli matrices:

\be{sigmamatrices}
\sigma_x &=& \left( \begin{array}{cc} 0&1\\1&0 \end{array}\right) \nonumber \\
\sigma_y &=& \left( \begin{array}{cc} 0&-i\\i&0 \end{array}\right) \nonumber \\
\sigma_z &=& \left( \begin{array}{cc} 1&0\\0&-1 \end{array}\right)
\ee

With our choice of spinors, see (\ref{spinor}),  the spin
transformations are more cumbersome to write down because of two reasons: first, we 
chose the spinors to have definite spin, $\Psi$ having spin $\frac{1}{2}$ and 
$\bar\Psi$
having $-\frac 1 2$ . Therefore, a spin rotation mixes $\Psi$ and $\bar\Psi$ 
and thus, it is not possible to represent it as a matrix as in \cite{Fisher}. 
Secondly, although $\Psi(\vec q)$ looks local in $\vec q$-space it is not. Note the 
convention, I used to define the fermionic operators (\ref{k-exp}). $\vec q$ was 
defined in such a way, that it gives the direction of the spin current. Spin 
rotations on the other hand are local operations. For example flipping the spin of 
a particle reverses its contribution to the spin current, changing $\vec q$ to 
$-\vec q$. 

In order to derive the correct form of the spin operators, it is convenient to write 
down the transformations for the original fermionic operators $c_\sigma(\vec k)$ and 
translate this expression to the linearized ones. The original operators transform 
in the usual way under rotations with angle $\alpha$: 
\be {spintransform}
(c_\uparrow, c_\downarrow)& \rightarrow &\exp{\left(i\frac \alpha 2 \sigma_j\right)} 
(c_\uparrow, c_\downarrow) 
\ee
The spin operators themselves are given by 
\be {oldspin}
S_j &=& \sum_{\vec k}
(c_\uparrow^\dagger(\vec k),c_\downarrow^\dagger(\vec k))\,\sigma_j\,
\left(\begin{array}{c} c_\uparrow(\vec k)\\ c_\downarrow(\vec k)
\end{array}\right)
\ee

Using equation (\ref{k-exp}), these expressions can be written in terms of 
the linearized fermion operators. After transforming to $\vec x$-space, one 
has to attach the singular phase to spin-up operators. Then, I use the 
relations (\ref{1spinor}) to (\ref{4spinor}), 
to obtain the final expression for the spin operators. 
I do this calculation explicitely for $S_x$. $S_y$ is obtained in a completely
analogous way, while $S_z$ is even simpler to derive.

Due to the definition of the fermionic field operators, the spin in the x- and 
y-directions looks nonlocal in the linearized version. Although the sign convention
may seem unnecessary complicated it is nevertheless crucial, to write 
(\ref{startingpoint}) as a Dirac equation. 

\be {linearizedspin}
S_x &=& \sum_{\veq}\psiudag{1}(\veq)\psid{1}(-\veq)
+\psiudag{3}(\veq)\psid{3}(-\veq)\nonumber \\
&+&\psiddag{1}(\veq)\psiu{1}(-\veq)+\psiddag{3}(\veq)\psiu{3}(-\veq)
\ee
and after transforming back to $\vec x$-space:

\be{linspinx}
S_x & = & \int d^2x \, \psiudag{1}(\vec x)\psid{1}(-\vec x) +\psiudag{3}(\vec x)
\psid{3}(-\vec x)\nonumber \\
&+&\psiddag{1}(\vec x)\psiu{1}(-\vec x)
+\psiddag{3}(\vec x)\psiu{3}(-\vec x)
\ee

The next step is to attach the singular flux to the spin-up field operators. I 
neglect regular gauge transformations, as their phase contributions cancel anyway. 
For clarity, I introduce a new notation for the fermionic field with attached flux:

\be{introxi}
 \chi_{j,\uparrow}(\vec x)& =& e^{-i\varphi_s(x)}\psi_{j,\uparrow}(\vec x)\\
\chi_{j,\downarrow}(\vec x) &=& \psi_{j,\downarrow}(\vec x)
\ee

The Fourier transforms of the $\xi$ fields are defined exactly as for the $\psi$ 
fields, namely that both spin components transform with opposite momentum. 

\be{xitrans}
\chi_{j,\uparrow}(\vec x) &=& \sum_{\veq} \, 
e^{i\veq \vec x} \chi_{j,\uparrow}(\veq) \nonumber \\
\chi_{j,\downarrow}(\vec x) &=& \sum_{\veq} \, 
e^{-i\veq \vec x} \chi_{j,\downarrow}(\veq)
\ee

However, as the singular phase is space dependent, it changes the coupling of the 
fermionic 
fields. Without a singular gauge transformation, spin-up and spin-down couple with 
"opposite" $\veq$ values. The presence of the gauge field shifts one of these 
momenta.

\be{afterfourtrans}
S_x &=& \sum_{\veq, \vec k} 
(\int d^2x \, e^{-i\varphi(x)}e^{-i(\veq + \vec k)\vec x})
    (\chi_{1,\uparrow}^\dagger(\vec k)\chi_{1,\downarrow}(\vec q)
     +\chi_{3,\uparrow}^\dagger(\vec k)\chi_{3,\downarrow}(\vec q))\nonumber \\
&+&(\int d^2x \, e^{i\varphi(x)}e^{i(\veq + \vec k)\vec x})
    (\chi_{1,\downarrow}^\dagger(\vec k)\chi_{1,\uparrow}(\vec q)
     +\chi_{3,\downarrow}^\dagger(\vec k)\chi_{3,\uparrow}(\vec q))
\ee

In order to obtain the shift, one needs to find the value of the functions 
\be {integrals}
 f_+(\varphi, \veq,\vec k)&=& \int d^2x \, 
e^{i\varphi_s(x)} e^{i(\veq + \vec k)\vec x}\nonumber \\
f_-(\varphi, \veq, \vec k) &=& \int d^2x \, 
e^{-i\varphi_s(x)} e^{-i(\veq + \vec k)\vec x}
\ee

You can take $\varphi_s(\vec x)$ to be given by:
$$ \varphi_s(\vec x) =-2 a_x x -2 a_y y$$
After inserting this expression into (\ref{integrals}), the integrals can 
be computed easily to give:

$$f_+(\varphi, \veq, \vec k)=f_-(\varphi, \veq, \vec k) = 
\delta_{\veq,-\vec k+2\vec a} $$
 
In the following discussion, I keep writing $\veq$ instead of $\vec k+ 2\vec a$ in 
order to shorten the notation.
The reader should, however, keep the equivalence in mind. 
Note, that the spin operator is now non-local in momentum space because of the 
attachment of the singular phase. 
The remaining step is to express the obtained expression for the spin in terms
of the quasiparticle operators, $a_i$ and $b_i$ which diagonalize the Hamiltonian. 
Even though the eigenspinors form an orthonormal basis for each momentum, scalar 
products such as $u_j^\dagger(\vec k) \cdot v_j(\veq)$ are generally non-zero. 

First note that $\chi_{j,\sigma}(\vec k)$ and $\chi_{j,\sigma}(\veq)$ are linear 
combinations of quasiparticle operators with the same energy:
\be{energyrelation}
E(\vec k)&=& \sqrt{ v_F^2(k_x-\ai x)^2 + v_\Delta^2(k_y - \ai y)^2} \nonumber\\
&=&\sqrt{ v_F^2(-k_x+\ai x)^2 + v_\Delta^2(-k_y + \ai y)^2} \nonumber\\
&=& E(\veq)
\ee

Assume that neither $k_y-\ai y$ nor $k_x-\ai x$ is equal to zero. To perform the 
computation, it is favorable to introduce 2-dimensional vectors: 
\be{helpvec}
w_1(\vec k)&=& c\left(-v_\Delta(k_y-\ai y), E-v_F(k_x-\ai x)\right) \nonumber\\
w_2(\vec k)&=& c\left(E-v_F(k_x-\ai x),v_\Delta(k_y-\ai y)\right)
\ee
with the same normalization constant c as in (\ref{1spinor}). The vectors fulfill 
the relations:
\be{helpvec2}
w_1(\vec k) \cdot w_1(\veq) &=& 0 \nonumber \\
w_2(\vec k) \cdot w_2(\veq) &=& 0 \nonumber \\
w_1(\vec k) \cdot w_2(\veq) &=& -\frac{k_y-\ai y}{|k_y-\ai y|}\nonumber \\
w_1(\vec k) \cdot w_1(\veq) &=& \frac{k_y-\ai y}{|k_y-\ai y|}
\ee
As the spin operator is summed over all momenta, we are allowed to relabel 
$\vec k\leftrightarrow \veq$ and it can be written as
\be {helpspin} 
\tilde{S}_x &=& \sum_{k_y-\ai y \ne 0}
\chi_{1,\uparrow}^\dagger (\vec k) \chi_{1,\downarrow}(\veq) 
- \chi_{3,\downarrow}(\vec k)\chi_{3,\uparrow}^\dagger(\veq) 
+\chi_{1,\downarrow}^\dagger (\vec k) \chi_{1,\uparrow}(\veq) 
- \chi_{3,\uparrow}(\vec k)\chi_{3,\downarrow}^\dagger(\veq)\nonumber \\ 
&=& \sum_{k_y-\ai y \ne 0}
(\chi_{1,\uparrow}^\dagger(\vec k), \chi_{3,\downarrow}(\vec k))\cdot
\left(\begin{array}{cc} \chi_{1,\downarrow}(\veq) \\ -\chi_{3,\uparrow}^\dagger(\veq)
\end{array}\right) \nonumber \\
&+& (\chi_{1,\downarrow}^\dagger(\vec k), \chi_{3,\uparrow}(\vec k))\cdot
\left(\begin{array}{cc} \chi_{1,\uparrow}(\veq) \\ -\chi_{3,\downarrow}^\dagger(\veq)
\end{array}\right) \nonumber \\
\nonumber \\
&=& \sum_{k_y-\ai y \ne 0}
(a_1^\dagger(\vec k) w_1(\vec k) + b_1(\vec k)w_2(\vec k))\cdot
(-a_2^\dagger(\veq)w_1(\veq) +b_1(\veq)w_2(\veq))\nonumber \\
&+& (-a_2(\vec k)w_1^\dagger(\vec k) + b_1^\dagger(\vec k)w_2^\dagger(\vec k))
\cdot (a_1(\veq)w_1(\veq)+b_2^\dagger(\veq)w_2(\veq))
\ee

By using the relations (\ref{helpvec2}), one obtains a rather simple result:
\be{spinx1}
\tilde{S}_x &=& \sum_{k_y-\ai y \ne 0} \mbox{sign}(k_y-\ai y)\left(
a_2^\dagger(\veq)b_1^\dagger(\vec k)+a_1^\dagger(\veq)b_2(\vec k)
+ b_1^\dagger(\vec k)a_2(\veq)\right.\nonumber \\
&+& \left. b_2^\dagger(\vec k)a_1(\veq)\right)
\ee

The same calculations can be done for (\ref{2spinor}) and (\ref{3spinor}). 
In the case that $k_y - \ai y \ne0$, you obtain the same result as for 
(\ref{1spinor}). 
In the case that only $k_y -\ai y = 0$ but not $k_x-\ai x$, 
you obtain the same result but with 
$x\leftrightarrow y$  interchanged. All these cases can be summarized by introducing 
\be {f}
f(\vec k)= \mbox{sign}(k_y-\ai y)& \mbox{if}& k_y -\ai y \ne 0\nonumber \\
f(\vec k)=\mbox{sign}(k_x-\ai x)&\mbox{if}& k_y-\ai y =0
\ee

In the last case, (\ref{4spinor}), I chose the 
eigenvectors in such a way, that the spin operator looks the same for all 
momenta:
$$ a_2^\dagger(\vec k_0)b_1^\dagger(\vec k_0)+a_1^\dagger(\vec k_0)b_2(\vec k_0)
+ b_1^\dagger(\vec k_0)a_2(\vec k_0)+b_2^\dagger(\vec k_0)a_1(\vec k_0)$$

The computation of $S_y$ and $S_z$ can be done in a similar way. The operators can 
be written in a compact way as:
\be{quasispin}
S_x &=& \sum_{k\ne k_0} f(\vec k) \left(
a_2^\dagger(\veq)b_1^\dagger(\vec k)+a_1^\dagger(\veq)b_2(\vec k)
+ b_1^\dagger(\vec k)a_2(\veq)+b_2^\dagger(\vec k)a_1(\veq)\right) \nonumber\\
&+& a_2^\dagger(\vec k_0)b_1^\dagger(\vec k_0)+a_1^\dagger(\vec k_0)b_2(\vec k_0)
+ b_1^\dagger(\vec k_0)a_2(\vec k_0)+b_2^\dagger(\vec k_0)a_1(\vec k_0)\nonumber \\
S_y&=&i\sum_{k\ne k_0} f(\vec k) \left(
-a_2^\dagger(\veq)b_1^\dagger(\vec k)-a_1^\dagger(\veq)b_2(\vec k)
+ b_1^\dagger(\vec k)a_2(\veq)+b_2^\dagger(\vec k)a_1(\veq)\right)\nonumber\\
&-& a_2^\dagger(\vec k_0)b_1^\dagger(\vec k_0)-a_1^\dagger(\vec k_0)b_2(\vec k_0)
+ b_1^\dagger(\vec k_0)a_2(\vec k_0)+b_2^\dagger(\vec k_0)a_1(\vec k_0) \nonumber \\
S_z&=& \sum_k
a_1^\dagger(\vec k)a_1(\vec k)+a_2^\dagger(\vec k)a_2(\vec k) 
- b_1^\dagger(\vec k)b_1(\vec k)-b_2^\dagger (\vec k)b_2(\vec k)
\ee

\renewcommand{\theequation}{B-\arabic{equation}}
\setcounter{equation}{0}

\section{Symmetries of the Effective Lagrangian}
The effective Lagrangian exhibits additional symmetries, which were not 
present at the original Hamiltonian. This is mostly due to that
the node pairs (1-3) and (2-4) are completely decoupled after expanding
$c_k$ and $c_k^\dagger$ around the nodes. 

Thus the Lagrangian will be invariant if we apply separate SU(2) spin rotations
on the two pairs of nodes $\Psi_1$ and $\Psi_2$. This is also true for U(1) 
charge transformations. As was pointed out in \cite{Fisher}, this implies two 
new conserved charges, called nodon charges, $\Psi_j^\dagger\Psi_j$.  It was 
also mentioned that additional interaction terms, e.g. Umklapp scattering 
processes, will not conserve the nodon charge while being consistent with the 
original U(1) and SU(2) symmetries.

In our case there is even one more symmetry due to our choice of distributing
the phase of the gap function $\Delta(\vec R)$. 
\be{phase}
\psiu j&\rightarrow&e^{i(\varphi_r/2+\varphi_s)}\nonumber\\
\psid j&\rightarrow&e^{i(\varphi_r/2)}
\ee
(\ref{phase}) introduces a new gauge symmetry, where the singular and the 
regular phase are transformed simultaneously:
\be{gauge}
\varphi_r &\rightarrow&\varphi_r-\beta\nonumber\\
\varphi_s &\rightarrow&\varphi_s+\beta
\ee
Written in the spinor fields $\Psi_j$ this looks like the U(1) gauge 
transformation. 
\be{gauge2}
\Psi_j&\rightarrow& e^{i\beta/2}\Psi_j
\ee
Again two different values of $\beta$ are allowed for the different pairs 
of nodes. The invariance under U(1) charge transformations implies that the $\Psi$ 
fields are neutral while this obviously does not apply to $\psi_j$.

\renewcommand{\theequation}{C-\arabic{equation}}
\setcounter{equation}{0}

\section{Analysis of Spin and Chirality}

Here I want to sketch, how to derive the structure in sector four. Due to the zero 
mode, there are 16 linearly independent states with zero energy: these consist of 
quasiparticles with momentum $\vec{k}_0 = (\frac{\pi}{L_x}, \frac{\pi}{L_y})$ and 
can be labeled by spin and chirality, see Table \ref{table}.

\begin{table}

\begin{displaymath}
\begin{array}{|c||c|c|c|c|c|}\hline 
\Gamma_z \setminus S_z 
& \mathbf{1} & \mathbf{\half} 
& \mathbf{0} &\mathbf{-\half} &\mathbf{-1} \\ \hline\hline
\mathbf{1} & - & - & a_1^\dagger b_2^\dagger\zero & - & - \\ \hline
\mathbf{\half} & - & 
\begin{array}{c} a_1^\dagger\zero \\ 
a_1^\dagger a_2^\dagger b_2^\dagger \zero \end{array} &-&
\begin{array}{c} b_2^\dagger\zero \\ a_1^\dagger b_1^\dagger 
b_2^\dagger \zero \end{array}&-\\
\hline \mathbf{0} & a_1^\dagger a_2^\dagger\zero & - & 
\begin{array}{c} \zero\\ a_1^\dagger b_1^\dagger\zero\\
a_2^\dagger b_2^\dagger\zero\\a_1^\dagger a_2^\dagger b_1^\dagger 
b_2^\dagger\zero \end{array} & - & 
b_1^\dagger b_2^\dagger\zero \\ \hline
\mathbf{-\half} & - & 
\begin{array}{c} a_2^\dagger \zero \\ a_1^\dagger a_2^\dagger 
b_1^\dagger \zero \end{array} &-&
\begin{array}{c} b_1^\dagger \zero \\ a_2^\dagger b_1^\dagger 
b_2^\dagger \zero \end{array} &-\\
\hline \mathbf{-1} & - &-&a_2^\dagger b_1^\dagger \zero &-&- \\\hline
\end{array}
\end{displaymath}

\caption{eigenvalues of the degenerate ground states in sector 4}
\label{table}
\end{table}
Both, spin and chirality, can be represented by one triplet, four doublets and five
singlets. There are four states with eigenvalue zero for spin and chirality. In 
order to see, which ones are singlets and which belong to the triplets, you need 
to compute the eigenvalues to the Casimir operators, $S^2$ and $\Gamma^2$:
\be{Casimir}
S^2 &=& S_x^2 + S_y^2 + S_z^2 \nonumber \\
\Gamma^2 &=& \Gamma_x^2 + \Gamma_y^2 + \Gamma_z^2 
\ee

Let's look at the spin part first. 
I rewrite the Casimir operator  in terms of the ladder operators, $S_+$ and $S_-$ 
\be{ladderspin}
S_+ &=&(S_x+iS_y)\nonumber\\
&=& \sum_k f(\vec k) \left(
a_1^\dagger(\veq)b_2(\vec k) +a_2^\dagger(\veq)b_1(\vec k)\right)\\
S_- &=& (S_x-iS_y)\nonumber\\
&=& \sum_k f(\vec k)\left(
b_1^\dagger(\vec k)a_2(\veq) + b_2^\dagger(\vec k) a_1(\veq)\right)
\ee
Singlets will be eigenstates to $S_+S_-$ with eigenvalue zero.
\be{spineigen}
S_+S_-\zero &=& 0 \nonumber \\
S_+S_-a_1^\dagger b_1^\dagger \zero &=&\left(a_1^\dagger b_1^\dagger
-a_2^\dagger b_2^\dagger\right)\zero\nonumber\\
S_+S_-a_2^\dagger b_2^\dagger \zero &=& \left( a_2^\dagger b_2^\dagger -
a_1^\dagger b_1^\dagger\right)\zero\nonumber \\
S_+S_-a_1^\dagger a_2^\dagger b_1^\dagger b_2^\dagger\zero &=&
0
\ee
Thus, three states are singlets in spin, namely: 
\be{spinsing}
&\zero&\nonumber \\ 
&\frac{1}{\sqrt{2}}\left(a_1^\dagger b_1^\dagger+a_2^\dagger b_2^\dagger\right)
\zero& 
\mbox{and} \nonumber \\ 
&a_1^\dagger a_2^\dagger b_1^\dagger b_2^\dagger \zero&
\ee
The remaining  state, 

$$\frac{1}{\sqrt{2}}\left(a_1^\dagger b_1^\dagger -a_1^\dagger b_2^\dagger\right)
\zero$$ 
forms a spin triplet, together with $a_1^\dagger a_2^\dagger\zero$ and 
$b_1^\dagger b_2^\dagger\zero$. 

The same analysis can be done for the chirality operators. They can be computed by 
$\Gamma_1=\frac{1}{2}\Psi^\dagger\gamma_3\gamma_5\Psi$ 
and $\Gamma_2=\frac{i}{2}\Psi^\dagger\gamma_3\Psi$. 
Explicitly, they are given by:
\be{chirinab}
\Gamma_1&=&\frac{i}{2}\sum_{k\ne k_0}
\left(-\ad{1}a_2(\vec k)+b_1(\vec k)\bd{2}-b_2(\vec k)\bd{1}+
\ad{2}a_1(\vec k)\right)\nonumber\\
&+& \frac i 2 \left(a_1^\dagger(\vec k_0)b_2^\dagger(\vec k_0)-
b_1(\vec k_0)a_2(\vec k_0) +a_2^\dagger(\vec k_0)b_1^\dagger(\vec k_0)
-b_2(\vec k_0)a_1(\vec k_0)\right) \nonumber \\
\Gamma_2&=& \frac{1}{2}\sum_{k\ne k_0} 
\left(-\ad{1}a_2(\vec k)+b_1(\vec k)\bd{2}+b_2(\vec k)\bd{1}-\ad{2} a_1(\vec k)
\right)\nonumber \\
&+& \frac 1 2 \left(a_1^\dagger(\vec k_0)b_2^\dagger(\vec k_0)-
b_1(\vec k_0)a_2(\vec k_0) -a_2^\dagger(\vec k_0)b_1^\dagger(\vec k_0)
+b_2(\vec k_0)a_1(\vec k_0)\right)
\ee
The ladder operators are computed analogously to the spin case:

\be{ladderchir}
\Gamma_+&=&-i\sum_{k\ne k_0} \bd{1}b_2(\vec k)+\ad{2}a_1(\vec k)
\nonumber\\
&+& i \left(a_1^\dagger(\vec k_0)b_2^\dagger(\vec k_0) -b_1(\vec k_0)a_2(\vec k_0)
\right) \nonumber \\
\Gamma_-&=&i\sum_{k\ne k_0} \ad{1}a_2(\vec k)+\bd{2}b_1(\vec k) \nonumber \\
&+& i\left( a_2^\dagger(\vec k_0)b_1^\dagger(\vec k_0) -b_2(\vec k_0)a_1(\vec k_0)
\right)
\ee

Again, one can act with the ladder operators on the states, computing the effect of 
$\Gamma_+\Gamma_-$. They pick another state to be the member of the (chirality) 
triplet:
\be{chireigen}
\Gamma_+\Gamma_-\zero&=& \left(1-a_1^\dagger a_2^\dagger 
b_1^\dagger b_2^\dagger \right)\zero \nonumber \\
\Gamma_+\Gamma_-a_1^\dagger b_1^\dagger\zero &=& 0 \nonumber\\
\Gamma_+\Gamma_-a_2^\dagger b_2^\dagger \zero &=& 0 \nonumber\\
\Gamma_+\Gamma_-a_1^\dagger a_2^\dagger b_1^\dagger b_2^\dagger \zero &=& 
\left(a_1^\dagger a_2^\dagger b_1^\dagger
b_2^\dagger -1\right)\zero 
\ee
Thus, there are also three chirality singlet states: 
\be{chirsing}
&\frac{1}{\sqrt{2}}\left(1+a_1^\dagger a_2^\dagger
b_1^\dagger b_2^\dagger \right)\zero & \nonumber \\
& a_1^\dagger b_1^\dagger\zero&  
\nonumber \\
&a_2^\dagger b_2^\dagger\zero&
\ee 

The triplet is given by 
\be{chirtrip}
&a_2^\dagger b_1^\dagger& \nonumber \\ 
&\frac{1}{\sqrt{2}}\left(1-a_1^\dagger a_2^\dagger 
b_1^\dagger b_2^\dagger \right)\zero& \nonumber \\
&a_1^\dagger b_2^\dagger \zero
\ee 
Now it is easy to see the structure of the ground state: there are two states, which 
are singlets, both in chirality and spin:
\be{doublesing}
&\frac{1}{\sqrt{2}}\left(1+a_1^\dagger a_2^\dagger
b_1^\dagger b_2^\dagger \right)\zero& \nonumber \\ 
&\frac{1}{\sqrt{2}}\left(a_1^\dagger b_1^\dagger
+a_2^\dagger b_2^\dagger \right)\zero&
\ee 
The state 
$$\frac{1}{\sqrt{2}}\left(1-a_1^\dagger a_2^\dagger
b_1^\dagger b_2^\dagger \right)\zero$$
is singlet in spin but not chirality and the state 
$$\frac{1}{\sqrt 2}\left(a_1^\dagger b_2^\dagger - a_2^\dagger  b_1^\dagger 
\right)\zero$$ 
is singlet in chirality but not in spin. 
These ground states can now be compared to the (unique) ground states in the other 
sectors.

\renewcommand{\theequation}{D-\arabic{equation}}
\setcounter{equation}{0}
\section{The Poisson resummation formula}
The Poisson resummation formula gives a relation between a function $f$ and its 
corresponding Fourier transform $\hat{f}$. 
\be{general}
\sum_{m=-\infty}^\infty f(x+m) = \sum_{n=-\infty}^\infty\ \hat{f}(n)e^{(2\pi i n x)}
\ee
where both m and n are integers. 
This equality can be used to transfer the sum over momenta to a sum over 
windings on the torus. First, I express the Green's function in terms of dimensionless 
variables : $\vec{n}$ and  $\vec{x}'$. 
 
\be{green}
\hat{G}(\vec{x},\vec{y}, \omega)&=& \sum_{\vec{p}} G(\vec{p},\omega)
\exp{(i\vec{p}(\vec{x}-\vec{y}))}\\
&=& \frac{1}{L_1L_2} \sum_{\vec{n}}
G(\vec{n},\omega)\exp{\left(2\pi i\vec{n}(\vec{x}'-\vec{y}')\right)} \nonumber \\
& &\cdot \hat{f}(\vec{x}',\vec{y}',\vec{a})
\ee 
where  $x_1' = \frac{x_1}{L_1}$ and $x_2 = \frac{x_2'}{L_2}$. The additional 
factor is given by:
$$\hat{f}(\vec{x}',\vec{y}',\vec{a}) =\exp{\left(i\left(\left(\frac{\pi}{L_1},\frac{\pi}{L_2}\right)-\vec{a}\right) (\vec{x}-\vec{y})\right)}$$
It is independent of $\vec{n}$ and can be transferred to the right hand side. 
It gives a modified  Fourier transform of $\tilde{G}(\vec{n},\omega)$ for each sector.
\be{greensector}
\frac{1}{L_1L_2}\sum_{\vec{n}} G(\vec{n},\omega)\exp\left(2\pi i\vec{n}(\vec{x}'-\vec{y}')\right) &=&\hat{G}(\vec{x},\vec{y},\omega)\hat{f}(\vec{x}',\vec{y}',\vec{a})
\ee
The next step is to calculate the Fourier transform of $\tilde{G}(\vec{n},\omega)$ 
(assuming $\vec{n}$ to be a continuous variable) and use Poissons resummation formula:
\be{greenbessel}
\int \frac{d^2n}{(2\pi)^2} \tilde{G}(\vec{n},\omega)e^{(2\pi i \vec{n}(\vec{x}'-\vec{y}'))}
 &=& \frac{1}{2\pi} \frac{L_1L_2}{v_Fv_\Delta}\Big(-i\omega \gamma_0 
 K_0(|\omega||\vec{x}''-\vec{y}''|)\nonumber\\
 & & +|\omega|\frac{\vec{\gamma}(\vec{x}''-\vec{y}'')}{|\vec{x}''-\vec{y}''|} K_1(|\omega||\vec{x}''-\vec{y}''|)\Big)\nonumber\\
 &=& G_0(\vec x,\vec y,\omega)
 \ee
here: $x_1''=\frac{x_1}{v_F}$ and $x_2''=\frac{x_2}{v_\Delta}$ for the nodal pair (1,3). For the other pair, $v_F$ and $v_\Delta$ have to be exchanged. 
The final expression is obtained by inserting (\ref{greenbessel}) into (\ref{green}) and using(\ref{general}): 
\be{fingreensec}
\hat{G}(\vec{x},\vec{y},\omega)&=& \frac{1}{L_1L_2} \sum_{\vec{n}}G(\vec{n},\omega)e^{2\pi i\vec{n}(\vec{x}'-\vec{y}')}  \hat{f}(\vec{x}',\vec{y}',\vec{a})\nonumber \\
&=& \sum_{\vec m}\hat{f}(\vec x+\vec m, \vec y, \vec a)G_0(\vec x+\vec m, \vec y, \omega)\nonumber\\
&=& \frac{1}{v_fv_\Delta}\frac{1}{2\pi}\sum_{\vec{m}}\hat{f}(\vec{m})\Big(-i\omega K_0(|\omega||\vec{x}''-\vec{y}''+\vec{m}'|) \nonumber\\
& & +|\omega|\frac{\vec{\gamma}(\vec{x}''-\vec{y}''+\vec{m}'}{|\vec{x}''-\vec{y}''+\vec{m}'|}K_1(|\omega||(\vec{x}''-\vec{y}''+\vec{m}'|)\Big)
\ee

The Poisson formula  restates the problem in terms of the free space Green's functions which are much easier to compute than the ones on a torus. The sums over windings can simply be regularized by omitting the term without any winding. This term is divergent and corresponds to the free space Green's function.

\renewcommand{\theequation}{E-\arabic{equation}}
\setcounter{equation}{0}

\section{S-wave Contribution}
In several experiments there has been found evidence that the pairing in high-$T_c$ 
Superconductors may not be of pure d-wave but rather a mixture of s- and d-wave 
symmetry. The s-wave contribution prohibits zero-modes: every energy eigenvalue 
is finite, due to the constant energy gap at the Fermi surface. Moreover, the 
s-wave gap  introduces further anisotropies between the direction perpendicular 
and the one parallel to the Fermi surface. In the d-wave case, the only anisotropy 
arises from different values for the characteristic velocities $v_F$ and $v_\Delta$. 
For the mixed case, the s-wave contribution permits additional excitations 
perpendicular to the Fermi surface but none in the other. Also, the energy gap 
affects the excitations differently. To describe the situation,
I replace the gap function by: 
$\Delta(\vec k)\rightarrow \Delta(\vec k)+s(\vec k)$, assuming s-wave symmetry for 
the function $s(\vec k)$. It can also be expanded around the nodes to first order:
\be{swaveexp}
s(\vec k_j+\veq)&=& \Delta_s\left((\vec k_j+\veq)_x^2+(\vec k_j+\veq)_y^2\right)
\nonumber \\
 &=& \frac 1 4 v_sk_F+v_sq_j
\ee
For simplicity I neglect vortices, so $v_s$ and $v_\Delta$ are assumed to be real. 
After inserting (\ref{swaveexp}) into (\ref{startingpoint}), the Hamiltonian is 
given by:
\be{swaveham}
\mathcal{H}&=& \mathcal{H}_d + \Psi_j^\dagger(v_si\partial_j\alpha_2+\frac 1 4
v_sk_F\tilde{\alpha})\Psi_j
\ee
with $\tilde{\alpha}$ given by: 
\begin{displaymath}
\tilde{\alpha}=
\left(\begin{array}{cc} \sigma_1&0\\0&-\sigma_1 \end{array}\right)
\end{displaymath}
$\mathcal{H}_d$ is the pure d-wave Hamiltonian. Although this looks very 
similar to the former case, there are some new aspects to discuss.

The s-wave contribution does not effect the conservation of the nodon charge. Thus, 
for simplicity I restrict the discussion again to the nodal pair 1/3.
The most important feature is that the new Hamiltonian does not have any zero modes. 
There are four distinct energy eigenvalues:
\be{swaveenergy}
E_{1,2} &=& \pm\sqrt{(v_fq_x)^2+(v_\Delta q_y-v_sq_x+1/4v_sk_F)^2}\nonumber\\
E_{3,4} &=& \pm\sqrt{(v_fq_x)^2+(v_\Delta q_y-v_sq_x-1/4v_sk_F)^2}
\ee
each leading to a unique ground state. It is easy to see, that the energy gap acts 
differently depending on the direction of the excitations. It appears as a mass term 
for the excitations in the x-direction, analogously to the normal s-wave case. 
However, it only shifts the excitation spectrum in the y-direction.
As the gap function is constant over the Fermi surface, excitations along the 
Fermi surface see the gap only as an overall shift of energy: a chemical potential. 

I can also be shown that the chiral symmetry is partially broken. The 3-direction of 
the chiral charge is given by the difference of the number of particles at node 1 and
3. This conservation is a consequence of the formation of Cooper pairs and is not 
changed by the introduction of an additional s-wave term. However, in the d-wave case,
all three components of the chirality were equal to each other. I could have chosen 
any of these to be diagonal. The s-wave contribution picks the 3-direction and 
breaks the symmetry of the other two.

\renewcommand{\theequation}{F-\arabic{equation}}
\setcounter{equation}{0}
\section{Frequency dependence of the Green's function}
We want to extract the frequency dependence of the Green's functions after the sums over windings are performed:

$$\sum_{\vec n} G(\vec x+\vec n, \vec x_0,\omega)$$

It can be divided into two parts, which I discuss separately:
\be{startomega}
(1)&=& \sum_{\vec n} \omega K_0(\omega\sqrt{\vec n^2}) \\
(2)&=& \sum_{\vec n} \omega K_1(\omega\sqrt{\vec n^2}) 
\ee
For simplicity, I set $\vec x=\vec x_0$, assumed $\omega\ge0$ and neglected all constants. Naively, you would think that there should not arise any problems as both, (1) and (2) are well-defined, finite expressions for each winding. In addition, the Besselfunctions decay exponentially for large arguments, thus the sum is truncated for large enough winding numbers. In fact, this argument holds for any finite $\omega$. However, for small $\omega$ there are approximately $\frac{1}{\omega^2}$ terms for which the argument of the Besselfunction has to be assumed as small, that is $\omega\sqrt{\vec n^2}\le 1$. So for decreasing frequency, the number of terms which which must be summed up is growing quadratically in $\omega$. Without the alternating factors, this results in a $\frac{1}{\omega^2}$ behaviour for small $\omega$. The Green's function is ill-defined at $\omega=0$ and the contribution of the zero mode to the energy shift has to be calculated in an alternative way. In the other sectors, there is no zero mode. So physically, there should be no contribution from $\omega=0$ to the energy as there is a low energy cut-off. Due to the alternating factors, the terms in the sum can take out each other partly and destroy the $\omega \rightarrow 0$ divergence. 

As the sum cannot be calculated analytically, I approximate it by an integral. This is a good approximation, as long as the frequency $\omega$ is small and gives the correct asymptotic behaviour for $\omega\rightarrow 0$.  The each alternating factor $(-1)^n$ is approximated by $\cos(n)$ functions.  It is enough to include one of these, as the problem becomes spherically symmetric when written as an integral: 

\be{nonzeroapprox}
(1) &\rightarrow & \int d^2n \, \cos(n_x) K_0(\omega n)\nonumber\\
&=& \int_0^{2\pi} d\phi\int dn \, n\cos(n\cos(\phi))K_0(\omega n)\nonumber\\
&=& \int_0^{2\pi}d\phi \, \frac{1}{\cos^2(\phi)+\omega^2}-\frac{|\cos(\phi)|\mbox{arcsinh}(\cos(\phi)/\omega)}{\sqrt{\cos^2(\phi)+\omega^2}^3}
\ee
For the second term, one can derive a similar expression:
\be{nonzeroapprox2}
(2) &\rightarrow & \int d^2n \, \cos(n_x) K_1(\omega n)\nonumber\\
&=& \int_0^{2\pi} d\phi\int dn \, n\cos(n\cos(\phi))K_1(\omega n)\nonumber\\
&=& \int_0^{2\pi}d\phi \, \frac{\omega\pi}{\sqrt{\cos^2(\phi)+\omega^2}^3}
\ee

Although the integrals look complicated, they can be solved exactly. Their asymptotic behaviour is the same in both cases for small and large $\omega$
\be{asymptotic1}
(1),(2)&\rightarrow&\left\{ \begin{array}{cc} \frac{1}{\omega},& \omega \mbox{ small}\\
                                                \frac{1}{\omega^2},& \omega \mbox{ large}
                                                \end{array}\right.
 \ee                                               
The approximation we made becomes very bad for large frequencies. In fact, for large $\omega$ the Green's function  shows an exponential decay, instead of the algebraic one in (\ref{asymptotic1}). However, the approximation becomes better the smaller the frequencies are.  The asymptotic behaviour for small frequencies is, therefore, accurate. The Green's function approaches a finite constant in the sectors without a zero mode.  When calculating the energy, one integrates over all frequencies and inserts an additional $\omega$ in the expression (\ref{vacuum energy}). In the end, there is indeed no contribution to the energy from $\omega=0$ in absence of a zero mode. 
The same asymptotic behaviour can be expected when the zero mode is excluded from the propagator. All remaining modes are gapped and the low-energy behaviour should therefore be analogous to the non zero mode sectors. 

\be{zeromodesubtraction}
\tilde{G}(\vec x,\vec x_0,\omega)&=& \sum_{\vec n}\{G(\vec x+\vec n,\vec x_0,\omega)-\frac{1}{L_xL_y}\int d^2y \,G(\vec y+\vec n,\vec x_0,\omega)\}
\ee

As the sum is responsible for the $\omega\rightarrow0$ divergence, I neglect the first terms and focus on winding numbers where $|\vec m|\gg|\vec x|$. For large $\vec m$ the difference whether you integrate over the torus or whether you choose a specific point on the torus becomes negligible and the two terms cancel. By doing a Taylor expansion around $\vec m$, I show that the subtraction of the zero mode effectively regularizes the Green's function. In fact, it is sufficient to show that $K_0(\omega|\vec n|)$ is regularized. The properties for $K_1(\omega|\vec n|$ follow out of it.  

The first non-vanishing contribution from the Taylor expansion is second order in $\vec x$. The first derivative, $\vec{\nabla} K_0(\omega|\vec{y}+\vec n)$ as well as the mixed term in the second order $\partial_1\partial_2 K_0(\omega|\vec{y}+\vec n)$ vanish because of symmetry reasons. Therefore, the first contribution comes from 
\be{secondOrderTaylor}
&&(\partial_1^2+\partial_2^2)K_0(\omega|\vec y+\vec n)\nonumber \\
&=& \{-\frac{\omega}{|\vec n|}K_1(\omega|\vec n|) + \omega^2K_0(\omega|\vec n|)\}\frac{L_x^3+L_y^3}{L_xL_y}
\ee

Performing the sum, one obtains that the Green's function actually goes to zero for $\omega=0$ which is even more convergent than one would have expected from the beginning. This is due to the symmetry on the torus.

\end{appendix}

\end{document}